\documentclass[fleqn,usenatbib]{mnras}

\usepackage{newtxtext,newtxmath}

\usepackage[T1]{fontenc}

\DeclareRobustCommand{\VAN}[3]{#2}
\let\VANthebibliography\thebibliography
\def\thebibliography{\DeclareRobustCommand{\VAN}[3]{##3}\VANthebibliography}

\usepackage{graphicx}	
\usepackage{amsmath}	
\usepackage{amssymb}	
\usepackage{siunitx}    
\usepackage{xcolor}
\usepackage{float}
\usepackage{placeins}   
\newcommand{\vmax}{V_{\mathrm{max}}}
\newcommand{\rmax}{r_{\mathrm{max}}}

\title[Halo Densities and Pericenter Distances of the Bright Milky Way Satellites]{Halo Densities and Pericenter Distances of the Bright Milky Way Satellites as a Test of Dark Matter Physics}

\author[K. E. Andrade et al.]{
Kevin E. Andrade,$^{1}$\thanks{E-mail: kandrad1@uci.edu (KEA)}
Manoj Kaplinghat$^{1}$,
and Mauro Valli$^{2}$
\\
$^{1}$University of California, Irvine, Irvine, CA 92697, USA\\
$^{2}$ INFN Sezione di Roma, Piazzale Aldo Moro 2, I-00185 Rome, Italy \\
}

\date{Accepted XXX. Received YYY; in original form ZZZ}

\pubyear{2024}

\begin{document}
\label{firstpage}
\pagerange{\pageref{firstpage}--\pageref{lastpage}}
\maketitle

\begin{abstract}
{We provide new constraints on the dark matter halo density profile of Milky Way (MW) dwarf spheroidal galaxies (dSphs) using the phase-space distribution function (DF) method. After assessing the systematics of the approach against mock data from the Gaia Challenge project, we apply  the DF analysis to the entire kinematic sample of well-measured MW dwarf satellites for the first time. Contrary to previous findings for some of these objects, we find that the DF analysis yields results consistent with the standard Jeans analysis. In particular, in the present study we rediscover: \textit{i)} a large diversity in the inner halo densities of dSphs (bracketed by Draco and Fornax), and \textit{ii)} an anti-correlation between inner halo density and pericenter distance of the bright MW satellites. Regardless of the strength of the anti-correlation, we find that the distribution of these satellites in density vs. pericenter space is inconsistent with the results of the high-res N-body simulations that include a disk potential. Our analysis motivates further studies on the role of internal feedback and dark matter microphysics in these dSphs.}
\end{abstract}




\section{Introduction}

{
The prevailing theory of the evolution of the Universe, $\Lambda$CDM, is quite successful in predicting the large scale structures we observe today. On sub-galactic scales, discrepancies between predictions and observations begin to emerge \citep{Bullock2017,Simon2019a}.
Among the so-called ``small-scale puzzles'', the too-big-to-fail (TBTF) problem~\citep{Boylan-Kolchin2011, Kaplinghat2019} for the observed bright dwarf spheroidal galaxies (dSphs), satellites of the Milky Way (MW) has received a lot of attention.

MW dSph galaxies are dark matter (DM) dominated objects~\citep{Walker2006}, primarily dispersion-supported~\citep{Wheeler2017}, and benefit from the availability of increasingly precise stellar data thanks to instruments such as the Gaia satellite \citep{2018A&A...616A...1G}. It follows that these galaxies may represent one of the most important laboratories in order to investigate and decipher the nature of DM~\citep[for example, see recent reviews][]{Buckley:2017ijx,2022NatAs.tmp..130S,2022arXiv220710638A}.

Stars in dSphs can be typically modeled as tracers in a collisionless system. By observing their position and velocity one can draw conclusions about the nature of the underlying potential, and consequently the distribution of DM. This kind of analysis commonly employs one of three  methods~\citep{Binney2008,Strigari2018,Battaglia2022}: (a) Jeans analysis, (b) Schwarzschild modeling, or (c) phase-space distribution function (DF) modeling.

Jeans equations relate second-order velocity moments to the density and total gravitational potential of a collisionless system \citep{1915MNRAS..76...70J, Binney2008}. Under the assumption of spherical symmetry, the three Jeans equations stemming from the collisionless Boltzmann equation collapse into a single one. Despite the simplification, even the spherical Jeans equation suffers from a well-known degeneracy between mass and velocity anisotropy profile of the system \citep{Binney1982}, and several ideas have been put forward to ameliorate this issue, see, e.g.,~\citet{Binney2008, Walker2011a, Diakogiannis2014, Pace2020}. Utilizing moments of velocity higher than second order is one method of addressing this issue, see, e.g.,~\citet{Lokas2003, Richardson2014, Read2019, Kaplinghat2019}. 

The spherical Jeans equation has been a playground for a multitude of studies on dSph kinematics \citep[e.g.,]{Strigari2007, Battaglia2008, Walker2009, Zhu2016, Diak2017, Hayashi2012, Hayashi2020, Strigari2008, Evans2009}. More recently, \citet{Read2019} used the Jeans equation solver \texttt{gravsphere} and fourth-order velocity moments to examine the inner densities of MW classical dwarfs. \citet{Kaplinghat2019} also performed a spherical Jeans analysis coupled with fourth-order velocity moments to predict the inner densities of bright MW dSphs. \citet{Chang:2020rem,Guerra2021} used the Jeans approach to examine the inner halo density profile in simulated dwarf galaxies, to eventually report that only with the radial velocity data from $\mathcal{O}(10^4)$ stars could cusps and cores could be easily distinguished.

Going back to the seminal paper of \citet{Schwarzschild1979}, orbit-based models consist of integrating particle paths in a given potential in order to create an ``orbit library''. Consequently, a numerical approximation to system's phase space distribution function (DF) can be obtained as a superposition of the orbit library elements \citep{Breddels2013, Breddels2013b, Kowalczyk2017, Kowalczyk2018, Kowalczyk2019, Hagen2019, Jardel2012, Jardel2013}. Contrary to the Jeans approach, the Schwarzschild one does not require any assumption on the orbital anisotropy profile of the system, that can be a-posteriori computed without any a-priori ansatz. For this reason, the Schwarzschild modeling has been adopted to some extent to analyze mock and observed MW dSph kinematic data \citep[as examples]{VanDenBosch2008, Breddels2012, Kowalczyk2017, Jardel2012, Jardel2013, Kowalczyk2022}. The Schwarzschild method is quite general, relying essentially on the assumption of dynamical equilibrium for the system and on the geometry of the problem. Nevertheless, it remains computationally demanding once a marginalization over unknowns related to the assumed total gravitational potential has to be performed. Additionally, it has the additional drawback of yielding an approximated DF intelligible only numerically.

Differently from (a) and (b), method (c) -- the phase-space DF approach -- requires an analytic ansatz in six dimensions (3 in position, 3 in velocity) for the probability distribution of the stellar system in their DM potential. The ansatz is typically obtained exploiting Jeans’ theorem, i.e. expressing the DF via the integrals of motion. This approach allows for flexible forms for the stellar distribution, and can also allow consideration of velocity moments above second order, potentially mitigating the mass-anisotropy degeneracy of the spherical Jeans analysis. As examples, \citet{Wu2006} used the DF method to derive the mass distribution of Messier~87 using its globular clusters as tracers. More recently, regarding the case of MW dSphs \citet{Strigari2017} used an approximate DF model to examine the DM profile of the Sculptor dwarf galaxy.

Recent examples of hybrid applications of the DF approach with the Jeans equation can be found in \citet{Lokas2005, Lokas2009, Strigari2010, Breddels2013, Battaglia2013, Ferrer:2013cla, Petac:2018gue, Lacroix:2018qqh, Li2020, Li2021, Read2021}. On top of that, examples of studies utilizing multiple chemo-dynamical populations are also present in the literature, see \citet{Agnello2012, Amorisco2012, Battaglia2008, Zhu2016,Strigari2017,Pascale2018}.\footnote{Regarding this point, in this work we will not entail any metallicity distinction in the stellar population of a MW satellite.} 

The central profiles and densities of dwarf galaxies have long presented challenges to the $\Lambda$CDM model of galaxy formation~\citep{Salucci2000, Hayashi2003, Governato2010, Weinberg2015}. More recently, it has been asserted that there is an anti-correlation between the central densities of the MW dSphs and their pericenter distances \citep{Kaplinghat2019}. In this work we reexamine that relationship. Here, we apply the DF method to the kinematic data of the bright MW dSphs with the aim of providing a new, theoretically broad study of the DM content in these galaxies that is completely decoupled and, hence, complementary to the Jeans approach, along the lines of what was carried out originally in~\cite{Strigari2017}. In order to validate our modeling method, we first examine 32 mock data sets of various configurations from the Gaia Challenge project \citep{GaiaChallenge}. We then analyze the bright dSphs of the MW: Draco, Fornax, Sculptor, Carina, Sextans, Leo I, Leo II, Ursa Minor and Canes Venatici I. We show that with the DF approach it is possible to constrain the DM halo of these galaxies in the inner regions with similar precision to the one previously obtained in literature with the Jeans approach \citep{Read2019,Kaplinghat2019}. We have chosen to use the DM density at 150 pc, $\rho_{150}$, as the key metric for inner density. While not perfect, it is a good parameter for encapsulating the inner density for halos in this size range, and is at a radial position that lends itself well for inferences by stellar data. Moreover it was used by several prior authors \citep{Read2018, Read2019, Kaplinghat2019, Hayashi2020} and thus facilitates comparisons.

A key result of this work is an inference of the inner density of the bright MW dSphs using a uniform set of priors and a generalized distribution function. It serves as a test of dark matter physics (see recent work, for instance, \citet{Valli:2017ktb,Read:2018pft,Kaplinghat2019,Nadler:2019zrb,2021arXiv210609050K,DES:2020fxi,Correa2021,Slone2021,Nadler:2021dft,2023ApJ...949...67Y}) and provides constraints on dark matter models (see the recent review of \cite{2022arXiv220710638A} in this regard). {\it In particular, we show that the central densities (inferred within 150 pc) of the bright MW dSphs vary by a factor of $\sim$5, with Fornax and Carina on the low end, and Draco and Leo I on the high end.}

When the inner densities of the MW dSphs are compared to their pericenter distances, an interesting anti-correlation emerges \citep{Kaplinghat2019}. We reexamine and confirm this relation. The distribution of the MW dSphs in the density-pericenter plane appears to be in stark conflict with the result of the "Phat Elvis" N-body simulation in \citet{Kelley2019}, which examined MW-like halos with a disk potential. This is our second key result: {\it We find that the distribution of the bright MW dSphs in density-pericenter space is starkly inconsistent with high-res $\Lambda$CDM N-body simulation results.}
Solutions to the TBTF problem and, in general, all particle physics models that predict deviations on sub-galactic scales from the $\Lambda$CDM model should include information about the orbits of dSphs when looking for consistency with dark matter density inferences.

This paper is organized as follows: Section~\ref{sec: DF models} develops the theory of the DF approach and lays the foundation for our statistical analysis. In Section~\ref{sec:mock data} we present the mock data validation. Section~\ref{sec:MW dwarf models} contains the results of applying the model to the bright MW dSphs, and, in particular, constraints on $\rmax$ and $\vmax$. In Section~\ref{sec: Draco and Fornax} we zoom on Draco and Fornax as representatives of the diversity among the bright MW dSphs emerging from our DF approach. In Section~\ref{sec:Comparing to other methods} we detail our inference for all the bright dSphs of $\rmax$, $\vmax$, $\rho_{150}$, as well as mass estimates within various radii, offering also a direct comparison with the recent studies on the subject based on the Jeans approach. We examine the anti-correlation of $\rho_{150}$ and pericenter distance in Section~\ref{sec:inconsistency with simulation}. We present our conclusions in Section~\ref{sec:conclusions}. 
Further details on our analysis and interesting cross checks related to our study can be found in Appendices~\ref{app:Strigari comparison}--\ref{app:mock data info}.

\section{Distribution Function Models}
\label{sec: DF models}

Let us start by describing our approach in modeling the stellar and DM distributions in dSphs. It is possible to describe the position and velocity of stars (or other objects) in a galaxy using a phase-space DF in six dimensions, three for position and three for velocity. Our intent is to use DFs to analyze the bright dSphs of the MW, using the stars as tracers to determine the DM distribution.

We define a Cartesian coordinate system, centered on the galaxy center, with the z-axis along the line of sight to the system. The projected radius of a star as seen from the observer is then $R=\sqrt{x^2 + y^2}$. An individual star will have a position coordinate $\mathbf{x}$, given by (x, y, z). The star will have a velocity vector $\mathbf{v}$, with components ($v_\mathrm{x}, v_\mathrm{y}, v_\mathrm{z}$). We also define $\eta$ to be the angle between $\mathbf{x}$ and $\mathbf{v}$.

We can introduce the distribution function $f$ such that $f(\mathbf{x}, \mathbf{v}, t) d^3\mathbf{x}~d^3\mathbf{v}$ is the probability of finding a star in the infinitesimal volume element  $d^3\mathbf{x}~d^3\mathbf{v}$. Under the assumption of dynamical equilibrium, the DF can be regarded as constant in time, $f(\mathbf{x}, \mathbf{v})$. We require DF to be normalized to one over all phase space according to the definition of probability distribution.

Motions of particles like stars in a stationary potential can be determined by the collisionless Boltzman equation. Under the approximation of  spherical symmetry, the Strong Jeans Theorem then tells us that solutions to the collisionless Boltzman equation depend only upon two integrals of motion, the orbital energy $E$ and the total angular momentum $L$ \citep{Binney2008}. 

Given a spherically symmetric potential $\Phi$, the energy of a star per unit mass is given by $E(r,v)=\Phi(r)+v^2 /2$, and the angular momentum per unit mass corresponds to $L(r,v)=rv~sin \eta $. Several useful quantities can be derived from the DF, including the density profile, the radial velocity dispersion profile and the tangential velocity dispersion profile \citep{Binney2008, Strigari2017}:
\begin{equation}
	\mu(r)=\int d^3\textbf{v} ~f(x,v)=2\pi \int_0^\pi d\eta~ sin\eta \int_0^{v_{\mathrm{lim}}} dv~ v^2~ f(E,L)
	\label{eq: nu}
\end{equation} 
\begin{equation}
	\sigma_\mathrm{r}^2(r)=\frac{2\pi}{\mu} \int_0^{\pi} d\eta ~sin\eta~cos^2\eta \int_0^{v_{\mathrm{lim}}} dv~ f(E,L)v^4
\end{equation} 
\begin{equation}
	\sigma_\mathrm{t}^2(r)=\frac{\pi}{\mu} \int_0^{\pi} d\eta ~sin^3\eta~ \int_0^{v_{\mathrm{lim}}} dv f(E,L)v^4
\end{equation} 

We define $\mu(r)$ as the probability per unit volume of finding a star at radius r. The number density of stars at radius r is then
\begin{equation}
\label{eq: number density}
	n(r)=w~\mu(r)
\end{equation}
where $w$ is the total number of stars in the population. We define the velocity above which stars become unbound as $v_{\mathrm{lim}}=\sqrt{2(\Phi_{\mathrm{lim}}-\Phi(r))}$, where $\Phi(r)$ and $\Phi_{\mathrm{lim}}$ are given explicitly in Section~\ref{sec:potentials} for specific cases that are relevant for our analysis. The total velocity dispersion can be found by combining the radial and tangential components:
\begin{equation}
	\sigma_\mathrm{tot}^2(r)=\sigma_\mathrm{r}^2(r)+2\sigma_\mathrm{t}^2(r).
\end{equation}

The projected stellar density $\Sigma_*$ at a radius R can be found by integrating over the line of sight (LOS): 
\begin{equation}
\label{eq: surface density}
    \Sigma_*(R)=2\int_0^\infty dz ~ n(r), 
\end{equation}
where $r=z^2 + R^2$. The LOS velocity dispersion can be found from 
\begin{equation}
\label{eq: dispersion}
\begin{split}
    \Sigma_*(R) \; \sigma_{\mathrm{LOS}}^2(R) =2\int_0^\infty dz ~ n(r) \frac{z^2\sigma_\mathrm{r}^2+R^2\sigma_\mathrm{t}^2}{z^2+R^2} \\
    = 2 \pi w \int_0^{\pi}d\eta \sin{\eta} \int_0^{v_{\mathrm{lim}}}dv v^4 \\
    \centerdot \int_0^\infty dz \frac{(2z^2\cos^2\eta +R^2\sin^2 \eta)}{z^2+R^2}f(E,L)
\end{split}
\end{equation}

Higher order moments of velocity can also be predicted by this method. We will use a virial shape parameter (VSP) that is the fourth moment of velocity in our analysis. For our purposes we opt to compute the global VSP rather than one that varies with radius, which helps to minimize noise in the calculation. The derivation of the VSP is presented in Appendix~\ref{app:vsp}.

\subsection{Halo DM Profiles}
\label{sec:potentials}  

We consider here the total stellar mass of the system to be negligible in comparison to that of the DM -- a good approximation for the study of MW dSphs -- and so the stars are tracers of the DM potential but do not influence it. We will consider three potential/density profiles: ``NFW", ``cored" and ``cNFW". The NFW and cored profiles can be completely described by two parameters, while the cNFW profile has one additional parameter, the core parameter ``c". The cNFW core parameter $c \equiv r_\mathrm{c}/r_\mathrm{s}$, where $\log_{10}[r_\mathrm{c}/\mathrm{kpc}]$ is the parameter used in the model (which we distinguish from the core radius $r_\mathrm{core}$, defined below). We also use the scale radius $r_\mathrm{s}$ and scale velocity $v_\mathrm{s}$ as specifying parameters for all three profiles. The scale density $\rho_\mathrm{s}$ and the scale potential $\Phi_\mathrm{s}$ are determined  via the relation $\Phi_\mathrm{s}=v_\mathrm{s}^2 =4 \pi  G r_\mathrm{s}^2 \rho_\mathrm{s}$, where G represents Newton's gravitational constant.

Let $x\equiv r/r_\mathrm{s}$. The NFW profile density and potential pair is then
\begin{equation}
	\rho(r)=\frac{\rho _s}{x(x+1)^2} \ , 
\end{equation}	
and the corresponding gravitational potential becomes
\begin{equation}
\Phi(r) =\Phi_\mathrm{s} \left(1-\frac{\log (x+1)}{x}\right) \ .
\end{equation}
Note that $\Phi$ has been defined so that it is non-negative everywhere, with a value of zero at $r=0$, and goes to $\Phi_\mathrm{s}$  as $r\rightarrow \infty$.

 Define the peak circular velocity in a potential as $\vmax$, and the radius at which the peak occurs as $\rmax$. For the NFW profile, it can be shown that $\rmax = 2.163~r_\mathrm{s}$, and $\vmax=0.465~v_\mathrm{s}$.

The "cored" profile is a generalized Hernquist profile \citep{Hernquist1990, Zhao1996} of the form
\begin{equation}
	\rho(r)=\frac{\rho _s}{(x+1)^3 } \ , 
\end{equation}
with underlying gravitational potential
\begin{align}\nonumber
	\Phi(r) =\Phi_\mathrm{s} \frac{x \left(x+2 \right) - 2 \left(x+1 \right)  \log \left(x+1 \right) }{2x \left(x+1 \right)} \ .
\end{align} 
The potential in the cored case has a zero value at $r=0$, and goes to $\Phi_\mathrm{s}/2$  as $r\rightarrow \infty$. For the cored case, $\rmax=4.4247~r_\mathrm{s}$, and $\vmax=0.3502~v_\mathrm{s}$. 

The cNFW profile is defined as:
\begin{equation}
\label{eq:cNFW}
    \rho = \frac{\rho_\mathrm{s} }{\left(x + c\right)\left(x + 1\right)^2} \ ,
\end{equation}
 with potential being
\begin{align}\nonumber
	\Phi(r) =\frac{\Phi_\mathrm{s}}{(c-1)^2}  \left( \frac{x \left(c - 1 \right)}{\left( x +1 \right)} + \left(1 - 2c \right) \log(x+1) + c^2 \log \left(\frac{c+x}{c} \right)  \right) \, .
\end{align}
This profile reduces to the canonical NFW form for $c \rightarrow 0$, and reduces to the cored form when $c \rightarrow 1 $. The relation for conversion between $r_\mathrm{s}$ and $\rmax$ (and similarly between $v_\mathrm{s}$ and $\vmax$) becomes nonlinear but can be solved numerically. 

For all profiles, we define the core radius $r_\mathrm{core}$ as the radius at which the DM density falls to 50\% of its central value. For the NFW profile there is no core radius. For the cored profile, $r_\mathrm{core}$ is 0.26 $r_\mathrm{s}$. For the cNFW profile, the core radius is a nonlinear function of c to be computed numerically. Finally, we define $\Phi_\infty$ as the value of the potential as $r \rightarrow \infty$.

\subsection{Stellar Distribution Function Form}
We take the form of the stellar DF to be the product of an energy function and an angular momentum function, following the ansatz
\begin{equation}
    \label{eq:h}
	h(E)=\begin{cases}
		 E^a (E_\mathrm{c}^q+E^q)^{d/q} (\Phi_{\mathrm{lim}} - E)^e, &E<\Phi _{\lim } \\
		0, &E\geq\Phi _{\lim }\\
	\end{cases}
\end{equation}

\begin{equation}
    \label{eq:g}
	g(L)=\Bigg( \frac{1}{2} \Big( \big( \frac{L}{L_{\beta}} \big)  ^{\frac{b_{\mathrm{in}}}{\alpha}} + \big( \frac{L}{L_{\beta}} \big)  ^{\frac{b_{\mathrm{out}}}{\alpha}} \Big) \Bigg)^{\alpha},
\end{equation}
with $\alpha$ nonnegative for $b_{\mathrm{in}} <= b_{\mathrm{out}}$,  and $\alpha$ negative for $b_{\mathrm{in}} > b_{\mathrm{out}}$.
The total DF is their normalized product:
\begin{equation}
\label{eq:f}
    f(E,L)= n_\mathrm{f} \; h(E) \; g(L),
\end{equation}
that multiplied by $w$ yields  the total phase-space density distribution of $w$ stars.
In these equations, $n_\mathrm{f}$ is a normalizing factor that ensures that the DF integrates to unity over all phase space, i.e.,

\begin{equation}
    n_\mathrm{f} = \Big( \int h(E)  g(L) \,  d^3x~d^3v \Big) ^{-1}.
\end{equation}

The normalization factor is required so that the DF can be interpreted as a probability density for finding a particle in a given location in phase space. It is computationally expensive, because it is a multidimensional integral that must be calculated at every iteration in a Monte Carlo Markov Chain (MCMC) analysis. It might be argued that $n_\mathrm{f}$ changes little as the chain converges, so that its calculation at every iteration is unnecessary. However, we found that changes did indeed impact on the results, possibly through an impact on the shape of the prior volume, and it is therefore necessary to calculate at every iteration of the parameters. 

Note that these expressions correspond closely to those reported in \cite{Strigari2017}, except we have inserted a factor of 1/2 in the angular momentum function to ensure that the function transitions smoothly as $\alpha$ changes sign, avoiding any parametric discontinuity in our ansatz. We compare the results of \cite{Strigari2017} with ours in Appendix~\ref{app:Strigari comparison}. The parameter $\Phi_{\mathrm{lim}}$ is a limiting potential beyond which no stars exist, analogous to a tidal cutoff potential, and we define $r_{\mathrm{lim}}$ as the radius at which this cutoff occurs for a particle with zero velocity. The $e$ parameter controls the shape of the tidal cutoff. The parameters $a$ and $d$ control the log-slope of the energy response. $E_\mathrm{c}$ is a cutoff energy, below which the log-slope is approximately $a$, and above which the log-slope is approximately $a+d$. We restrict $d$ such that $d<0$.

The parameter $L_{\beta}$ characterizes the angular momentum scale, and the parameters $b_{\mathrm{in}}$ and $b_{\mathrm{out}}$ control the inner and outer log-slopes of the angular momentum function, respectively. At angular momenta $\gg L_{\beta}$, the log-slope is approximately $b_{\mathrm{out}}$, and for angular momenta $\ll L_{\beta}$ the slope is approximately $b_{\mathrm{in}}$. As a result, the parameters $b_{\mathrm{in}}$ and $b_{\mathrm{out}}$ determine the anisotropy of the system. The anisotropy parameter $\beta$ is given by
\begin{equation}
    \beta(r) = 1 - \sigma_\mathrm{t}^2(r) / \sigma_\mathrm{r}^2(r).
\end{equation}
If $b_{\mathrm{out}} \approx 0$, then $\beta \approx -b_{\mathrm{in}}/2$ for $L \ll L_{\beta}$. Similarly, if $b_{\mathrm{in}} \approx 0$, $\beta \approx -b_{\mathrm{out}}/2$ for $L \gg L_{\beta}$.  

\begin{table*}
	\centering
	\caption{Parameter limits for the top-hat priors in our MCMC analysis. Units are kpc for $r_\mathrm{s}$ and $r_\mathrm{c}$, and km/s for $v_\mathrm{s}$. The units for w are the number of stars in the population. The other parameters are dimensionless. The parameters $\tilde{E_\mathrm{c}}$ and $\tilde{\Phi}_{\mathrm{lim}}$ are made dimensionless by dividing by $\Phi_\infty$ for the distribution being used, and $\tilde{L_{\beta}}$ is made dimensionless by dividing by $r_\mathrm{s} \sqrt{\Phi_\infty}$.}
	\label{tab:parameter limits}
	\renewcommand{\arraystretch}{1.2}
	\begin{tabular}{lcccccccccccccc} 
	        & $\log_{10}(\mathrm{\frac{r_\mathrm{s}}{kpc}})$ & $\log_{10}(\mathrm{\frac{v_\mathrm{s}}{km/s}})$ & $\log_{10}(\mathrm{\frac{r_\mathrm{c}}{kpc}})$ & a & q & $\tilde{E_\mathrm{c}}$ & d & $\tilde{\Phi}_{\mathrm{lim}}$ & e & $\tilde{L_{\beta}}$ & $b_{\mathrm{in}}$ & $b_{\mathrm{out}}$ & $\alpha$ & $\log_{10}(\mathrm{\frac{w}{stars}})$ \\
		\hline
		lower limit & -2.5  & 0 & -2  & -4 & 0.1 & 0.01 & -12 & 0.01 & 0.1 & 0.01 & -10 & -10 & 0.1 & 1 \\
		upper limit & 1     & 2.5 & 1 & 5  & 25  & 1    & 0   & 1    & 10  & 1    & 10  & 10 &  10 & 7 \\
		\hline
	\end{tabular}
	\renewcommand{\arraystretch}{1.}
	
\end{table*}

\subsection{Approximate Likelihood Function}
\label{sec: approximate likelihood}
From the DF method, one can perform a statistical analysis to extract the halo parameters and constrain the DM profile based on the full likelihood function discussed in Appendix~\ref{app:full likelihood}. A significant problem with the full likelihood function is its intensive computation requirement. For each star, we are required to perform a multi-dimensional integration of our DF. In particular, for data sets with hundreds or even thousands of stars, the time to compute the normalized likelihood to perform a Monte Carlo Markov Chain (MCMC) analysis becomes computationally prohibitive. To make the model faster to calculate, we therefore employ an approximation of the full likelihood as described below.

Using the equations in Section~\ref{sec: DF models} and Appendix~\ref{app:vsp}, the DF can be used to make predictions of the radial profiles of surface density and velocity dispersion, and a prediction of the (global) VSP. We can compare these predictions to observed values from photometry (in the case of surface brightness) or from spectroscopy (in the cases of velocity dispersion and VSP). The surface density and dispersion observations use binned data, with bins at 8 to 25 radial locations, typically. The $\chi^2$ for each characteristic is calculated by comparison of the predicted points with the observed values, relative to the uncertainty in the observation:
\begin{equation}
    \chi^2=\frac{(\mathrm{data}- \mathrm{prediction})^2}{\mathrm{uncertainty}^2 }.
\end{equation}
The total $\chi^2$ is then the result of:
\begin{equation}
\label{eq: chisq components}
    \chi^2_{\mathrm{tot}}=\chi^2_{\mathrm{SD}}+\chi^2_\mathrm{disp} + \chi^2_{\mathrm{VSP}},
\end{equation}
where the subscripts refer to surface density, dispersion and virial shape parameter, respectively. We construct the log likelihood according to $\log{\mathcal{L}} =-\chi^2_\mathrm{tot} / 2$.  We perform a Bayesian analysis to derive parameter posteriors. The model employs sampling via the \textrm{emcee} package~\citep{Foreman-Mackey2019}.  Table~\ref{tab:parameter limits} shows the upper and lower parameter limits for the uniform priors adopted. 

 As described above, it is necessary to bin the data to make use of this approximation method. For surface density data, the binning is straightforward, because the uncertainty in the measurement is determined by Poisson statistics. However, for the dispersion data, the uncertainty is a combination of spectroscopic measurement uncertainty and the intrinsic random variations of velocities of the stars in each bin. As such, the binning process can make nontrivial differences in the data and resulting inferences. We discuss the binning process in detail in Appendix~\ref{app:binning}. Importantly, we found that using the logarithm of the dispersion resulted in Gaussian distributions of the binned data values, while using the dispersion itself did not. We use $\log_{10}$ velocity dispersion as the variable of interest for $\chi^2_\mathrm{disp}$. 

 To perform the multidimensional integrations, we used the \textsc{Vegas} integration routine \citep{LePage1978}, which employs adaptive importance sampling and is quite fast. We found that we had to carefully check the convergence of the integrations, as some parameter combinations would cause pathological problems. 

\subsection{Derived Parameters}
Once the parameters specifying the DM potential and the DF are inferred we can calculate distributions of surface density and velocity dispersion at a range of radii, and we can derive other quantities of interest such as the half-light radius $r_{1/2}$, the stellar orbital anisotropy $\beta$, the DM density at 150 pc $\rho_{150}$ and the DM halo mass $M_{200}$.

Since the DF model makes a smooth prediction for surface density, calculation of the half-light radius $r_{1/2}$ is relatively straightforward. The 2D half light radius $R_{1/2}$ satisfies the equation
\begin{equation}
    \label{eq:rhalf}
    \frac{\int_0^{R_{1/2}} \Sigma_*(R) R d\!R} {\int_0^{R_\mathrm{max}} \Sigma_*(R) R d\!R} =\frac{1}{2},
\end{equation}
where $R_{\mathrm{max}}$ is the radius of the outermost surface density data point. We verified that using $R_{\mathrm{max}}$ rather than an infinite limit did not have a significant effect on the result. This equation can be solved numerically; we then multiply the result by 1.33 to derive the 3D half-light radius. \cite{Wolf2010} found that the ratio of 1.33 is valid for a variety of stellar profile shapes, and we confirmed this to be a very good approximation for our own mock data sets. We also verified that for the mock data sets, the value obtained by this method was very close to the median radius of the stars in the data set. We use the photometry integration method to calculate the half-light radius posteriors directly from the density predicted by the distribution function (see Equation~\ref{eq: nu}). In what follows, we will also calculate $M(<r_{1/2})$, the mass enclosed within the half-light radius.

\section{Mock Data Modeling}
\label{sec:mock data}

Testing the model with mock data allows us to validate our approach and provides an indication of what we can reliably infer via our DF method. We use mock data from the Gaia Challenge spherical data sets \citep{GaiaChallenge}. The Gaia Challenge data were developed for the express purpose of modeling collisionless stellar systems such as dwarf galaxies. We use the spherical versions to match our modeling assumptions. Gaia Challenge employs two types of DM distributions: cuspy (“NFW”) and cored. There is also a variety of stellar and anisotropy profile configurations, as we describe below. 

\subsection{Mock Data Characteristics}

\begin{figure}
    \centering
    \caption{Typical fits to surface density (top), velocity dispersion (bottom) and VSP (bottom, inset), in this case for mock data set 15 (ID bdaO\_2677). The data is shown in red and the best fit DF solution is shown in green.}
    \label{fig:mock_typ_fits}
    \includegraphics[width=0.47\textwidth]{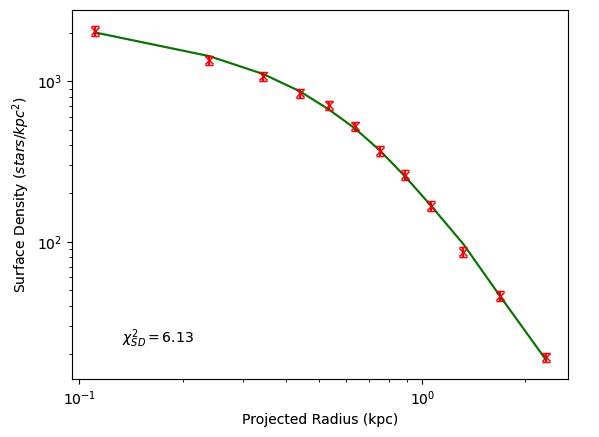} \\
    \includegraphics[width=0.47\textwidth]{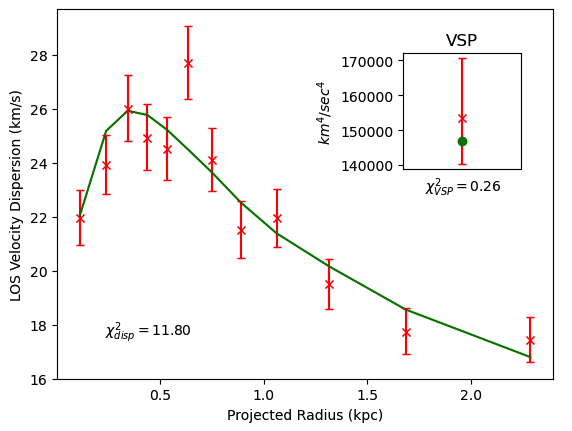}
\end{figure}

The stellar density profile in the mock data is given by a generalized Hernquist profile \citep{Hernquist1990, Zhao1996}:

\begin{equation}
	\nu_*(r) = \nu_0 \Big( \frac{r}{r_*} \Big) ^{-\gamma_*} \bigg(1 + \Big( \frac{r}{r_*} \Big)^2 \bigg)^{ \frac{(\gamma_* - 5)}{2}}
\end{equation}

The parameter $\gamma_*$ is set to 0.1 for the cored stellar profile and 1.0 for the cuspy stellar profile. The parameter $r_*$ determines how embedded the star population is placed in the DM potential, and was varied among four values: 0.1 kpc, 0.25 kpc, 0.5 kpc and 1.0 kpc.

The DM potential in the mock data is either "cored" or "NFW", as described in Section~\ref{sec:potentials}. The DM central density $\rho_0$ is also determined by this choice, with $\rho_0= \num{400e6} M_\odot \mathrm{kpc}^{-3}$ for the cored case and $\rho_0= \num{64e6} M_\odot \mathrm{kpc}^{-3}$ for the NFW case. All of the mock data sets have scale radius $r_\mathrm{s}=1$ kpc. The scale velocity $v_\mathrm{s}$ is 147.1 km/s in the cored case and 58.8 km/s in the NFW case.

\begin{figure*}
    \centering
    \caption{Posteriors for $r_{\mathrm{max}}$ vs. $\vmax$ for the 32 mock data sets. Left: 16 Cored profiles. Right: 16 NFW profiles. The 68\% and 95\% levels are shown, with the 68\% level in a darker color. The black "x" indicates the true value. The data sets are color-coded by their embeddedness in the DM halo. There are 4 sets for each value of embeddedness in each plot, sharing the same color. }
    \label{fig:mock_rmax_vmax}
    
    \includegraphics[width=0.47\textwidth]{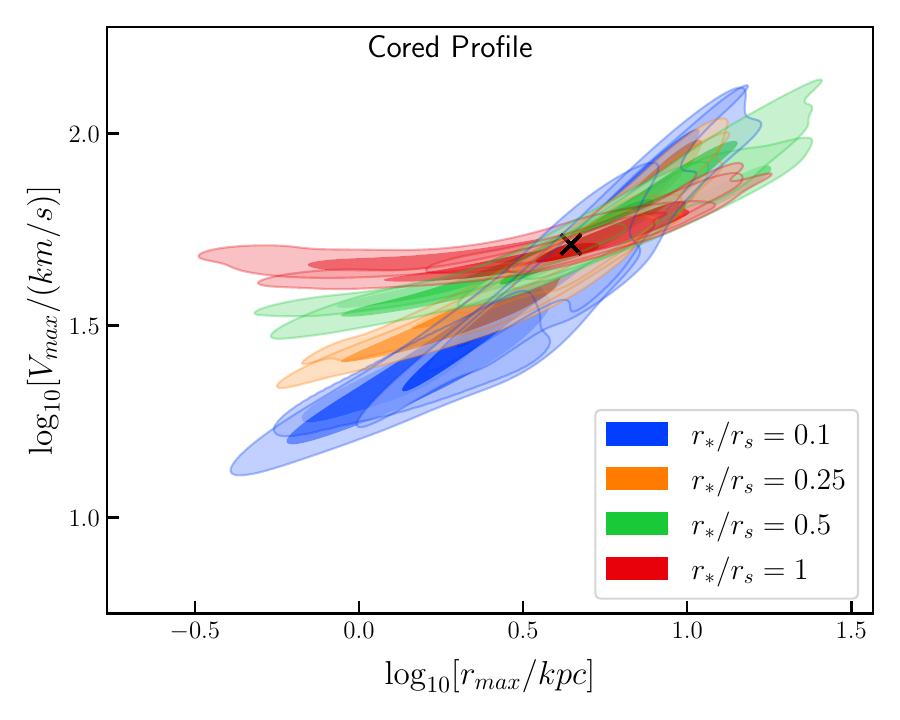}
    \includegraphics[width=0.47\textwidth]{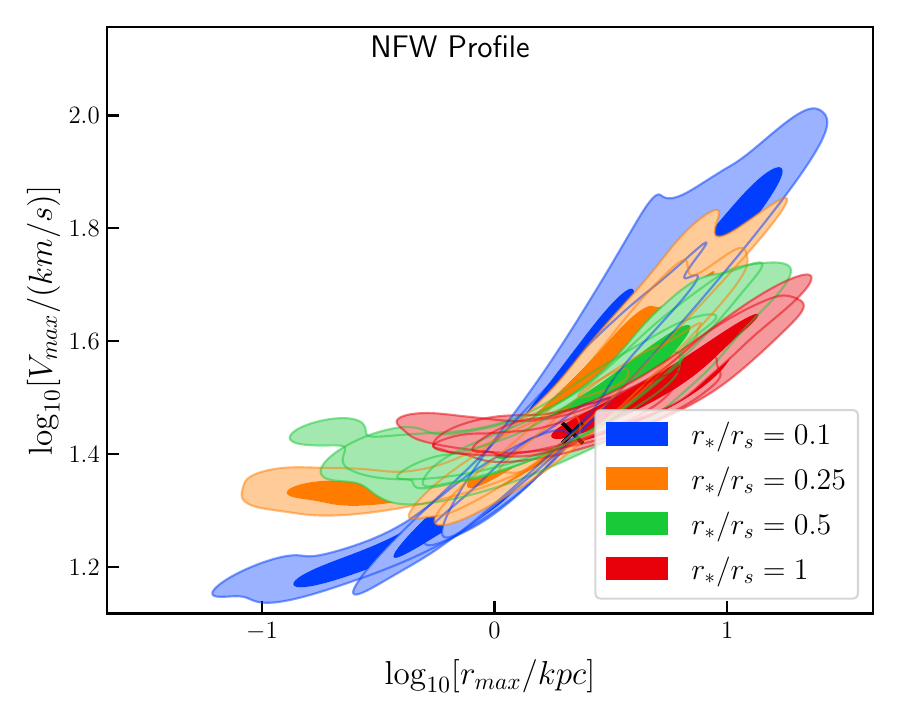}
\end{figure*}

The stellar velocity anisotropy profile is also varied among two cases. The orbital anisotropy profile is varied according to an Osipkov-Merrit form \citep{Binney2008}:
$\beta(r) = r^2 / (r^2 + r_a^2)  $, where $r_a$ is the anisotropy radius. The parameter $r_a$ takes the values of either 1 kpc or 10,000 kpc. A value of 1 kpc creates a profile in which $\beta$ rises from 0 in the center to 1 in the outer parts, reaching 0.5 at a radius of 1 kpc. A value of $r_a>10^2$ kpc creates essentially isotropic profiles with $\beta=0$ everywhere. The mock data sets therefore have $2 \times 4 \times 2 \times 2 = 32$ possible unique configurations. 

The Gaia Challenge data sets provide good model validation cases for our model, since certain key parameters are known: $r_\mathrm{s}$, $v_\mathrm{s}$ and w. The data sets contain multiple populations. We selected stars from only one population in each set, and did not include non-member foreground stars. The stars were binned into bins with equal number of stars. We found that the data sets typically had a small fraction of stars with very large orbital radii, which made the outer bins very wide and presented computational challenges. To address this, we opted to exclude the outermost stars from the data sets. Stars farther than 5 half-light radii from the center were excluded. Less than 10\% of the stars from any data set were excluded in this fashion, typically about 5\%. To simulate measurement error in the line-of-sight velocities, Gaussian error was added with a standard deviation of 2 km/s. The data set characteristics are summarized in Appendix~\ref{app:mock data info}, Table~\ref{tab:mock data characteristics}.

\subsection{Mock Data Modeling Results}

The approximate DF model was applied to the 32 mock data sets, the results of which are presented below. Since we wish to simulate that we do not have a priori knowledge of the DM profile, we used the cNFW profile in the model, in which the core size is a varying parameter. The model found very good fits to the surface density curves, disperion curves and VSP values in all cases, with $\chi^2$ per degree of freedom $\le 1.3$ for all data sets. A typical fit is shown in Figure~\ref{fig:mock_typ_fits}.

Figure~\ref{fig:mock_rmax_vmax} shows the posterior inferences in $r_{\mathrm{max}}$ versus $\vmax$ for the 32 mock data sets, with the true value shown as an "x" near the centers. We used GetDist \citep[]{Lewis2019} for two-dimensional plots. The models have a wide diversity of shapes in the $r_{\mathrm{max}}$ - $\vmax$ plane, depending on the various profiles for DM density, stellar density, anisotropy and "embeddedness" (i.e., the depth of the stars in the DM potential). The figure is color-coded by embeddedness, and shows how the embeddedness impacts the shape of the posteriors, the degeneracy characteristics between the two parameters, and the inference capability. We found that the highly embedded data sets ($r_*/r_\mathrm{s}=0.1$) were the least accurate in their inferences of $\rmax$ and $\vmax$, and that tendency carried over into inferences of many other parameters. The model made reliable inferences for the data sets with $r_*/r_\mathrm{s}>=0.25$. The reason for the difference is that the highly embedded data sets do not trace the potential near the scale radius $r_\mathrm{s}$, and so have limited accuracy in that region. 

\begin{figure}
    \centering
    \caption{Predicted and true values for the half-light radius for the 32 mock data sets, color-coded by embeddedness. The predictions are determined from the DF (see Equation~\ref{eq: nu}) by calculating and integrating the surface density curve, finding the radius that yields have the total value (see Equation~\ref{eq:rhalf}). The error bars indicate the 68\% confidence interval. The true value is taken to be the median radius of the stars in the given data set.}
    \label{fig:mock_rh_diag}
    \includegraphics[width=0.47\textwidth]{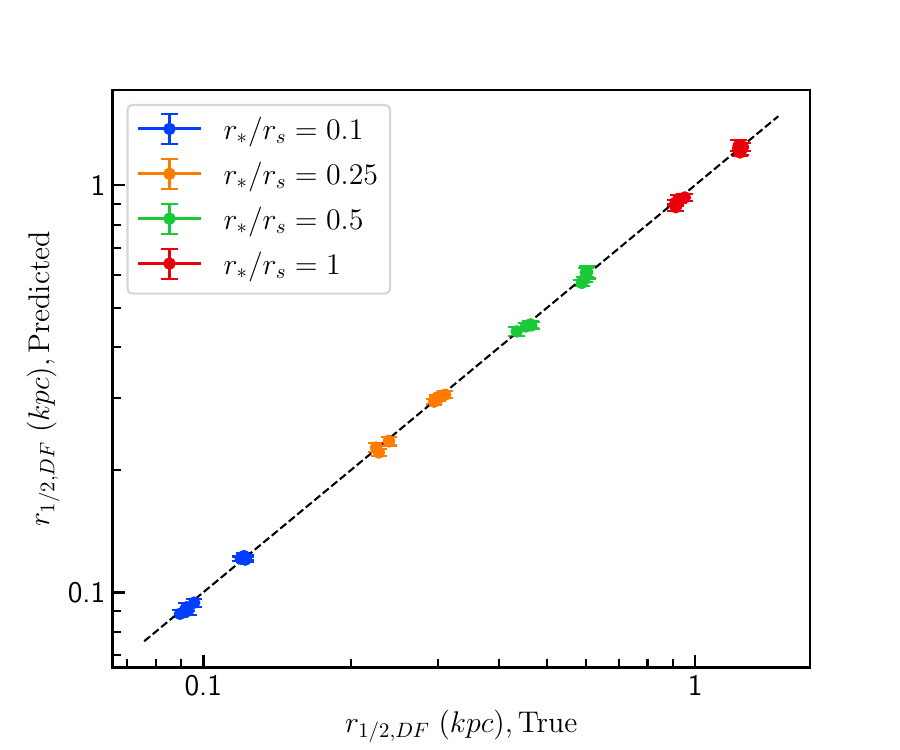}
\end{figure}

 Figure~\ref{fig:mock_rh_diag} compares the posterior for the calculated half-light radius to the true value, which is taken to be the median radius of the stars in the data set. The accuracy is very good, with a difference of less than 2\% between the median prediction and the true value for all data sets.

\begin{figure}
    \centering
    \caption{Predicted and true values for the mass within the half-light radius for the 32 mock data sets, color-coded by their embeddedness in the DM halo. The error bars indicate the 68\% confidence interval. The dashed diagonal line indicates equality, with dotted lines indicating $\pm 0.1$ dex.}
    \label{fig:mock_mrhalf_diag}
    \includegraphics[width=0.47\textwidth]{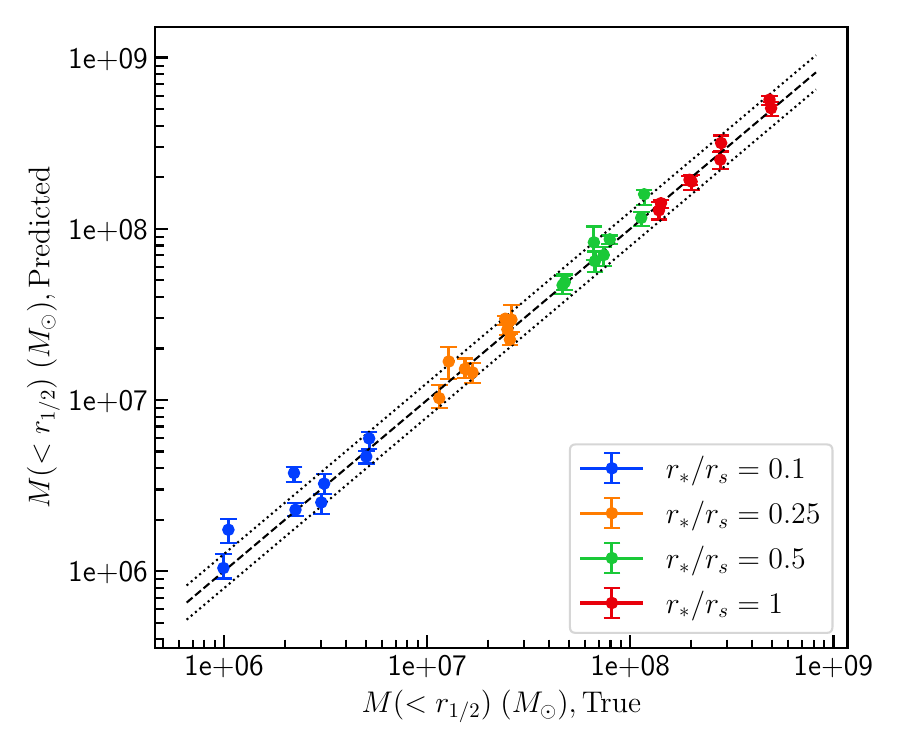}
\end{figure}

The mass within the inferred half-light radius can be determined for the cNFW profile by using the posterior values for $r_\mathrm{s}$, $v_\mathrm{s}$ and $r_\mathrm{c}$. Figure~\ref{fig:mock_mrhalf_diag} shows the true and predicted values for the mass within the half-light radius for the mock data sets. The predictions are fairly accurate for the data sets with $r_*/r_\mathrm{s} \ge 0.25$, i.e., those not deeply embedded in the DM potential. For the data sets with the lowest mass enclosed (and correspondingly very deeply embedded in the DM halo), the model tends to systematically overestimate the mass enclosed.  

\begin{figure}
    \centering
    \caption{Predicted and true values for DM density at 150 pc for the 32 mock data sets, color-coded by embeddedness. The error bars indicate the 95\% confidence interval. The labels for each data set are shown on the left and correspond to those in Table~\ref{tab:mock data characteristics}}
    \label{fig:mock_rho150_err_list}
    \includegraphics[width=0.47\textwidth]{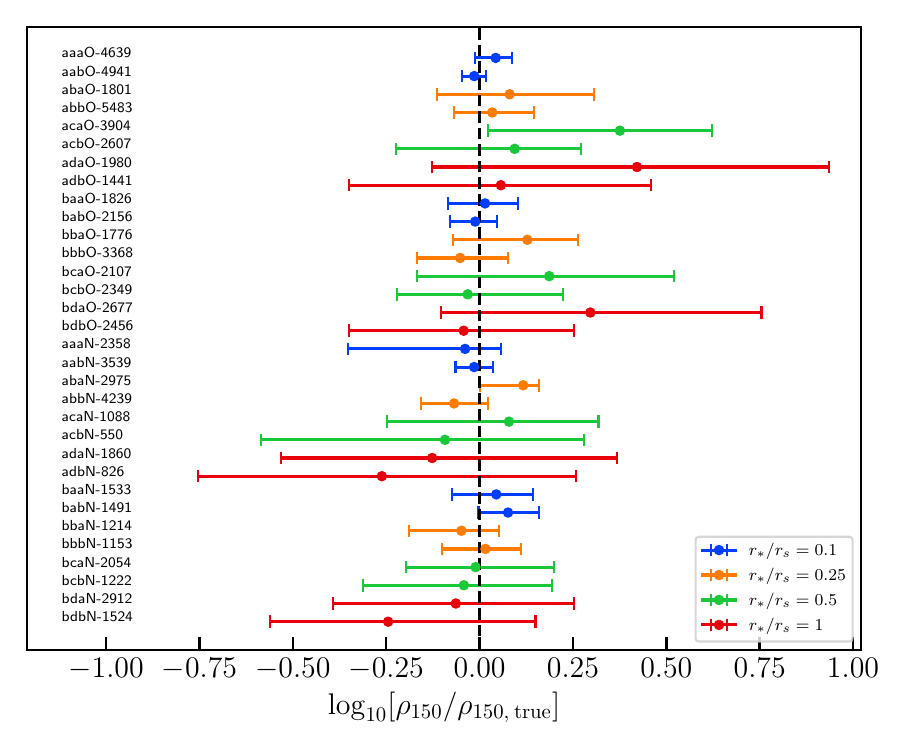}
\end{figure}

Predictions for the density at 150 pc as compared to their true values are shown in Figure~\ref{fig:mock_rho150_err_list}. The median predictions are generally within 0.3 dex of the true value, with one case near 0.5 dex. In three cases the true values were outside the 95\% confidence level of the posterior, all of which were over-estimations of the density.

In Appendix~\ref{app:core radii} we show the details for the inference of the core radius for the mock data set, with comparison to the inferences of the observed dSphs. In Appendix~\ref{app:anisotropy at rhalf} we provide details of the models prediction performance for the anisotropy parameter $\beta$ at the half-light radius.

\subsection{Summary of Model Performance with Mock Data}

The approximate DF model makes accurate predictions in the $\rmax$ - $\vmax$ plane and for half-light radius of the data sets (Figures~\ref{fig:mock_rmax_vmax} and~\ref{fig:mock_rh_diag}, respectively). The mass within the half-light radius is predicted well for those data sets that are not too deeply embedded in the DM potential. For the highly embedded data sets, there is a modest tendency to overestimate the mass (Figure~\ref{fig:mock_mrhalf_diag}). The density at 150 pc ($\rho_{150}$) is accurate to within 0.5 dex in all cases, and within 0.3 dex in most cases (Figure~\ref{fig:mock_rho150_err_list}). 

The model shows some ability to distinguish between NFW and cored profiles, which is evident in Figure~\ref{fig:core radii}. The key difference lies in how sharply the posterior gets small at small values of core radius. Our mock data analyses reveals that if the posterior is peaked in core radius, then it is likely a sign of a non-zero core radius if the stars are not too deeply embedded. For deeply embedded stellar profiles, it seems difficult to make this distinction. We also looked at the the predictions for $\beta(r_{1/2})$ and found them to be of limited accuracy. The inferences become progressively less robust for cases that are deeply embedded and that have rising $\beta$ profiles (Figure~\ref{fig:mock_beta_rhalf_diag}).

\begin{table*}
	\centering
	\caption{The Dwarf Galaxy Sample. Adopted distance, 2D half-light radius $R_{1/2}$ and V-band magnitude $M_V$ are from \citet{Simon2019a}. Ellipticity $\epsilon$ and position angle $\theta$ are from \citet{Munoz2018}. Center coordinates are from The NASA/IPAC Extragalactic Database (NED). Pericenter distances are from the Light MW model (no LMC) of \citet{Battaglia2022b}. Stellar masses $M_*$ are from  \citet{McConnachie2012}.}
	\label{tab:observed_dSph_list}
	\renewcommand{\arraystretch}{1.4}
	\begin{tabular}{lccccccccc} 
		\hline
                       &  Adopted  & Center & Center &    &     \\
                       & Distance & RA & DEC & $R_{1/2}$ &  $\epsilon$ & $\theta$ & Pericenter  & $M_V$ & $M_*$  \\
         Name          & (kpc) &    (deg.) & (deg.) & (pc) &  & (deg.) & (kpc) &  & ($10^6 \; M_{\sun} $)   \\
        \hline
        Draco              & 82.0  & 260.051625  & 57.915361  & 237 $\pm$ 17 & 0.29 & 87 & $51.7_{-6.1}^{+4.1}$ & $-8.88_{-0.05}^{+0.05}$ & 0.29  \\
        Fornax             & 139.0  & 39.997200 &  -34.449187  & 792 $\pm$ 18 &  0.29 & 45 & $89_{-26}^{+31}$ & $-13.34_{-0.14}^{+0.14}$ & 20  \\
        Carina             &  106.0 & 100.402888  & -50.966196  & 311 $\pm$ 15 & 0.36 & 60  & $106.7_{-5.4}^{+6.4}$ & $-9.45_{-0.05}^{+0.05}$ & 0.38   \\
        CnV I         & 211.0   & 202.014583  &  33.555833 & 437 $\pm$ 18 & 0.44 & 80  & $68.09_{-42.17}^{+71.49}$ &  $-8.73_{-0.06}^{+0.06}$ & 0.23  \\
        Leo I              & 254.0  & 152.117083  & 12.306389  & 270 $\pm$ 17 & 0.30 & 78  & $46.53_{-26.54}^{+30.50}$  &  $-11.78_{-0.28}^{+0.28}$ & 5.5  \\
        Leo II             & 233.0  & 168.370000  & 22.151667  & 171 $\pm$ 10 & 0.07 & 38  & $115.55_{-58.87}^{+88.35}$ & $-9.74_{-0.04}^{+0.04}$ & 0.74  \\
        Sculptor           & 86.0 & 15.038984  & -33.709029	  & 279 $\pm$ 16 & 0.33 & 92 & $63.7_{-3.1}^{+4.5}$ & $-10.82_{-0.14}^{+0.14}$ & 2.3  \\
        Sextans            & 95.0  & 153.262319  & -1.614602  & 456 $\pm$ 15 & 0.30 & 57  & $74.45_{-5.68}^{+4.38}$ &  $-8.94_{-0.06}^{+0.06}$ & 0.44  \\
        Ursa Minor         & 76.0  & 227.285379 & 67.222605   & 405 $\pm$ 21 & 0.55 & 50  & $48.9_{-3.3}^{+3.4}$ & $-9.03_{-0.05}^{+0.05}$ & 0.29   \\
        
		\hline
	
	\end{tabular}
	\renewcommand{\arraystretch}{1.}
\end{table*}

\begin{figure*}
    \centering
    \caption{Constraints on $r_\mathrm{max}$ and $V_\mathrm{max}$ from the $\chi^2$ components of Surface Density, Velocity Dispersion and VSP. {\it Top Panel:} Draco. {\it Bottom Panel:} Fornax. The black contour lines indicate the 68\% and 95\% confidence levels. The first column shows the posterior distrubtion when only surface density is used in $\chi^2$. The second column corresponds to only using velocity dispersion in $\chi^2$. The third column corresponds to using surface density and velocity dispersion, but not the virial shape parameter (VSP). The fourth column corresponds to using all three components in the calculation of $\chi^2$.}
    \label{fig: Draco and Fornax chisq components}
    \includegraphics[width=0.98\textwidth]{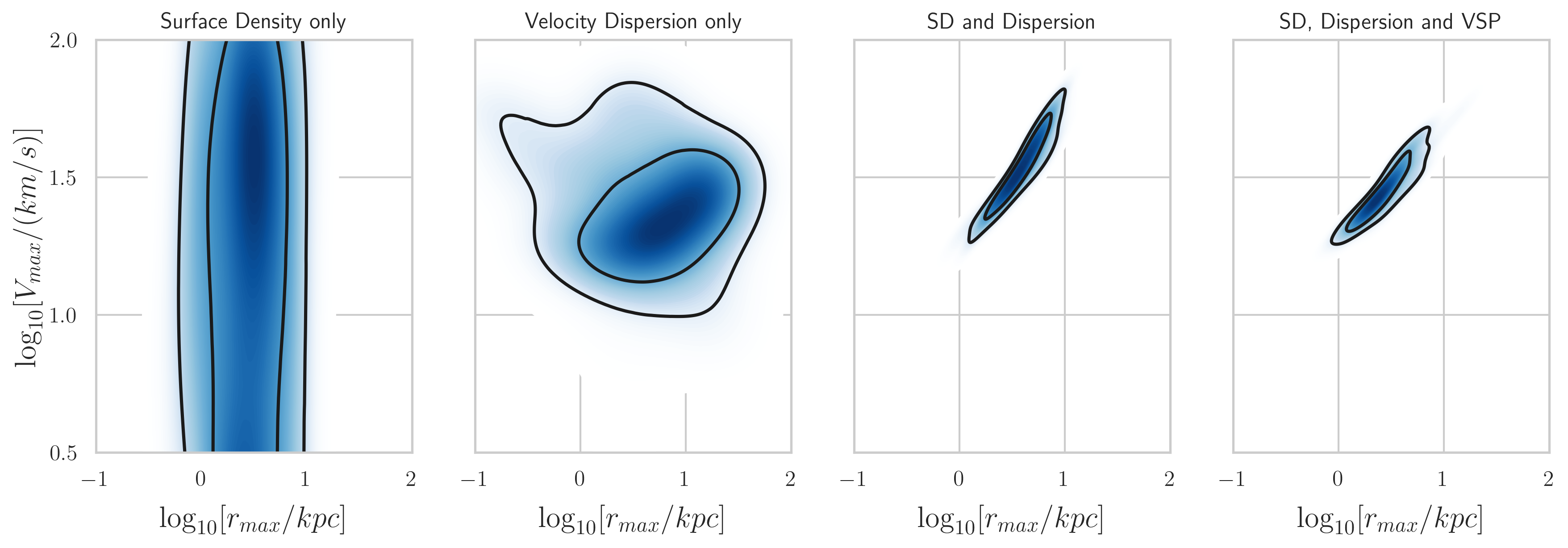}
    \includegraphics[width=0.98\textwidth]{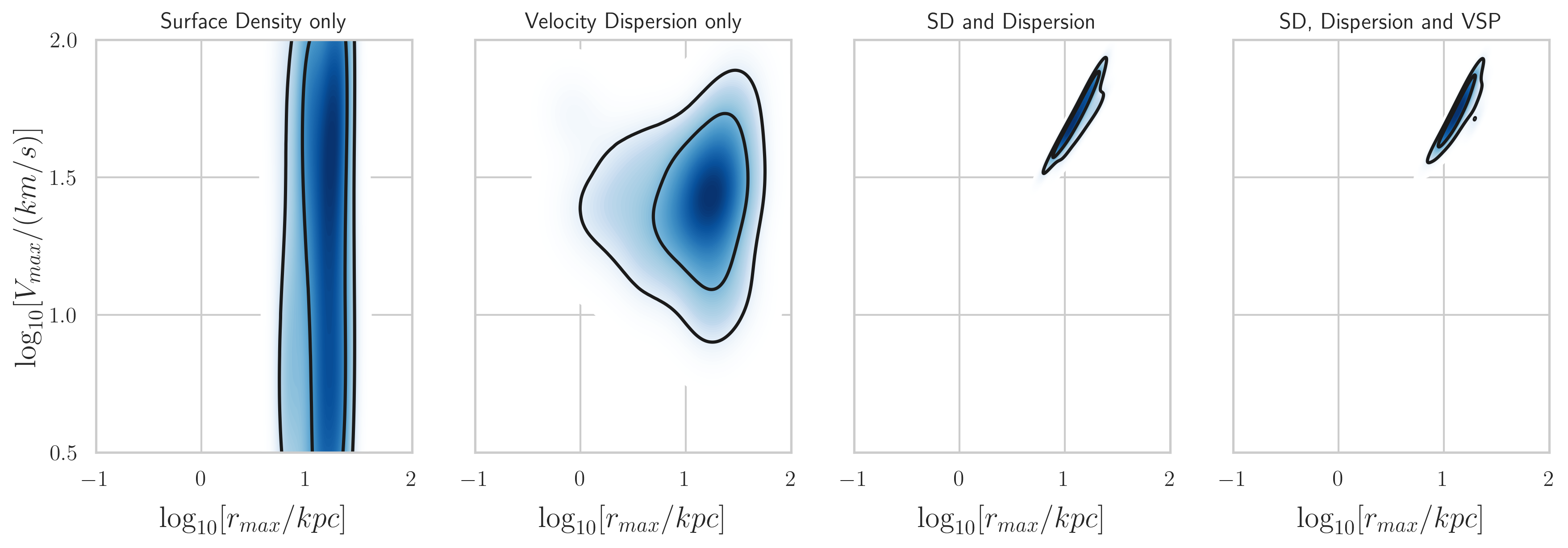}
\end{figure*}

\section{Bright Dwarf Spheroidal Models: Constraints on the halo parameters}
\label{sec:MW dwarf models}

We selected as our sample the eight classical dSphs of the MW, plus Canes Venatici I, as shown in Table~\ref{tab:observed_dSph_list}. The results from applying the DF model are described here. We use cNFW as the DM profile, as it is the most general of our profiles. For the distance to each target we adopt the median value of the distance shown in the second column of Table~\ref{tab:observed_dSph_list}. We use surface density data from \citet{Munoz2018}. Dispersion data is from \citet{Walker2009, Walker2015, Spencer2017, Mateo2008} and M. Walker, private communication. VSP data is from \citet{Kaplinghat2019}. The results from the analysis are discussed here.

\subsection{How Surface Density and Velocity Data Constrain \boldmath$r_\mathrm{max}$, \boldmath$V_{\mathrm{max}}$ and DM Density}
\label{sec:How Constraints}

Here we ask: How do the various components of the data set put constraints on key parameters such as $r_\mathrm{max}$, $V_{\mathrm{max}}$, and (indirectly) the DM density $\rho_{150}$? The parameters $r_\mathrm{max}$ and $V_{\mathrm{max}}$ are related in a straightforward way to the scale radius $r_\mathrm{s}$ and the velocity scale $v_\mathrm{s}$, so let us turn our attention to these. The prediction for surface density is given by Equations~\ref{eq: nu}, \ref{eq: number density}, and \ref{eq: surface density}, which in turn depends on the potential, which is defined in terms of $v_\mathrm{s}$. Therefore, at first blush, surface density appears to depend intimately on $v_\mathrm{s}$. However, it can be demonstrated numerically that there is very little dependence. This can be explained as follows. Assume for the moment that a particle is not near the tidal limit, so we can ignore the term $(\Phi_{\mathrm{lim}} - E)^e$ in Equation~\ref{eq:h}. Note that the energy of a particle is given by $E(r,v) = \Phi(r) + \frac{1}{2}v^2$. For stars near the center of the galaxy, the second term is dominates, and $E \propto v^2$, independent of $v_\mathrm{s}$. For stars far from the center (but not near the tidal limit), the potential term dominates, and $E \propto \Phi(r) \propto v_\mathrm{s}^2$. The energy term of the DF is given by Equation~\ref{eq:h}. If $E \ll \Phi_{\mathrm{lim}}$, then $h(E) \propto E^p$, where the exponent p takes a value $p \approx a$ for small energies, with $p \approx a+d$ at large energies. Since the star is far from the center, its potential energy will be large and the star will likely be in the region $p \approx a+d$. Recall that d must be negative, and in fact all the mock data and observed dSph models prefer solutions with $(a+d) < 0$. This then gives the energy function $h(E) \propto v_\mathrm{s}^{2(a+d)} + \frac{1}{2}v^2$. The negative exponent in the first term causes that term to be small compared to the second, and again the result is insensitive to $v_\mathrm{s}$. If a particle is near the tidal limit, the term $(\Phi_{\mathrm{lim}} - E)^e$ will be small by definition, and there will be very few stars in that area of parameter space.

\begin{figure*}
    \centering
    \caption{The $r_\mathrm{max}, V_\mathrm{max}$ plane for Draco ({\it top row}) and Fornax ({\it bottom row}). The black contour lines indicate the 68\% and 95\% confidence levels. {\it Left Column}: $\rho_{150}$ shown in color. {\it Right Column}: $M(<r_{1/2})$ shown in color.  Lines of constant $\rho_{150}$ and $M(<r_{1/2})$ are roughly parallel to the long axis of the posterior, allowing relatively strong constraints on both parameters. }
    \label{fig: Draco and Fornax rmax vmax 3D}
    \includegraphics[width=0.44\textwidth]{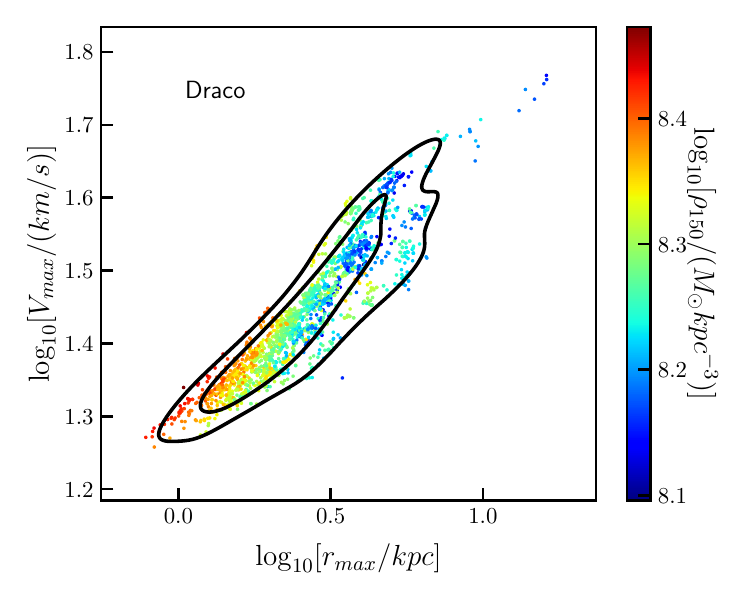}
    \includegraphics[width=0.44\textwidth]{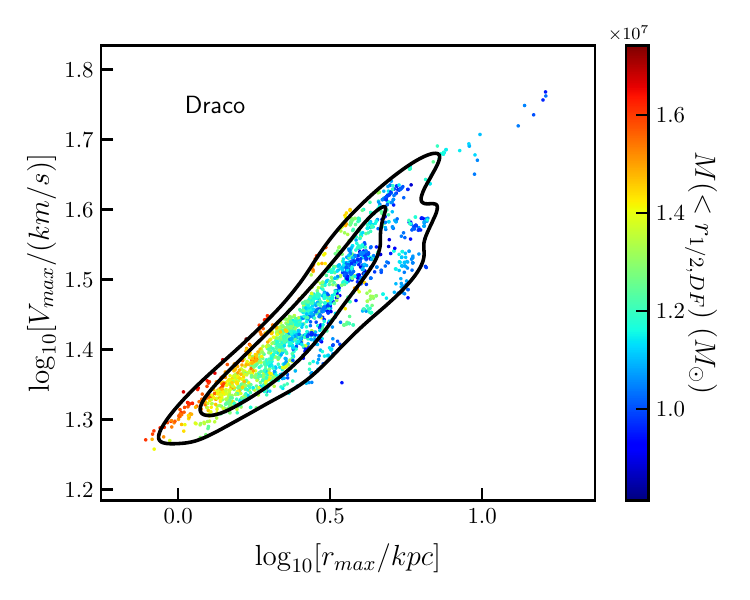} \\
    \includegraphics[width=0.44\textwidth]{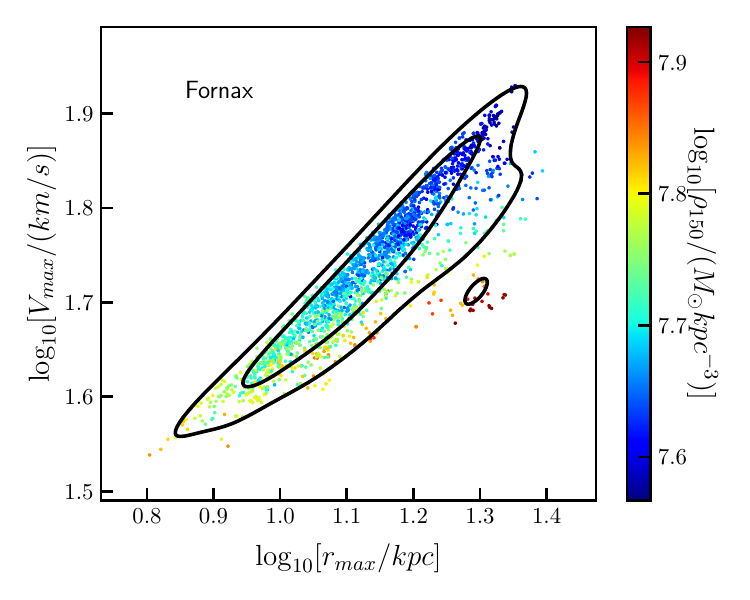}
    \includegraphics[width=0.44\textwidth]{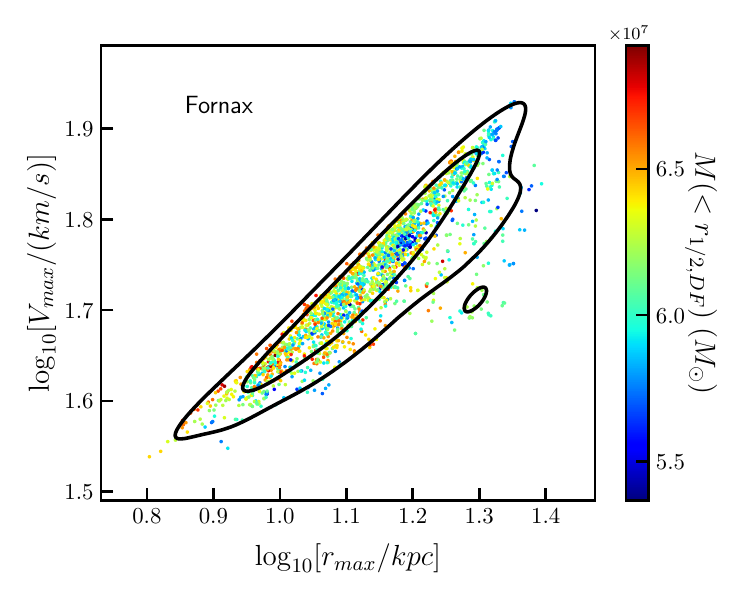}
\end{figure*}

To illustrate the constraining power of the various $\chi^2$ components of the DF model (ref. Equation~\ref{eq: chisq components}) we examine the results of the Draco and Fornax dwarfs as typical examples. Figure~\ref{fig: Draco and Fornax chisq components} shows how the three components of $\chi^2$ put restrictions on $r_\mathrm{max}$ and $V_\mathrm{max}$ for those dSphs. The surface density data strongly constrains $r_\mathrm{max}$ but has virtually no constraining power for $V_\mathrm{max}$, as expected from the above discussion. This also matches the intuitive notion that without stellar velocity information it is difficult to characterize the velocity scale of the DM potential. The velocity dispersion grossly constrains both $r_\mathrm{max}$ and $V_\mathrm{max}$, but it is the combination of surface density and dispersion data that results in a tight constraint in the ($r_\mathrm{max},V_\mathrm{max}$) plane. The fourth-order moment (VSP) adds a modest additional constraint (see also Figure~\ref{fig:obs_rmax_vmax_grid}). The constraint features illustrated here for Draco and Fornax are very similar for the other dSphs as well.

It is also interesting to examine how the constraints on $r_\mathrm{max}$ and $V_\mathrm{max}$ translate to $\rho_{150}$ (the DM density at 150 pc) and $M(<r_{1/2})$ (the mass within the half-light radius). Figure~\ref{fig: Draco and Fornax rmax vmax 3D} shows the dependence of those parameters on $r_\mathrm{max}$ and $V_\mathrm{max}$ for the Draco and Fornax dSphs. It illustrates that the lines of constant $\rho_{150}$ and $M(<r_{1/2})$ for these models tend to run parallel to the long axis of the posterior, which allows a strong constraint on those parameters even given the wide range of possible solutions in the ($r_\mathrm{max}, V_\mathrm{max}$) plane.

\subsection{Inference of Mass within Key Radii: \\Comparisons with Dispersion-Based Mass Estimators}
\label{sec: mhalf and mass estimators}

It is informative to examine the inferences for the mass of the dSphs enclosed within key radii, as such inferences can be readily compared with dispersion-based estimators. These key radii are the half-light radius ($r_{1/2}$) and $\mathcal{O}(1)$ multiples of it, which are good places to measure the mass and density of DM, since there is usually good luminosity and dispersion data there, and the inferred density can tell us something about the cores of the subject halos. \citet{Wolf2010} used the luminosity-weighted LOS velocity dispersion to derive an estimate for the mass within the half-light radius ($r_{1/2}$) that was relatively immune to the mass-anisotropy degeneracy problem. Other authors followed suit, notably \citet{Errani2018}, who found that the mass enclosed within $1.8 \; r_{1/2}$ was even better insulated from mass-anisotropy fluctuations. Note that as described more fully in Section \ref{sec: half light radius observed}, we use the spherical radius, and therefore convert the results of other authors from elliptical radius to its sphericalized equivalent.

In Figure~\ref{fig:Errani and Wolf comparison} we compare the mass enclosed within $1.8 \; r_{1/2}$, corresponding to the mass estimator of \citet{Errani2018}, and also the mass enclosed within $r_{1/2}$, corresponding to the mass estimator of \citet{Wolf2010}, for the observed dwarfs. The DF method predicts masses that are fairly consistent with those predicted by the mass estimator methods. The only substantial disagreement is for the Fornax dSph, where our inference of $M(<1.8 \; r_{1/2})$ is somewhat higher than that derived by \citet{Errani2018}, although our inference for $M(<r_{1/2})$ is consistent with that of \citet{Wolf2010}.

\begin{figure}
    \centering
    \caption{Comparison of mass estimators from \citet{Errani2018} and \citet{Wolf2010}, which utilize luminosity-weighted velocity dispersion, with the results of this work. The result of the DF method from this work are shown in black. \textit{Top:} $\log_{10}[M(<1.8 r_{1/2}) / M_{\odot}]$, which is the mass estimator of \citet{Errani2018}, shown in red.  \textit{Bottom:} $\log_{10}[M(<r_{1/2}) / M_{\odot}]$, which is the mass estimator of \citet{Wolf2010}, shown in red.}
    \label{fig:Errani and Wolf comparison}
    \includegraphics[width=0.47\textwidth]{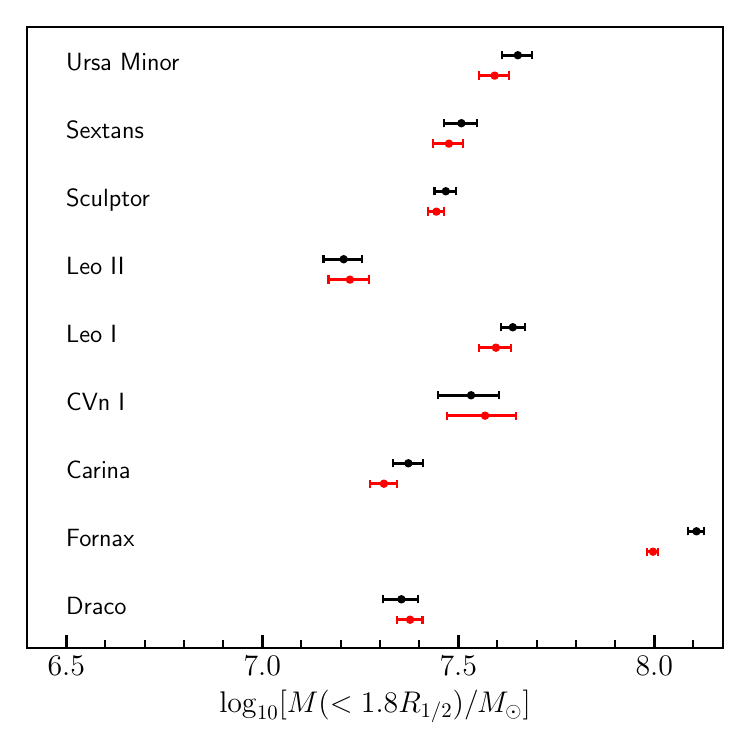}
    \includegraphics[width=0.47\textwidth]{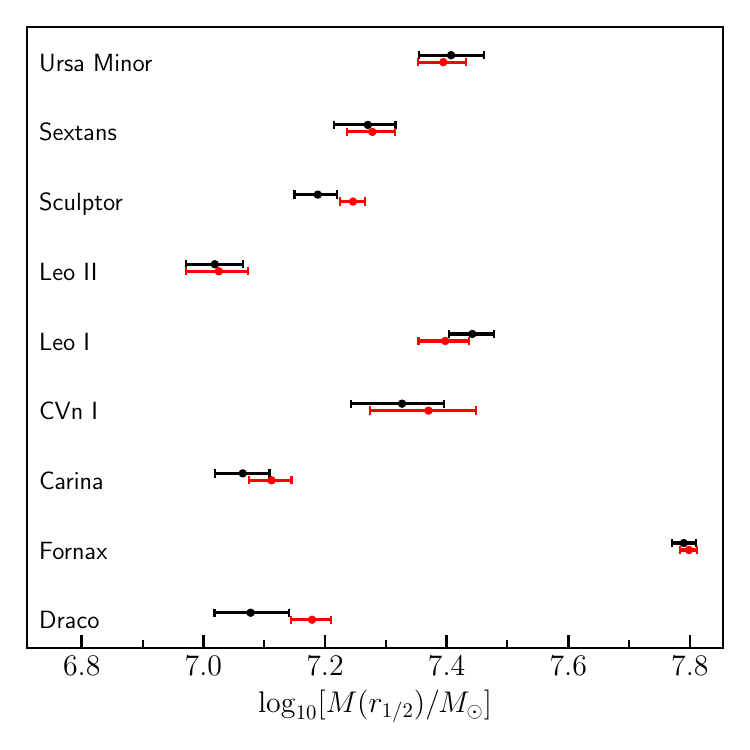}
\end{figure}

\subsection{Inferences for \boldmath$\rmax$ and \boldmath$\vmax$}

\begin{figure*}
    \centering
    \caption{Posterior inferences for $r_{\mathrm{max}}$ vs. $\vmax$ for the observed sample. The 68\% and 95\% levels are shown, with the 68\% level in a darker color. The dotted black lines indicate the posterior result without the VSP $\chi^2$ component. The black triangles represent the 10 subhalos from the Phat Elvis simulation (halo 1107, with disk) that are more than 50 kpc from the center of the halo and have the highest $\vmax$. The grey triangles show the subhalos with 10th through 20th highest $\vmax$.}
    \label{fig:obs_rmax_vmax_grid}
    \includegraphics[width=0.99\textwidth]{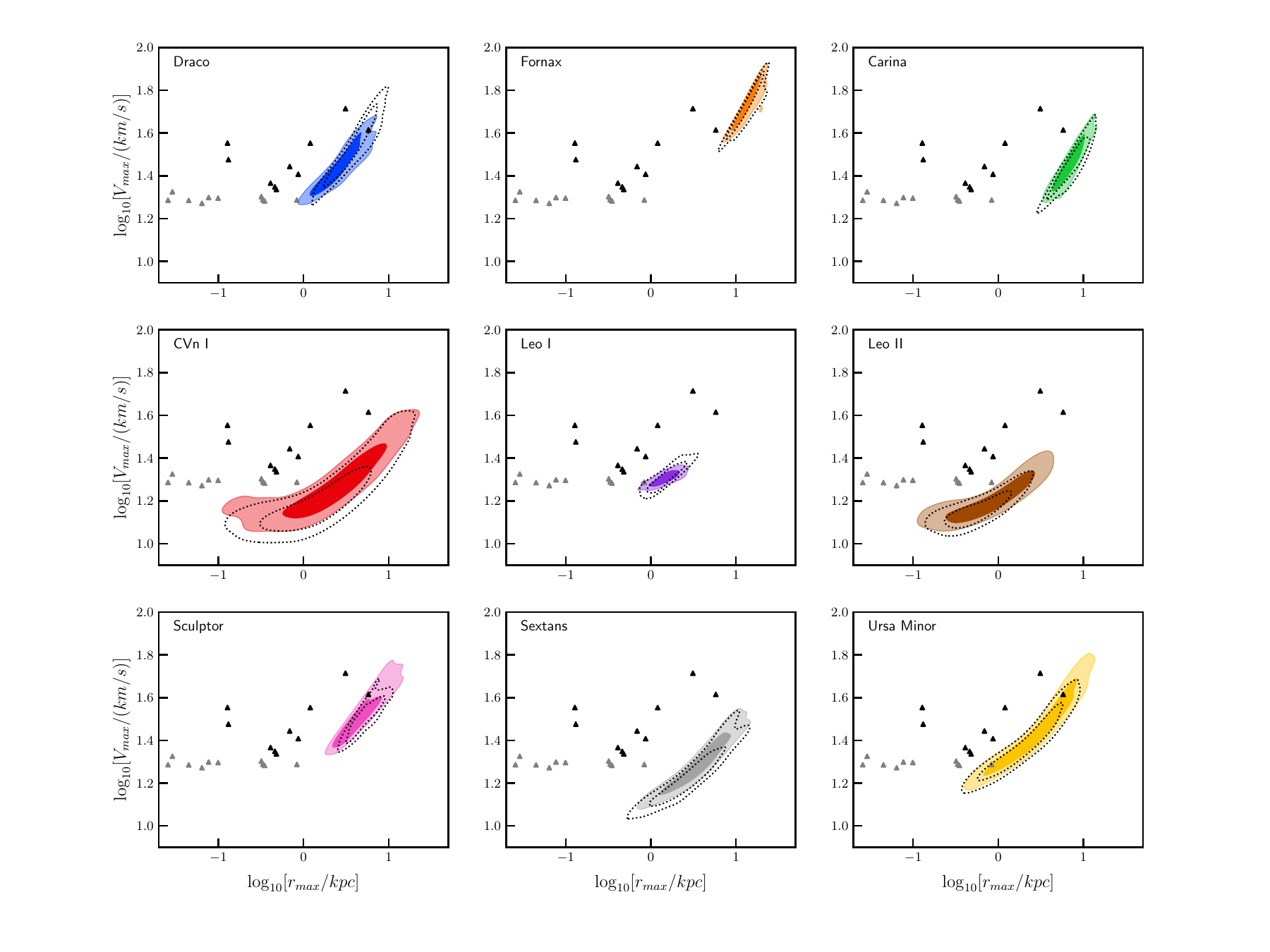}
\end{figure*}

Predictions for $r_{\mathrm{max}}$ and $\vmax$ for the observed sample are presented in Figure~\ref{fig:obs_rmax_vmax_grid}. The two parameters show strong positive correlation. Because the halo scale density $\rho_\mathrm{s} \propto r_\mathrm{s}^2 / v_\mathrm{s}^2$, this type of degeneracy is approximately along lines of constant density, so that the density is relatively well constrained, as discussed previously. To demonstrate the effect of the VSP, the posteriors that result from excluding the VSP component in the analysis are shown in the figure with dotted black lines. The VSP does indeed add some predictive power, making the posteriors somewhat smaller and in some cases shifting them modestly.
 
In Figure~\ref{fig:obs_rmax_vmax_grid}, the black triangles show the 10 most massive subhalos from the fiducial Phat Elvis halo (1107, which has a halo mass of $\num{8.9e11}M_\odot$) and restricted to those subhalos that are more than 50 kpc from the center.  We chose this as our fiducial halo because it is closest in mass to the light MW model used by \citet{Battaglia2022b}.

If we use a more massive Phat Elvis halo for fiducial comparison, the triangles of the 10 largest subhalos will tend to shift upward and to the right (i.e., they will have larger $\rmax$ and $\vmax$). The TBTF problem then becomes even more pronounced, i.e, the simulation predicts a large number of dense and massive subhalos, inconsistent with what is seen in the MW. Moreover, such a choice of fiducial halo mass for the Phat Elvis simulation would be inconsistent with the MW models used by \citet{Battaglia2022b} in our analysis. The simulated Vmax values would be systematically larger than those we infer for the MW satellites. (Comparing the $\rmax$ and $\vmax$ inferences for the bright MW dwarfs to those of all of the subhalos in the Phat Elvis suite of simulations, the results are similar: they are generally consistent in $\vmax$, but the inferences for the $\rmax$ of the bright MW dwarfs are generally higher than those in Phat Elvis.)

 Since our sample represents the brightest 9 MW dSphs, one would expect these to be of comparable Vmax to those in the Phat Elvis simulation. This is generally true; the posteriors for Fornax and Sculptor are centered near the top of the range and indeed extend above the top. The $\vmax$ posteriors for Draco, Carina and Ursa Minor straddle the middle range, while the others are closer to the bottom and indeed extend beyond the lowest $\vmax$ of the 10 most massive subhalos. In contrast, the $\rmax$ posteriors for many of the 9 bright MW dSphs seem to be systematically at larger values than those of the Phat Elvis subhalos, especially Draco, Fornax, Carina and Sculptor. This might be expected if the halos are cored, as may be indicated for Fornax and perhaps Carina. The posteriors of the MW dSphs all have the familiar degeneracy between $\rmax$ and $\vmax$ (i.e., they are positively correlated), very similar to that observed in the mock data in Figure 2. We note that, in the mock data tests, there was no systematic overprediction of $\rmax$. This also presents itself as a generally lower central density inference of the subhalos in the sample as compared to the simulated subhalos, as can be seen in Section~\ref{sec:inconsistency with simulation}.

\subsection{Half-Light Radius of the Observed Sample}
\label{sec: half light radius observed}
\begin{figure}
    \centering
    \caption{Posteriors for the half-light radius and mass within that radius for the observed sample. The 68\% and 95\% levels are shown, with the 68\% level in a darker color. Isodensity contours are shown as dotted lines, with the density value indicated, in units of $10^7 M_\odot \; \mathrm{kpc}^{-3}$.}
    \label{fig:obs_rh_mrh_2D}
    \includegraphics[width=0.47\textwidth]{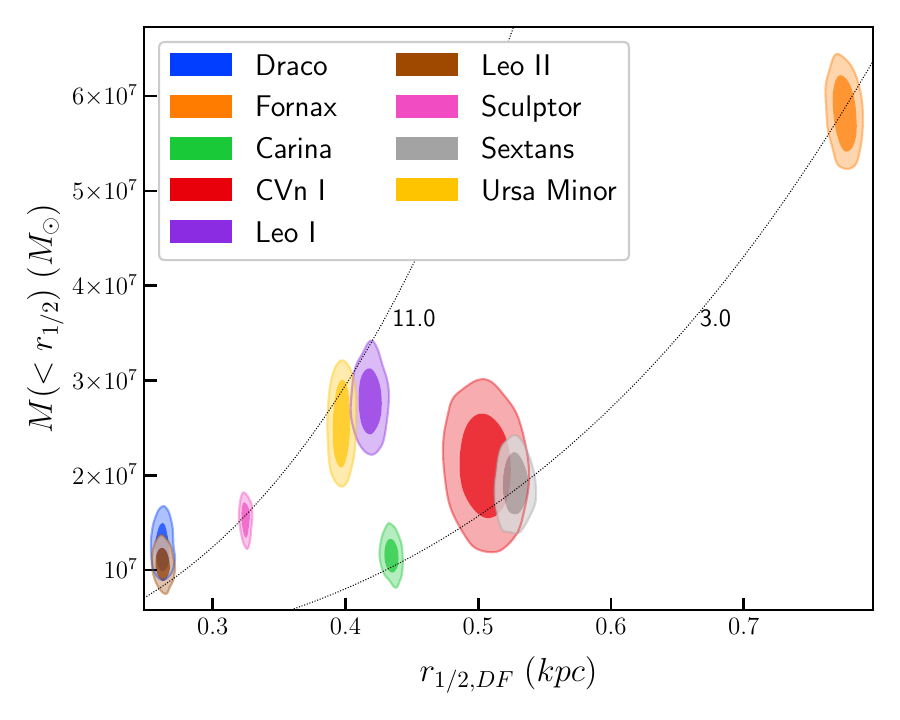}
\end{figure}

\begin{table}
	\begin{center}
	\caption{Comparison of 2D projected half-light radius ($R_{1/2}$), in kpc. The "DF" column is the $r_{1/2}$ posterior result from the DF, converted to 2D projected $R_{1/2}$ by dividing by 1.33. The "Plummer Fit" is the result from the best fitting 2-parameter Plummer profile, as applied to the (one-dimensional) sphericalized surface density data. The rightmost column is the half-light radius reported by \citet{Munoz2018} for a Sersic profile fit to 2D surface density maps, sphericalized as described in the text.}
	\renewcommand{\arraystretch}{1.1}
	\begin{tabular}{ lcccccccccccccc } 
	\hline
	   dSph Name & DF & Plummer Fit & \citet{Munoz2018} \\
	\hline
	Draco      &  0.197 $\pm$ 0.003  & 0.235  & 0.183  \\
Fornax     &  0.574 $\pm$ 0.004  & 0.688  & 0.668  \\
Carina     &  0.327 $\pm$ 0.003  & 0.344  & 0.277  \\
CVn I      &  0.381 $\pm$ 0.010  & 0.445  & 0.357  \\
Leo I      &  0.315 $\pm$ 0.004  & 0.308  & 0.204  \\
Leo II     &  0.200 $\pm$ 0.002  & 0.206  & 0.162  \\
Sculptor   &  0.243 $\pm$ 0.002  & 0.276  & 0.244  \\
Sextans    &  0.397 $\pm$ 0.005  & 0.470  & 0.370  \\
Ursa Minor &  0.299 $\pm$ 0.004  & 0.325  & 0.257  \\
\hline
	\end{tabular}
	\renewcommand{\arraystretch}{1.}
	\label{tab:obs rhalf comparison}
	\end{center}
\end{table}

The half-light radius posteriors for the observed sample are shown in Table~\ref{tab:obs rhalf comparison}. Comparison with previous authors is not straightforward, because of fundamentally different approaches in computation. We compare to \citet{Munoz2018}, who fit a Sersic profile to 2-dimensional data maps of the dwarfs. We multiply the \citet{Munoz2018} result by the axis ratio of its elliptical profile, $\sqrt{1 - \epsilon}$, to convert from elliptical radius to a spherical one. The data used for our models is a 1-dimensional equivalent of their data, also adjusted for ellipticity. We also show the $R_{1/2}$ resulting from a 2-parameter Plummer profile \citep{Plummer1911} fit of our input data. The DF approach does not rely on any profile shape; we simply find the radius that encloses half of the stars. As can be seen in the table, there can be substantial differences between the various methods. One notable difference is in Leo I, for which the DF predicts a median value of 0.315 kpc while the Plummer fit to the same data yields 0.308 kpc, and \citet{Munoz2018} find 0.204 kpc. Possible reasons for the difference are (a) the surface density map for Leo I is quite boxy, with ellipticity that appears to change with position angle, and (b) the surface density plateaus considerably at larger radii, making it a poor fit for most parameterized profiles. We note that \citet{Read2019} used Jeans analysis combined with virial shape parameters to examine these objects and found 2D half-light radii of 0.298 kpc and 0.194 kpc for Leo I and Leo II, respectively, consistent with our findings.

Figure~\ref{fig:obs_rh_mrh_2D} shows 2D posteriors for the half-light radius of the observed sample versus the mass enclosed within that radius. The distribution of masses enclosed within the half-light radii seems to split into two groups. Fornax stands out with the largest half-light radius and largest mass enclosed, however it is in the group with the lowest average density within the half-light radius, accompanied by Carina and Sextans. At the other end of the spectrum are Draco and Leo II, which are the most compact, enclose the least mass but have the highest density within $r_{1/2}$. 

We compare the results of our DF model to those of other approaches in 
Section~\ref{sec:Comparing to other methods} and find that our inferences for $\rmax$, $\vmax$ and $\rho_{150}$ are generally consistent with the other methods, with a few exceptions. The model inferences for core parameter ($c=r_\mathrm{c}/r_\mathrm{s}$), anisotropy ($\beta$) and embeddedness ($r_{1/2}/r_\mathrm{s}$) are discussed in Appendices ~\ref{app:core parameter}, ~\ref{app:anisotropy} and ~\ref{app:embeddedness}, respectively.  As the MW has strong tidal forces, we investigate the possible impacts of tidal truncation in Appendix~\ref{app:Tidal truncation}, and conclude that the likely impacts on our inferences for $\rmax$, $\vmax$ and $\rho_{150}$ are small.

\section{The Diversity of dSphs}
\label{sec: Draco and Fornax}

A convincing theory of DM will have to explain the diverse density profiles seen in the MW's dwarf spheroidal galaxies, with Draco and Fornax at the extreme ends. We find Draco to be the smallest and densest of the observed sample, with 3D half-light radius approximately 260 pc, while Fornax is the largest and among the least dense, with half-light radius of approximately 760 pc (see Section~\ref{sec: half light radius observed} for a full discussion.)  Carina looks similar to Fornax, though not as extreme, with a low DM core density and preference for a relatively large core. While it is difficult to predict the core radius from our method accurately, the shape of the posteriors for Carina and Fornax are clearly inconsistent with a cuspy profile (Fig.~\ref{fig:core radii bottom}). We refer the reader to Appendix~\ref{app:core radii} for more details on posteriors for the core radii. Conversely, Leo I and Leo II prefer small core radii or cuspy profiles, and high $\rho_{150}$. The posteriors of Draco and Sculptor are consistent with those dSphs being hosted by cored dark matter halos, but with core sizes smaller than those of Fornax and Carina. For all dSphs, the inferred core radii are smaller than or comparable to the respective half-light radii. We note that core collapse can occur in SIDM halos, with a time scale sensitive to the (possibly velocity-dependent) cross section per unit mass \citep{Elbert2015, yang_strong_2023, shah_abundance_2023, zeng_till_2023}, although we do not explore core collapse in this work.

Figure~\ref{fig:Draco_Fornax_Comparison} shows the this work's DF model inferences for the DM density as a function of radius compared to the Jeans analysis inferences of the cored isothermal and NFW cases of \citet{Kaplinghat2019}. Noted on the plots are lines for logarithmic slopes of 0 and -1, corresponding to cored and cuspy DM distributions, respectively. For both dwarfs, the density profiles are similar to the cored isothermal cases of \citet{Kaplinghat2019}, showing a cusp (or a very small core) in Draco and a core (or a very mild cusp) in Fornax. Cuspy DM halos are found in standard CDM only simulations \citep{Navarro1996}, whereas cored DM halos require either non-gravitational DM microphysics such as self-interactions, or explanations via baryonic mechanisms such as supernova feedback \citep{Loeb2012MNRAS.423.3740V, Bullock2017, 2019MNRAS.488.2387B, Rocha2012,  Elbert2015, DiCintio2014, penarrubia_tidal_2008, Sawala2016, Benitez-Llambay2019, Jenkins2014MNRAS.444.3684V, Zavala2022MNRAS.516.4543D}.
Note that while most of the MW dSphs are highly DM dominated, Fornax has a stellar mass of approximately $\num{2e7} M_\odot$ (see Table~\ref{tab:observed_dSph_list}), by far the largest in the sample, amounting to a few percent of the dynamical mass. This may suggest that baryonic effects could be responsible for the cored profile in Fornax. Further comparisons with prior works are noted in \citet{Battaglia2022}.

\begin{figure}
    \centering
    \caption{Inferences for dark matter density versus radius compared to those of \citet{Kaplinghat2019}, for Draco (top panel) and Fornax (bottom panel). The shaded bands indicate the 68\% confidence interval. The DF inferred (3-dimensional) half-light radius is shown as a dotted vertical line.}
    \label{fig:Draco_Fornax_Comparison}
    \includegraphics[width=0.47\textwidth]{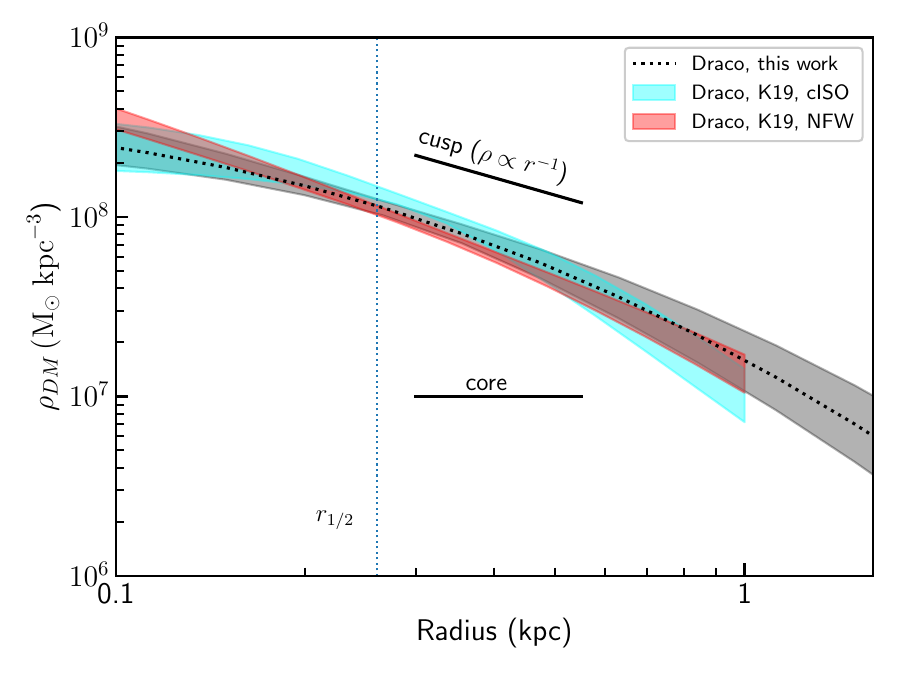}
    \includegraphics[width=0.47\textwidth]{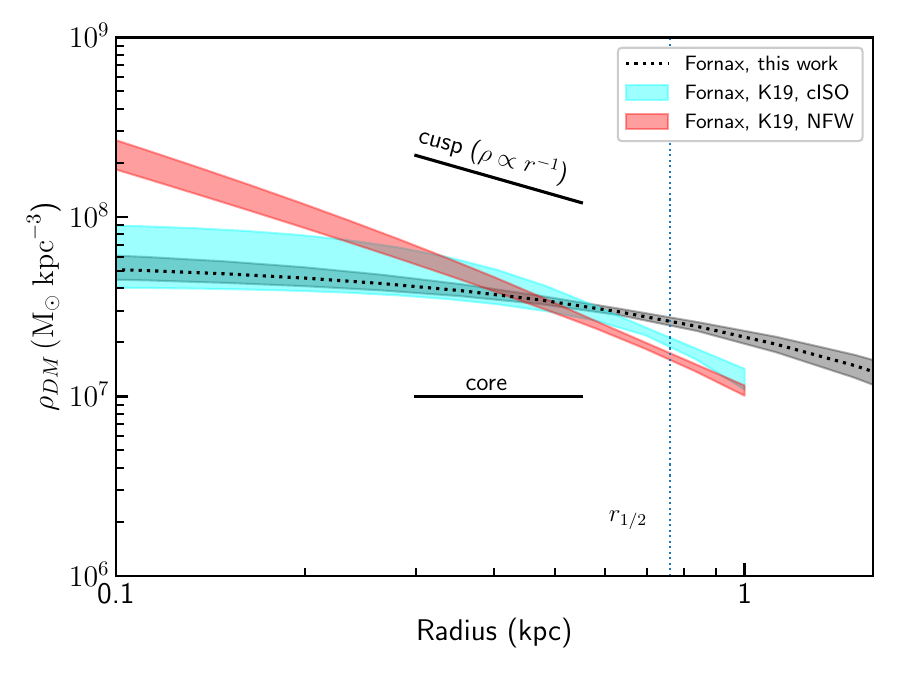}
\end{figure}

\section{Comparing the DF Method to Other Methods}
\label{sec:Comparing to other methods}

In Figure~\ref{fig:rmax vmax comparison} we compare the $\rmax$ and $\vmax$ inferences to those of \citet{Kaplinghat2019} and \citet{Errani2018}.  \citet{Kaplinghat2019} used Jeans analysis for their inference and also utilized the VSP. They analyzed two cases, one for an NFW profile and a second for a cored isothermal profile. Their results are similar to ours, with inferences from the DF and Jeans methods generally overlapping at their $1\sigma$ boundaries. The exceptions are for the $\vmax$ of Carina, Fornax and Draco, and the $\rmax$ of Fornax. In those, the DF predictions are larger than those from either profile in the Jeans analysis. The two methods have fundamental differences, namely the different modeling of stellar velocity anisotropy and the assumption of a Plummer surface density profile in the Jeans analysis. Our analysis is more general, as the DF approach accommodates a wide variety of distributions for the stellar population. Possible other reasons for the differences could include (1) different prior assumptions between the two methods and (2) for Fornax, that \citet{Kaplinghat2019} accounted for the stellar mass in the potential, in contrast to this work where we have assumed that the stars are massless tracers of the DM potential. We note that for Fornax, we infer $\vmax>45$ km/s at the $1\sigma$ level, substantially higher than either of the Jeans analysis cases. 

\citet{Errani2018} derived $\rmax$ and $\vmax$ by using the observed kinematics of the dwarfs in combination with a population of simulated subhalos. \citet{Errani2018} used spherical Plummer profiles for the stellar populuation. For the DM, they used an NFW profile for their cuspy case. For the cored case, they use
 \begin{equation}
     \rho(r) = \rho_\mathrm{s} [1 + (r/r_\mathrm{s})]^{-5}.
 \end{equation}

The inferences from their cuspy and cored cases can be seen in the orange solid lines and orange dashed lines, respectively, of Figure~\ref{fig:rmax vmax comparison}. Our results are consistent with their cuspy cases, except for Carina and Fornax (and, to a lesser extent, Sextans), where their cored case is a better match.

\begin{figure*}
    \centering
    \caption{Inferences for $r_\mathrm{max}$ and $V_\mathrm{max}$ for the observed sample from this work, shown in black, compared to the Jeans analysis results of \citet{Kaplinghat2019}, shown in blue, with their NFW case shown as blue solid lines and their cored isothermal case as blue dashed lines.  We also compare to the analysis of \citet{Errani2018}, in orange, with their cuspy case as orange solid lines and their cored case as orange dashed lines.}
    \label{fig:rmax vmax comparison}
    \includegraphics[width=0.47\textwidth]{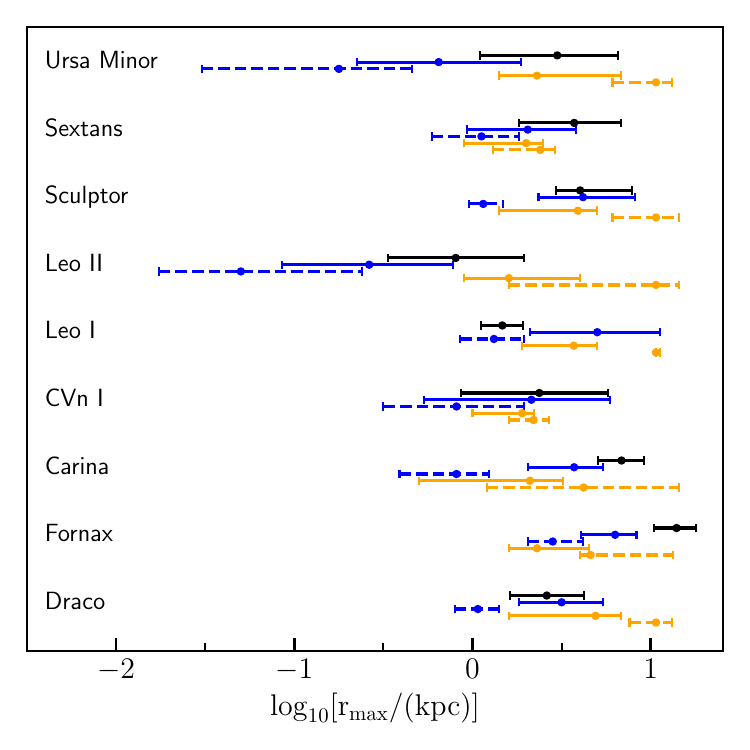}
    \includegraphics[width=0.47\textwidth]{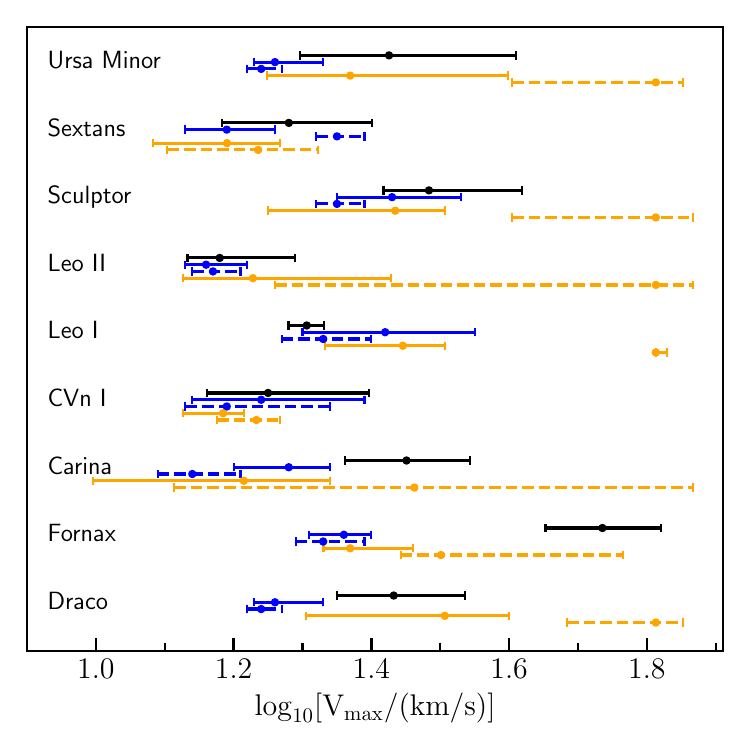}
\end{figure*}

\begin{table}
	\begin{center}
 	\caption{Comparison of findings for DM density at 150 pc ($\rho_{150}$), in units of $10^7 M_{\sun} \mathrm{kpc}^{-3}$. The median posterior value and the 68\% confidence intervals are indicated. The references for comparison are (A) \citet{Read2019}, (B) \citet{Kaplinghat2019}, isothermal case, (C) \citet{Kaplinghat2019}, NFW case, and (D) \citet{Hayashi2020}. Note that CVn I was not studied in references (A) or (D).}
	\renewcommand{\arraystretch}{1.3}
	\begin{tabular}{ lccccccccccccccc } 
	\hline
	   Name & This Work & Ref. A & Ref. B & Ref. C & Ref. D \\
	\hline
	
Draco          & ${18.8}_{-2.7}^{+3.7}$ & ${23.6}_{-2.9}^{+2.0}$ & ${21.2}_{-4.6}^{+5.4}$ & ${21.7}_{-2.2}^{+2.5}$ & ${23.5}_{-6.3}^{+12.8}$ \\
Fornax         & ${4.8}_{-0.5}^{+0.9}$ & ${7.9}_{-1.9}^{+2.7}$ & ${3.4}_{-1.3}^{+1.7}$ & ${7.5}_{-1.4}^{+2.0}$ & ${12.2}_{-2.3}^{+3.2}$ \\
Carina         & ${4.7}_{-0.7}^{+0.9}$ & ${11.6}_{-2.2}^{+2.0}$ & ${5.7}_{-1.7}^{+3.2}$ & ${10.2}_{-0.9}^{+1.1}$ & ${10.9}_{-3.2}^{+8.2}$ \\
CVn I          & ${10.5}_{-3.7}^{+5.4}$ & -- & ${13.5}_{-6.6}^{+5.3}$ & ${13.4}_{-2.4}^{+3.1}$ & -- \\
Leo I          & ${20.7}_{-2.9}^{+2.9}$ & ${17.7}_{-3.4}^{+3.3}$ & ${14.1}_{-4.5}^{+5.5}$ & ${15.0}_{-2.4}^{+3.3}$ & ${26.4}_{-9.1}^{+23.3}$ \\
Leo II         & ${17.1}_{-2.0}^{+2.1}$ & ${18.4}_{-1.6}^{+1.7}$ & ${13.4}_{-1.6}^{+4.3}$ & ${17.0}_{-3.7}^{+2.3}$ & ${20.2}_{-6.1}^{+12.7}$ \\
Sculptor       & ${14.7}_{-2.0}^{+1.8}$ & ${14.9}_{-2.3}^{+2.8}$ & ${16.1}_{-3.3}^{+2.9}$ & ${17.1}_{-2.1}^{+2.1}$ & ${21.4}_{-6.3}^{+12.6}$ \\
Sextans        & ${6.8}_{-1.9}^{+2.8}$ & ${12.8}_{-2.9}^{+3.4}$ & ${8.6}_{-3.5}^{+5.1}$ & ${10.9}_{-1.8}^{+2.9}$ & ${5.2}_{-2.3}^{+3.6}$ \\
Ursa Minor     & ${17.5}_{-5.2}^{+7.3}$ & ${15.3}_{-3.2}^{+3.5}$ & ${25.4}_{-5.6}^{+6.2}$ & ${25.1}_{-4.3}^{+3.0}$ & ${23.8}_{-7.2}^{+38.6}$ \\
\hline
	\end{tabular}
	\renewcommand{\arraystretch}{1.}
	\label{tab:obs rho150 comparison}
	\end{center}
\end{table}

Table~\ref{tab:obs rho150 comparison} and Figure~\ref{fig:rho_150 comparison} compare our findings for $\rho_{150}$ to those of \citet{Read2019}, \citet{Kaplinghat2019} and \citet{Hayashi2020}. The results are generally comparable within errors. However, our finding for Carina at $4.7_{-0.7}^{+0.9} \; M_{\sun} \mathrm{kpc}^{-3}$ is lower than the others, inconsistent with that of \citet{Read2019}, \citet{Hayashi2020} and the NFW case of \citet{Kaplinghat2019}, but compatible with their isothermal case. Our finding is consistent with a cored halo, as is suggested by the posterior for $r_\mathrm{core}$ (see Figure~\ref{fig:core radii}). We checked to see if excluding large values of c in the cNFW profile would change this inference significantly, but it does not; we found that if the core parameter is restricted so that $0<c<1$, the inference for $\rho_{150}$ increases only approximately 0.1 dex. 

\begin{figure}
    \centering
    \caption{Inference of $\rho_{150}$ from this work (black solid lines), compared to those of \citet{Read2019} (red solid lines), the Jeans analysis of \citet{Kaplinghat2019} (NFW case: blue solid lines, cored isothermal case: blue dashed lines) and axisymmetric Jeans modeling of \citet{Hayashi2020} (gray solid lines).}
    \label{fig:rho_150 comparison}
    \includegraphics[width=0.47\textwidth]{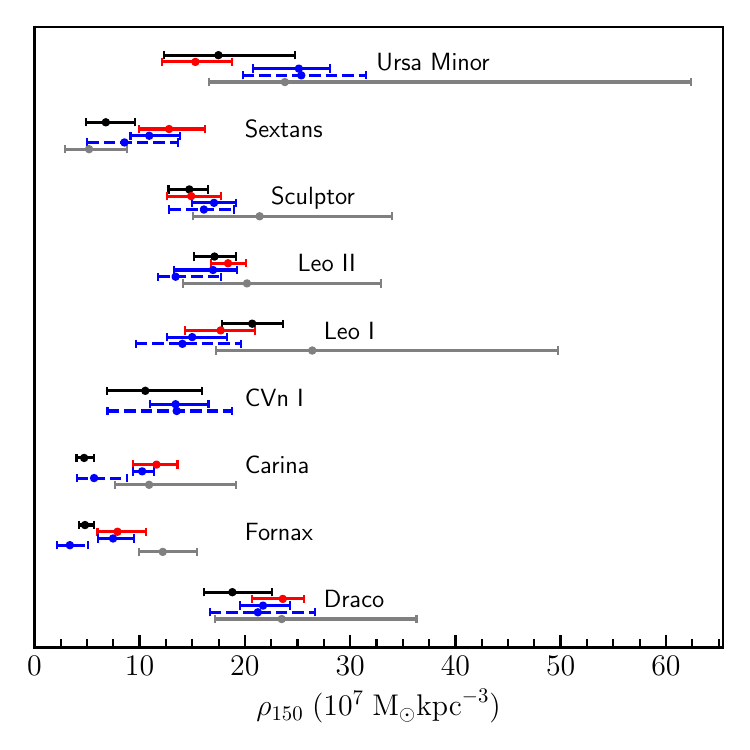}
\end{figure}

\section{Inconsistency with Simulation: Density at 150 pc versus Pericenter Distance}
\label{sec:inconsistency with simulation}

\begin{figure*}
    \centering
    \caption{DM density at 150 pc ($\rho_{150}$) inferred from the DF fits for the bright MW dSphs vs orbital pericenter distance ($r_\mathrm{p}$), in red, using the light MW model of \citet{Battaglia2022b} and comparing to host halo 1107 of the Phat Elvis simulation \citep{Kelley2019}. The right panel uses the light MW model without the LMC, while the left panel uses the light MW model including the effect of the LMC. The error bars indicate the 68\% confidence interval. The best fit line through the observations is shown in dashed red, with the 68\% confidence interval in light red. The black circles indicate the 10 subhalos with current radial positions greater than 50 kpc and with the largest $V_{\textrm{peak}}$ for host halo 1107 of the Phat Elvis simulation. The gray circles denote the 10th through 20th largest $V_\mathrm{peak}$ subhalos. The best-fit regression line for the Phat Elvis points is shown as a black dashed line. The MW dSphs are numbered as follows: 1:Draco, 2:Fornax, 3:Carina, 4:CVn I, 5:Leo I, 6:Leo II, 7:Sculptor, 8:Sextans, 9:Ursa Minor.}
    \includegraphics[width=0.495\textwidth]{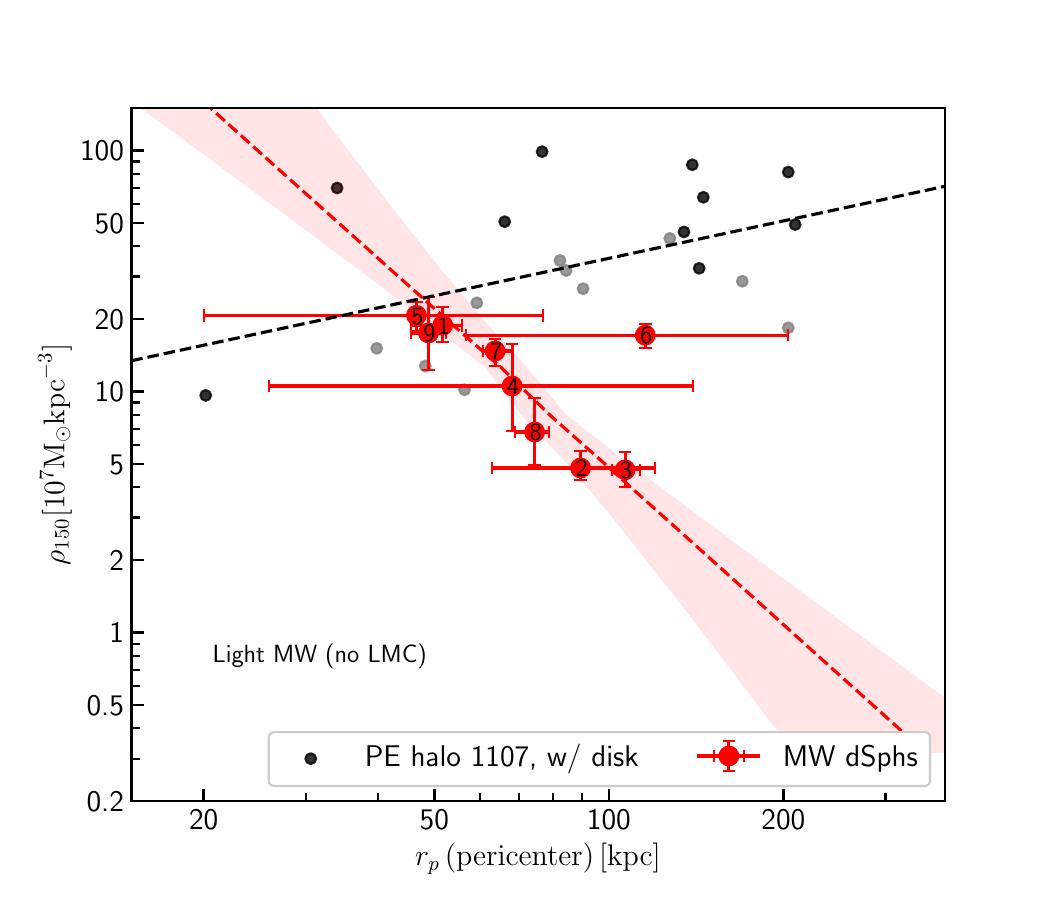}
    \includegraphics[width=0.495\textwidth]{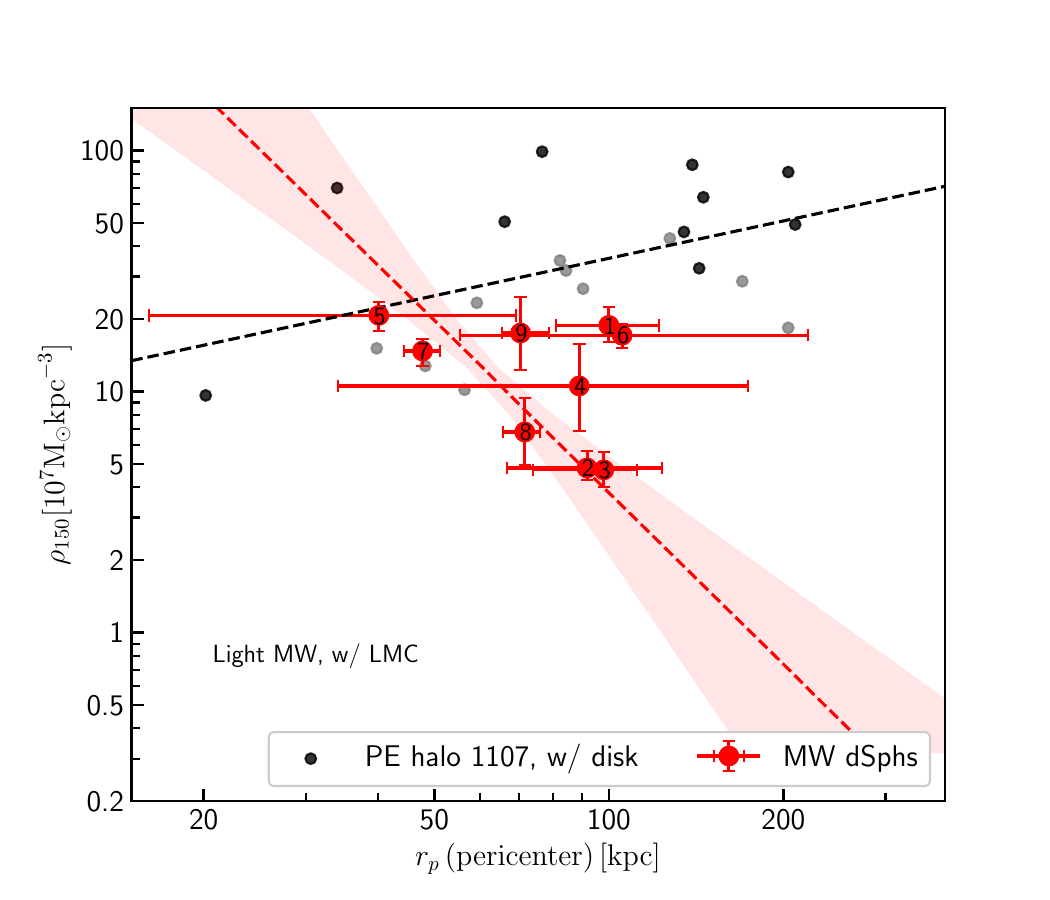}
    \label{fig: PE pericenter comparison 1107 Battaglia}
\end{figure*}

An anti-correlation between the density at 150 pc ($\rho_{150}$) and the orbital pericenter distance ($r_\mathrm{p}$) for the MW dSphs was noted in \citet{Kaplinghat2019}, and is the subject of some debate \citep{Hayashi2020, Cardona2023}. A closely related and perhaps more cogent question is whether the $\rho_{150}$- $r_\mathrm{p}$ relationship is consistent with N-body simulations of MW analogues, for if they are not, it is a challenge for $\Lambda$CDM that could require more sophisticated physics in such simulations, or could point to new physics such as DM self-interaction \cite{Correa2021}.

For orbital pericenter data we turn to the work of \citet{Battaglia2022b}, which calculated the pericenter distances for the MW dwarfs using Gaia data release 3 and which attempts to account for the impact of the Large Magellanic Cloud (LMC) on the potential and orbits. They examined two MW mass scenarios, a "light" version with mass $\num{0.88e12}M_{\sun}$, and a "heavy" version with mass $\num{1.6e12}M_{\sun}$. They also examine the light version without the LMC. We use their light model (both with and without the LMC) for our comparisons, although we check the result against the heavy model in Appendix~\ref{app:comparisons}. Note that the pericenter distances quoted are those of the last calculated pericenter passages, although the orbit integration calculations are carried backward in time to approximately 8 Gyr ago. Prior to that study, \citet{Patel2020} published an analysis accounting for the effect of the LMC in five of the MW dwarf pericenters: Carina, Draco, Fornax, Sculptor and Ursa Minor. Subsequent to our main analysis, another study was published that attempts to account for the LMC in their pericenter projections: \citet{Pace2022}. The various sources for pericenter are compared in Appendix~\ref{app:comparisons}. Although there are some differences, there is a fair amount of consistency between them after considering their stated uncertainties. \citet{dsouza2022} showed that care must be taken when back-integrating the orbits of MW satellites in parametric potentials, and that the LMC does have a substantial effect on the projection, a result that is underlined by the differences in the pericenters obtained in the with-LMC and without-LMC models of \citet{Battaglia2022b}.

The posteriors for the density at 150 pc ($\rho_{150}$) for the observed sample are shown in Figure~\ref{fig: PE pericenter comparison 1107 Battaglia}, plotted against the orbital pericenter distance ($r_\mathrm{p}$) of each dwarf, in the left panel without considering the LMC, and in the right panel accounting for the LMC. In both figures there is a clear anti-correlation between the pericenter distance and the density at 150 pc, as was also noted in \citet{Kaplinghat2019}, however the correlation appears somewhat stronger in that work than it does here. The best fitting line is shown in dashed red in Figure~\ref{fig: PE pericenter comparison 1107 Battaglia}. We infer that the slope of the best fit line is negative, as detailed in Appendix~\ref{app:line fit}. We examined this correlation using a variety of alternative sources for pericenter distances, including \citet{Fritz2018, Patel2020, Battaglia2022b} (including the "heavy" MW variations in \citet{Patel2020} and \citet{Battaglia2022b}; see Appendix~\ref{app:comparisons}}), and also using \citet{Read2019} data for $\rho_{150}$ rather than our own. The negative correlation between $\rho_{150}$ and pericenter distance persists in all cases. \citet{Hayashi2020} also found an anti-correlation in their work, although their analysis is not as directly comparable because they use an axisymmetric model for their DM halo, leading to more parameters, more degrees of freedom and large uncertainties in parameter inferences. We have used more recent pericenter data than \citet{Kaplinghat2019} and \citet{Hayashi2020}. \citet{Cardona2023} examines the correlation in some detail for various data sets, and concludes that the anti-correlation is statistically significant at the $3 \sigma$ level in only a minority of the various combinations. We discuss their comparison in Appendix~\ref{app:comparisons}. 

For comparison with simulation, we turn to the Phat ELVIS simulations \citep{Kelley2019}, a suite of 12 MW-similar halos with a disk potential, with masses ranging from $\num{7.1e11} M_\odot$ to $\num{1.95e12} M_\odot$. We use their host halo 1107 for our fiducial comparison, which has a mass of $\num{8.88e11} M_{\odot}$ (the most similar to the light MW model of \citet{Battaglia2022b}), but the results are similar for all 12 host halos (see Appendix~\ref{app:Pericenter vs density for all PE halos} and Figure~\ref{fig: Phat Elvis grid}). The Phat ELVIS simulation did not attempt to account for the effect of such a large satellite as the LMC, so we present comparisons to both the with- and without-LMC models of \citet{Battaglia2022b}. Shown in Figure~\ref{fig: PE pericenter comparison 1107 Battaglia} are the 20 subhalos in the fiducial host halo with the largest $V_\mathrm{peak}$ (i.e., the largest $\vmax$ since their infall) that are currently located more than 50 kpc from galactic center, plotted as black and gray circles. There is a clear discrepancy between the simulated and observed halos, with the simulated halos at distances greater than 50 kpc exhibiting a positive correlation between $\rho_{150}$ and pericenter. Note that significant negative correlation between $\rho_{150}$ and pericenter is not a requirement for inconsistency here; even an absence of correlation would appear to be inconsistent with the simulated halos. We note that \citet{Hayashi2020} did a similar analysis but did not restrict their regression to the largest halos. Smaller halos tend to show some survivor bias in that less dense subhalos are more likely to be disrupted by tides, thus removing halos from the lower left of the plot, as described in \citet{Kaplinghat2019}. Artifical numerical disruption of subhalos on orbits with small pericenters is also a crucial factor to consider here but the most massive subhalos should be the ones that are the least impacted by this \citep{Madau2007ApJ...667..859D, dsouza2022}. 
In addition, if we choose to populate the bright MW dwarfs in lower mass subhalos, then we will be left with an even more pronounced too-big-to-fail problem. 
For these reasons, we restrict the analysis to the 20 largest subhalos. 
Our results show that the density-pericenter data still remains a challenge that  be met by galaxy formation models.  
In this regard, it is useful to note that the orbital radii and densities are expected to have an anti-correlation in SIDM models with large cross sections \citep{Nishikawa2020,Sameie2020PhRvL.124n1102S,Correa2021,2023ApJ...949...67Y}, and that baryonic effects may also indirectly impact this \citep{Read2019}.

One might wonder if using the heavier MW models would alter the conclusion, but it does not; see Appendix~\ref{app:comparisons}, Figure~\ref{fig: PE pericenter comparison 609 heavy}.
The pericenter projections of \citet{Patel2020}, \citet{Battaglia2022b} and \citet{Pace2022} are the current state-of-the-art for the MW dSph pericenters but rely on static, axisymmetric potentials for the MW. We look for possible biases in this approach in Appendix~\ref{app:Pericenter reprojection}, by performing a reprojection of Phat Elvis pericenters using the z=0 positions and velocities of the subhalos and a static MW-like potential. We conclude that pericenters calculated using this approach usually have good agreement with the true pericenter, albeit with a minor tendency to underestimate the pericenter and with occasional outliers.

\section{Conclusions}
\label{sec:conclusions}

In this work, we presented a comprehensive study of the internal dynamics of the brightest dSphs of the MW based on a flexible distribution function model. Going beyond the standard Jeans analysis often adopted for these systems, our method relies on a separable DF~\citep{Strigari2017} that describes the phase space of stellar tracers via 10  parameters, shaping the energy and angular momentum functional form. 
The DF approach we follow here is completed by the modeling of the gravitational potential of the system, for which we adopted a 3-parameter cNFW distribution. This distribution is suitable for an investigation of both cuspy and cored DM halos. For the first time in literature, we apply such a general approach to the set of 9 bright dSphs with well-measured kinematics, and perform a data-driven Bayesian analysis on the photometric and spectroscopic data available for these objects.

Our analysis via DF modeling is validated by the use of mock data extracted from the Gaia Challenge project. In particular, we adopted mock data sets to test the predictive capability of our approach both for cuspy and cored DM profiles, for cuspy and cored stellar profiles, for different level of embeddedness of the stellar distribution within the DM halo of the system and for spatially-varying stellar orbital anisotropy profiles. From the study of the mock data we find that our DF approach is able to recover the true values of the $\vmax$ and $\rmax$ shape parameters of the underlying DM profile remarkably well, usually within the 68\% posterior probability region (see Figure~\ref{fig:mock_rmax_vmax}). It also has high accuracy for the recovery of key dynamical quantities such as the total mass within the half-light radius, $M_{1/2}$ (see Figure~\ref{fig:mock_mrhalf_diag}) and the inner local density of the system at 150~pc, $\rho_{150}$ (Figure~\ref{fig:mock_rho150_err_list}). In contrast, the mock data show us that with this approach it remains difficult to reliably determine the size of the core of the DM inner halo or to obtain robust information about the orbital anisotropy profile of stellar tracers, both of which are difficulties also suffered by Jeans analysis. The accuracy of these predictions is higher for the cases where the stellar population is not too deeply embedded with the DM halo.

Equipped with these findings, our detailed study of the MW dSphs allowed us to revisit, reiterate and reinforce some well-known conclusions already drawn in literature within the standard Jeans analysis. Our study of the Classical dSphs via DF modeling provides a state-of-the-art inference of $\rho_{150}$ in these objects. In particular, we find a low inner density for systems like Carina and Sextans, in contrast to galaxies like Draco and Leo II, characterized by inner densities approximately four times larger (Figure~\ref{fig:obs_rh_mrh_2D}). With the DF approach, we are then able to confirm the large diversity in the dark matter densities of these dark-matter dominated objects. These inferences of the inner density constitute key dynamical information that needs to be captured by any successful model of galaxy formation within the $\Lambda$CDM cosmological model, or another model where the dark matter is not made up of cold and collisionless DM particles.

We have reexamined the anti-correlation between dwarf spheroidal pericenters and density at 150 pc found in \citet{Kaplinghat2019}, using our method rather than Jeans analysis and using more recent assessments of the pericenter determination by \citet{Battaglia2022b}. We also observe a negative correlation. This is inconsistent with both the dark-matter-only and disk versions of the Phat Elvis N-body simulation of \citet{Kelley2019} (see Figure~\ref{fig: PE pericenter comparison 1107 Battaglia}). This inconsistency remains a compelling clue for investigating dark matter microphysics. 

We observe that for Fornax and Carina, the results of our analysis with the cNFW profile point to the presence of a large core in these systems (Figure~\ref{fig:core radii}). Most of the other dSphs have smaller cores or show no evidence for cores. 
Some care should be taken in considering this inference given the limited ability in inferring the core sizes in the mock data sets. 
Overall, our results argue that the DM core sizes are smaller than the respective half-light radii, which could be a further clue. 

The results of our work are promising in the regard that the DF modeling has a similar constraining power to that of the spherical Jean analysis and other methods, despite varying a larger set of parameters needed for a broad description of the tracer phase-space distribution function.
Natural extensions of this work will involve DF models that allow for multiple populations with separate metallicity distributions and  non-sphericity in the stellar profiles. 

This could allow for more robust inferences of the sizes of constant density cores in MW dSphs, and provide significant new constraints on proposed solutions to the too-big-to-fail problem.

\section*{Acknowledgements}

We thank Matthew Walker and Andrew Pace for providing velocity dispersion data. We gratefully acknowledge a grant of computer time from TACC allocation AST20027. MK acknowledges support from National Science Foundation grant PHY-2210283.

\section*{Data Availability}

The data underlying this article will be shared on reasonable
request to the corresponding author.



\bibliographystyle{mnras}
\bibliography{MyCollection.bib} 




\appendix
\onecolumn
\FloatBarrier
\section{Comparison to Strigari et al (2017) for Sculptor}
\label{app:Strigari comparison}

\begin{figure}
    \centering
    \caption{Comparison of the results of the modified model (shown in black) for $r_\mathrm{max}$ vs. $V_\mathrm{max}$ to the NFW results of \citet{Strigari2017} for Sculptor. The lines indicate the 68\% and 90\% confidence levels. \textit{Top Panel:} Metal poor stellar population. The results from \citet{Strigari2017} are in blue. \textit{Bottom Panel:} Metal rich stellar population. The results from \citet{Strigari2017} are in red. Note that for this figure only we follow the convention from \citet{Strigari2017} that the outer contour lines represent the 90\% confidence level rather than 95\%.}
    \label{fig:comparison with Strigari}
    \includegraphics[width=0.48\textwidth]{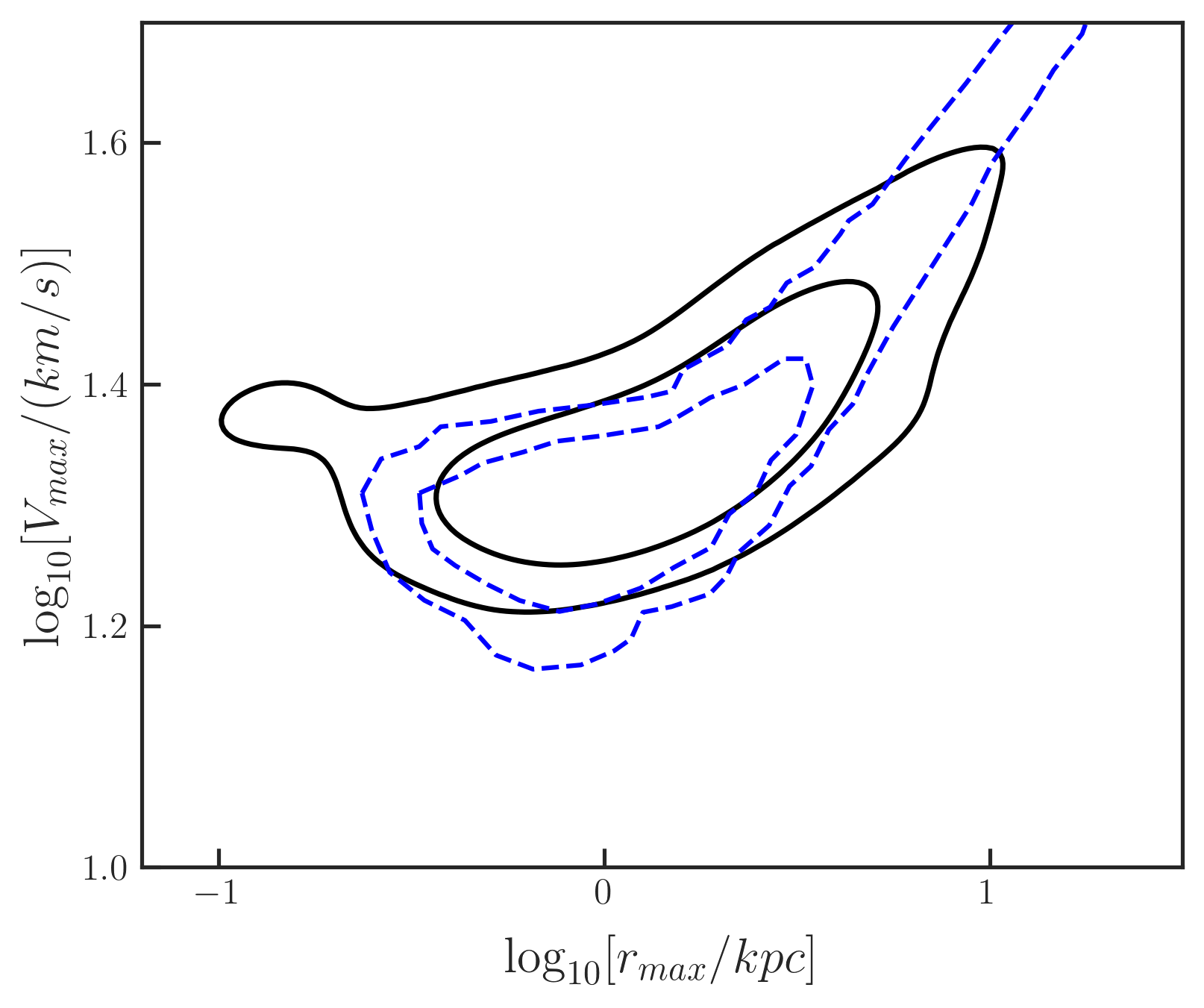}
    \includegraphics[width=0.48\textwidth]{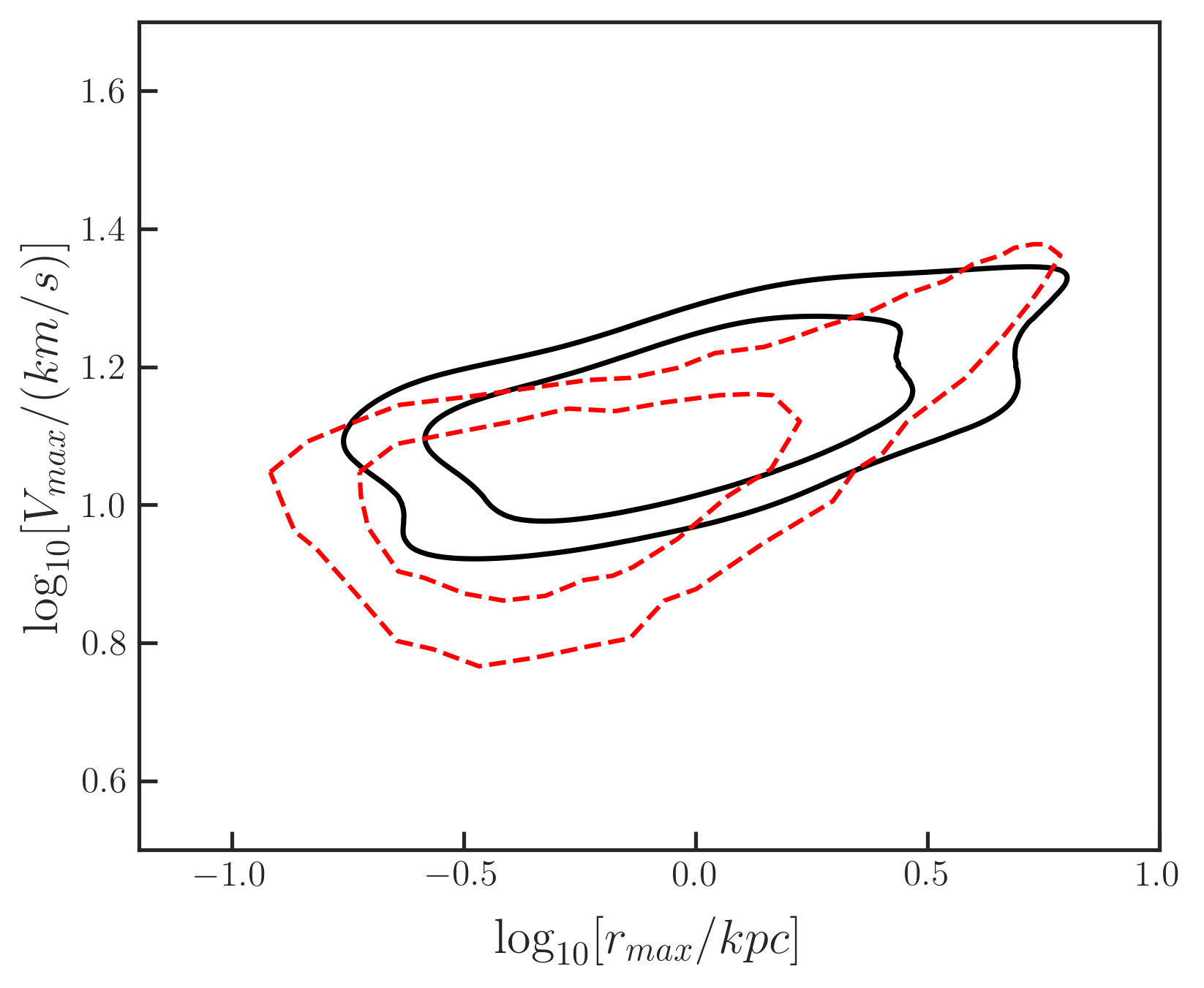}
\end{figure}

Because our approach is similar to that of \citet{Strigari2017}, we created a modified version of our model that maximizes its comparability to the one in that work, and compare the results of the two models here. To maximize comparability, we make the following changes to our model: (a) remove the factor of $\frac{1}{2}$ from Equation~\ref{eq:g}, (b) remove the normalization term $n_\mathrm{f}$ in Equation~\ref{eq:f} and allow the parameter w to vary freely in the MCMC analysis, (c) set $b_{\mathrm{in}}=0$, (d) set $\alpha=1$, (e) set $c=0$, which forces the DM profile to be an NFW profile, and (f) remove the VSP chi square component. We then run our model on the same Sculptor metal poor and metal rich population data as was used in \citet{Strigari2017}, i.e., the surface density and dispersion data from \citet{Battaglia2008}. The results are shown in Figure~\ref{fig:comparison with Strigari}. The top panel shows the results for the metal poor case, the bottom panel shows the metal rich case. Our results are shown in black and the results from Figure 4 of \citet{Strigari2017} are shown in blue (metal poor) and red (metal rich), respectively. Their result correspond fairly closely with ours. 

\FloatBarrier
\section{Virial Shape Parameter}
\label{app:vsp}

The virial shape parameter is derived from the fourth-order projected virial theorem \citep[]{Merrifield1990}, and for approximately spherical systems it can take two forms \citep[]{Richardson2014}. Following \citet{Kaplinghat2019}, we utilize the first form, which we label here the VSP:
\begin{eqnarray}
    \mathrm{VSP}= \frac{1}{2} \int_0^{\infty} d\!R^2 \Sigma \langle v_{\mathrm{los}}^4 \rangle
    = \frac{G}{5} \int_0^\infty d\!r^2 M(5-2\beta) \nu \sigma_\mathrm{r}^2,
\end{eqnarray}
where M denotes the mass distribution function, $\beta$ is the anisotropy parameter, $\nu$ is the stellar density and $\langle v_{\mathrm{los}}^4 \rangle$ is the fourth moment of the line-of-sight velocity distribution.
To calculate the VSP from the DF, we integrate the fourth velocity moment as follows:
\begin{eqnarray*}
(2\pi) \mathrm{vsp} = \int d^3r d^3v \ v_\mathrm{z}^4 f(E,L) \ .
\end{eqnarray*}
Note that there is no factor of $N_*$ because the DF is normalized to unity over the entire phase space. Now write $$v_r=v \cos\eta, v_\theta=v \sin\eta \cos\psi.$$
Then, 
\begin{eqnarray*}
v_\mathrm{z}=\hat{z}\times \vec{v} = v_r \cos\theta -v_\theta \sin\theta = v(\cos\theta \cos\eta - \sin\theta \sin\eta \cos\psi)
\end{eqnarray*}
and
\begin{eqnarray*}
\mathrm{VSP} = \int r^2 dr\  d\!\cos\!\theta \ v^2 dv \ d\psi d\!\cos\!\eta \ v_\mathrm{z}^4 f(E,L)
\end{eqnarray*}
Since $E=\Psi-v^2/2$ and $L=r v_\mathrm{t} = r v \sin\eta$, we can first do the $\theta$ and $\psi$ integrals over $v_\mathrm{z}^4$.
It can be shown that 
\begin{eqnarray*}
\int_0^{2\pi} d\psi \int_{-1}^{1} d\!\cos\!\theta (\cos\theta \cos\eta - \sin\theta \sin\eta \cos\psi)^4 = 4\pi/5 \,
\end{eqnarray*}
hence,
\begin{eqnarray}
 \mathrm{VSP} = (4\pi/5) \int_{0}^{r_\mathrm{t}} r^2 dr\ \int_0^{\sqrt{2\Psi(r)}} v^6 dv \ \int_{-1}^{1} d\!\cos\!\eta \ f(E,L) \ .
\end{eqnarray}
For data sets with measured line-of-sight velocities, the VSP can be calculated as follows. In our coordinate system, the z-axis is the line-of-sight. First, the mean value of $v_\mathrm{z}$ is subtracted from each $v_\mathrm{z,i}$ to remove bulk motion of the galaxy. The VSP is then
\begin{eqnarray}
    \mathrm{VSP} = \frac{1}{2 \pi N_*} \sum_{i=1}^{N_*} v_\mathrm{z,i}^4  \ .
\end{eqnarray}
For the mock data sets, we wish to find the an estimate of the distribution of the VSP given the one set of sampled velocities. We do so by generating 10,000 ensembles of binned velocity data, each with length $N_*$, from a Pearson distribution of Type VII, with the same star count and velocity dispersion in each bin as the original data set. To simulate measurement uncertainty, we add Gaussian error with a standard deviation of 2 km/s. The kurtosis of the Pearson distribution is adjustable via a parameter, and that parameter is iteratively varied until the kurtosis of the entire ensemble matches that of the original data set. We then tabulate the 15.9, 50 and 84.1 percentile values of the VSP of the entire ensemble, which are used as estimators for the mean and standard deviation of the VSP. Those values are used as data for the DF model and are tabulated for the mock data sets in Table~\ref{tab:mock data characteristics}.

\FloatBarrier
\section{Full Likelihood Function}
\label{app:full likelihood}
Consider a population of w stars in a potential $\Phi$ and with a distribution function $f$. Our goal is to estimate $\Phi$ and $f$ based on the star population. For star $i$, we have position coordinates $R_\mathrm{i}=\sqrt{x_\mathrm{i}^2+y_\mathrm{i}^2}$, and we have velocity coordinate $v_\mathrm{z, i}$ (but we do not generally know $z_\mathrm{i}, v_\mathrm{x, i}$ or $v_\mathrm{y, i}$). The best estimate of $\Phi$ and $f$ is the one that maximizes the likelihood function
\begin{eqnarray}
	LH(\Phi, f|[R_\mathrm{i}, v_\mathrm{z, i}]).
\end{eqnarray} \ 
By Bayes Theorem, we instead estimate the posterior and prior probabilities
\begin{eqnarray}
	LH(\Phi, f|[R_\mathrm{i}, v_\mathrm{z, i}])=LH([R_\mathrm{i}, v_\mathrm{z, i}]|\Phi, f)\frac{P(\Phi, f)}{P([R_\mathrm{i}, v_\mathrm{z, i}])}	\ ,
\end{eqnarray}
where $LH([R_\mathrm{i}, v_\mathrm{z, i}]|\Phi, f)$ is the posterior probability of observing the given data with a particular $\Phi$ and $f$, and $P(\Phi, f)$ is the prior probability for observing $\Phi$ and $f$, and incorporates any prior beliefs. The probability of observing the data for our model, $P([R_\mathrm{i}, v_\mathrm{z, i}])$, also known as the "evidence", is not generally known, but as it is a constant factor it will not affect our attempts to maximize the likelihood function. 

We wish to employ the distribution function as a probability of finding a star $i$ at radius R and line-of-sight velocity $v_\mathrm{z}$. The probability can be written as
\begin{eqnarray}
	p_\mathrm{*, i} (R_\mathrm{i}, v_\mathrm{z, i})= 2 \pi \int_{-\infty}^\infty v_R dv_R \int_{-\infty}^\infty dz~f(E_\mathrm{i},L_\mathrm{i}) \ .
\end{eqnarray}
The composite likelihood for all stars in the data set is then
\begin{eqnarray}
	LH([R, v_\mathrm{z}]|\Phi, f)=\prod_{i=1}^w p_\mathrm{*, i}(R_\mathrm{i}, v_\mathrm{z, i}) \ ,
\end{eqnarray}
and the log likelihood is then
\begin{eqnarray}
	LLH\triangleq log(LH([R, v_\mathrm{z}]|\Phi, f)))=\sum_{i=1}^w log(p_\mathrm{*, i}(R_\mathrm{i}, v_\mathrm{z, i})) \ .
\end{eqnarray}
Computationally, we have a vector of parameters $\mathbf{p}=\{r_\mathrm{s}, v_\mathrm{s}, \Phi_{\mathrm{lim}}, e, a, q, E_\mathrm{c}, d, L_{\beta}, b_{\mathrm{in}}, b_{\mathrm{out}}, w \}$ for which we want to calculate a given likelihood. The normalization factor $n_\mathrm{f}$ may be factored out of the sum, and LLH becomes
\begin{eqnarray}
	LLH(\mathbf{p})= n_\mathrm{f}(\mathbf{p}) \sum_{i=1}^w \log \Big(\int_{-\infty}^\infty dv_R \int_{-\infty}^\infty dz~h\big(\mathbf{p}, E(\mathbf{x_\mathrm{i}, v_i})\big) 
  g\big(\mathbf{p}, L(\mathbf{x_\mathrm{i}, v_i})\big)\Big) \ ,
\end{eqnarray}
where the functions $h$ and $g$ are given in Equations~\ref{eq:h} and \ref{eq:g}.

\FloatBarrier
\section{Binning of Velocity Dispersion Data}
\label{app:binning}

Here we describe our procedure for binning the velocity dispersion data. For the observed sample, the data consists of the right ascension and declination coordinates for each star, the LOS velocity for each star, and the uncertainty of the LOS velocity measurement. The position data is converted to physical $\delta x$ and $\delta y$ coordinates using the adopted distance to the galaxy specified in Table~\ref{tab:observed_dSph_list}. The centroid is calculated as the coordinates that minimize the sum of the squared distances from each star to the center. These correspond closely to the galaxy coordinates cited in the NASA/IPAC Extragalactic Database (https://ned.ipac.caltech.edu). To account for the ellipticity of the galaxies we draw elliptical bins based on the position angle and ellipticity noted in Table~\ref{tab:observed_dSph_list}.  We use Sturges' Rule to determine the number of bins, i.e., $B=\log_2(N_\mathrm{stars}) + 1$. The bin boundaries are chosen so that there are an equal number of stars in each bin to the maximum extent possible. For the Gaia Challenge data, the same process is used, but is simplified because the data center coordinates are known, and the data were generated with spherical symmetry so no adjustment for ellipticity is required.

Our method for estimating the binned velocity dispersions closely follows the maximum likelihood approach described in \citet{Walker2006}.  We let $v_\mathrm{ij}$, $u_\mathrm{ij}$ and $\sigma_\mathrm{ij}$ be the measured LOS velocity, the true LOS velocity and the measurement uncertainty, for star i of $N_j$ stars in bin j of B bins. Then $v_\mathrm{ij}=u_\mathrm{ij} + \sigma_\mathrm{ij} \epsilon_\mathrm{ij}$, and the $\epsilon_\mathrm{ij}$ have a standard Gaussian distribution. The variability in $v_\mathrm{ij}$ comes from two sources: the intrinsic LOS velocity dispersion in the $u_\mathrm{ij}$, which we denote $\sigma_\mathrm{j}$, and the measurement uncertainties $\sigma_\mathrm{ij}$. We assume that the $v_\mathrm{ij}$ have a Gaussian distribution with mean equal to the mean true velocity $\langle u \rangle$. The joint probability over all of the observations is therefore:
\begin{eqnarray}
    LH=\prod_{j=1}^B \prod_{i=1}^{N_j} \frac{1}{\sqrt{2 \pi (\sigma_\mathrm{ij}^2 + \sigma_\mathrm{j}^2)}} \exp{\Bigg( -\frac{(v_\mathrm{ij} - \langle u \rangle)^2}{2(\sigma_\mathrm{ij}^2 + \sigma_\mathrm{j}^2)} \Bigg)} \ .
\end{eqnarray}
We use MCMC analysis to determine posterior distributions for $\langle u \rangle$ and the $\sigma_\mathrm{j}$. We use the Emcee sampler \citep{Foreman-Mackey2019}. We explored using velocity dispersion directly as the parameter of interest, as well as using $\log_{10}$ of the velocity dispersions, and found that using the logarithm resulted in Gaussian distributions for the posterior distributions, while using the dispersions themselves did not. We therefore use $\langle u \rangle$ and $\log_{10}(\sigma_\mathrm{j})$ as parameters in the MCMC analysis. The resulting binned data values and their uncertainties are available in the online material.

\FloatBarrier
\section{Core Radii Inferences for Mock Data and Observed Dwarfs}
\label{app:core radii}

\begin{figure*}
    \centering
    \caption{Posterior histograms of $\log_{10}[r_\mathrm{core}/\mathrm{kpc}]$, with mock data sets. \textit{Left:} The 16 mock data sets with NFW DM profiles. The true value of $\log_{10}[r_\mathrm{core}/\mathrm{kpc}]$ is $-\infty$ (corresponding to $r_\mathrm{core} = 0$ kpc), although we limit the parameter to -2 in $\log_{10}$ space (corresponding to $r_\mathrm{core}=0.01$ kpc.). \textit{Right:} The 16 mock data sets with cored DM profiles. The true value is -0.585 (corresponding to $r_\mathrm{core} = 0.26$ kpc) and is indicated by the black dotted line.}
    \label{fig:core radii}
    \includegraphics[width=0.46\textwidth]{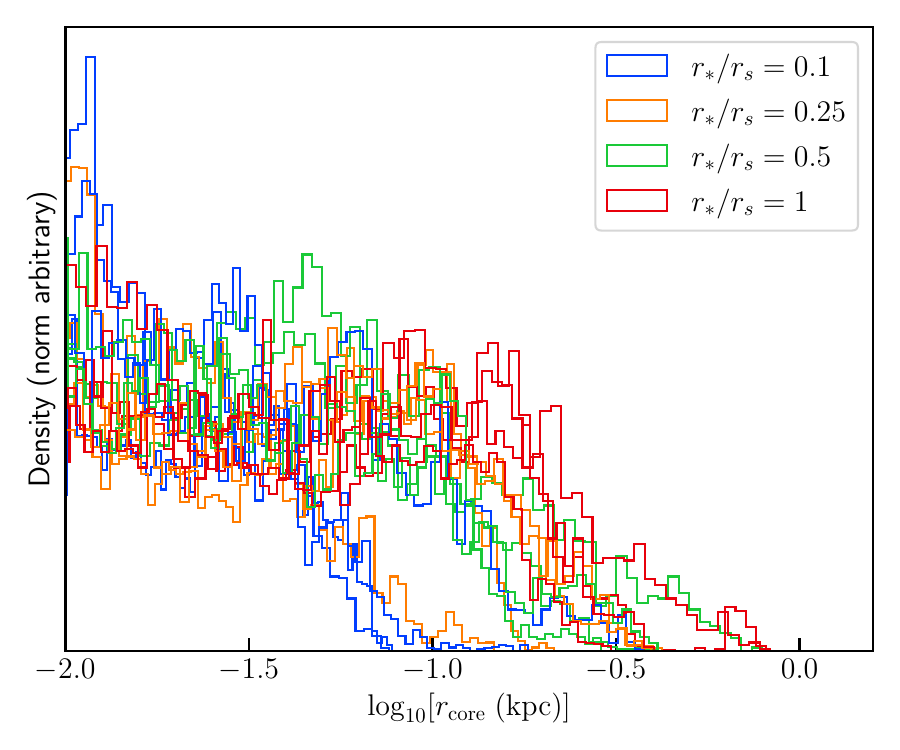}
    \includegraphics[width=0.46\textwidth]{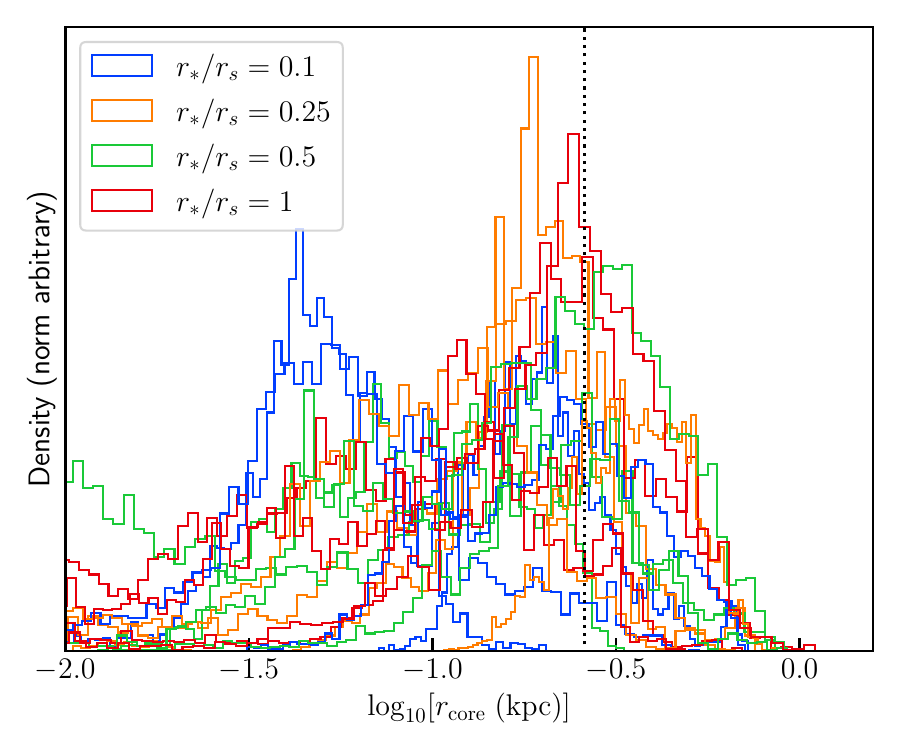}
    
\end{figure*}

The parameter $\log_{10}[r_\mathrm{c}/\mathrm{kpc}]$ is allowed to vary in the model to explore the best fitting value, with prior limits $-2 < \log_{10}[r_\mathrm{c}/\mathrm{kpc}] < 1$. As described in Section~\ref{sec:potentials}, the core radius $r_\mathrm{core}$ is calculated as the radius at which the density falls to one-half its central value. The true core radii of the mock data sets are either 0 kpc (NFW) or 0.26 kpc (cored), corresponding to $\log_{10}$ values of $-\infty$ and $-0.585$, respectively, although we use -2 as a practical lower limit, corresponding to $r_\mathrm{core}=0.01$ kpc. The top panels of Figure~\ref{fig:core radii} shows a composite plot of the posteriors for the 16 NFW data sets in the left panel and the 16 cored data sets in the right panel. The model shows some ability to distinguish between the two profiles, with an uncertainty of approximately 0.5 to 1.0 dex, although there is bias towards lower values for the cored profiles. The NFW data sets uniformly prefer small cores. The cored data sets generally prefer large core solutions, except for some of the most deeply embedded data sets with $r_*/r_\mathrm{s}=0.1$. 

\begin{figure*}
    \centering
    \caption{Posterior histograms of $\log_{10}[r_\mathrm{core}/\mathrm{kpc}]$, with observed dSphs in the three rows. \textit{Left:} The 16 mock data sets with NFW DM profiles. The true value of $\log_{10}[r_\mathrm{core}/\mathrm{kpc}]$ is $-\infty$ (corresponding to $r_\mathrm{core} = 0$ kpc), although we limit the parameter to -2 in $\log_{10}$ space (corresponding to $r_\mathrm{core}=0.01$ kpc.). \textit{Right:} The 16 mock data sets with cored DM profiles. The true value is -0.585 (corresponding to $r_\mathrm{core} = 0.26$ kpc) and is indicated by the black dotted line. \textit{Bottom three rows:} Inferences of $\log_{10}[r_\mathrm{core}/\mathrm{kpc}]$ for the observed sample. The modes for Fornax and Carina are at approximately $\log_{10}[r_\mathrm{core}/\mathrm{kpc}]$ of -0.3 and -0.5, respectively, which correspond to $r_\mathrm{core}$ of 0.3 kpc and 0.5 kpc. The other galaxies have modes at or less than $\sim$ 0.1 kpc.}
    \label{fig:core radii bottom}

    \includegraphics[width=0.79\textwidth]{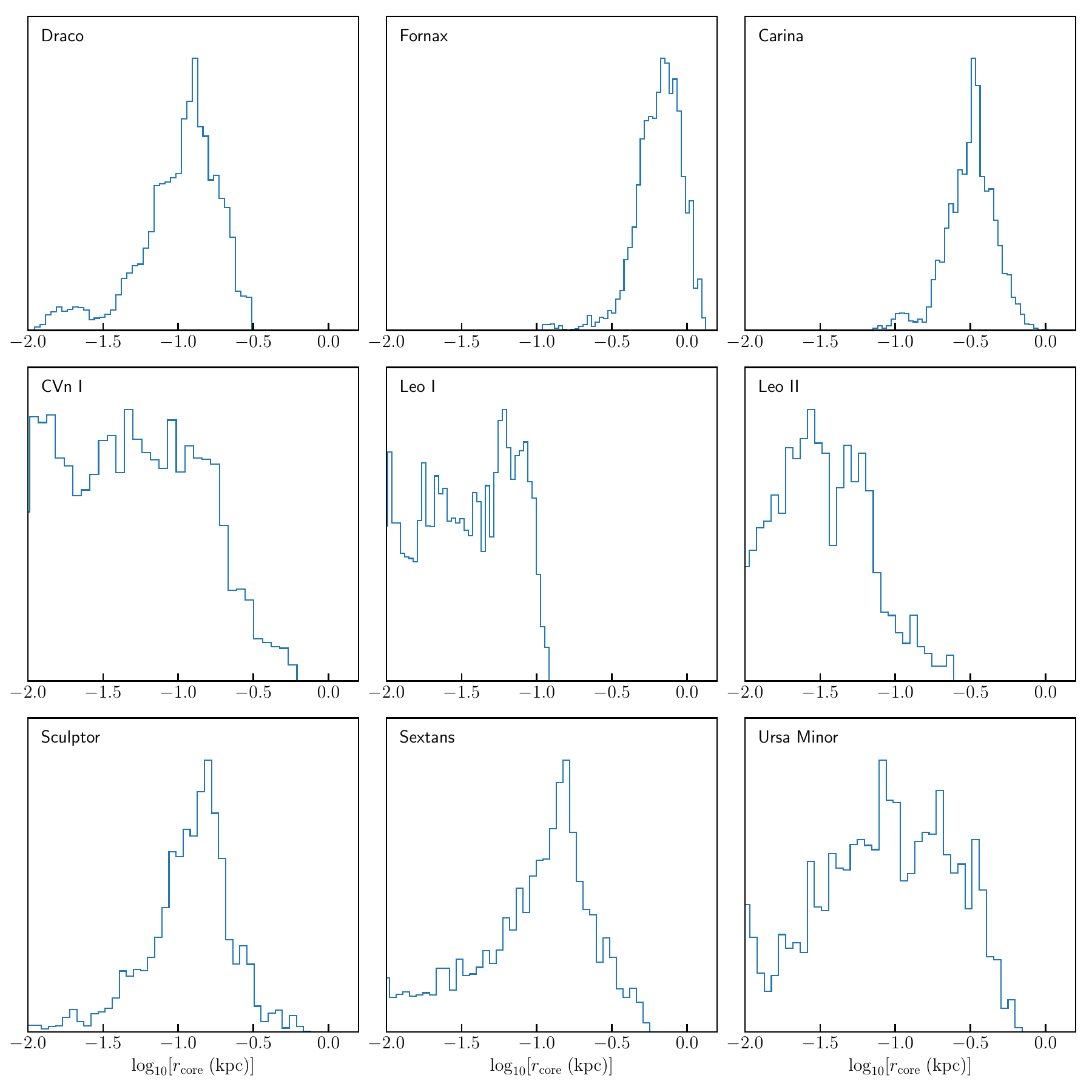}
\end{figure*}

We also present the core radius posteriors of the observed sample here, for easier comparison to the mock data results. As shown in the bottom panels of Figure~\ref{fig:core radii}, many of the observed sample prefer small cores, consistent with the NFW profile. However, Fornax and Carina prefer nonzero cores with large radii of approximately 0.5 kpc and 0.3 kpc, respectively. We note that \citet{penarrubia2013} found cores in Fornax and Sculptor by exploiting separate chemo-dynamical subcomponents, although \citet{Strigari2017} found only a weak preference for a core in Sculptor and that both cored and NFW profiles were good fits. \citet{Hayashi2020} found that Carina, Sextans, Sculptor, and Fornax favor smaller (core-like) DM inner slopes, using axisymmetric Jeans analysis. Other authors finding a likely core in Fornax include \citet{Walker2011a, Jardel2012, Pascale2018}. A key difference between our work and previous work is the use of mock data to validate our inferences. 

On the other hand, Draco, Sculptor and Sextans show evidence for a small core of about 100 pc. The left panel of Figure~\ref{fig:core radii} demonstrates that if the true profile is an NFW profile, the model posterior is unlikely to resemble those of Draco, Fornax, Carina, Sculptor and Sextans. Draco has been thought to be cuspy in prior works \citep{Read2018, Read2019, Hayashi2020}; our finding is for a small core but with a high density, so in that regard all studies seem to agree with each other.

\FloatBarrier
\section{Anisotropy at Half-light Radius}
\label{app:anisotropy at rhalf}

\begin{figure*}
    \centering
    \caption{True and predicted values for $\beta(r_{1/2})$ for the 32 mock data sets, segregated by embeddedness. The diagonal line indicates equality between the true and predicted values.}
    \label{fig:mock_beta_rhalf_diag}
    \includegraphics[width=0.98\textwidth]{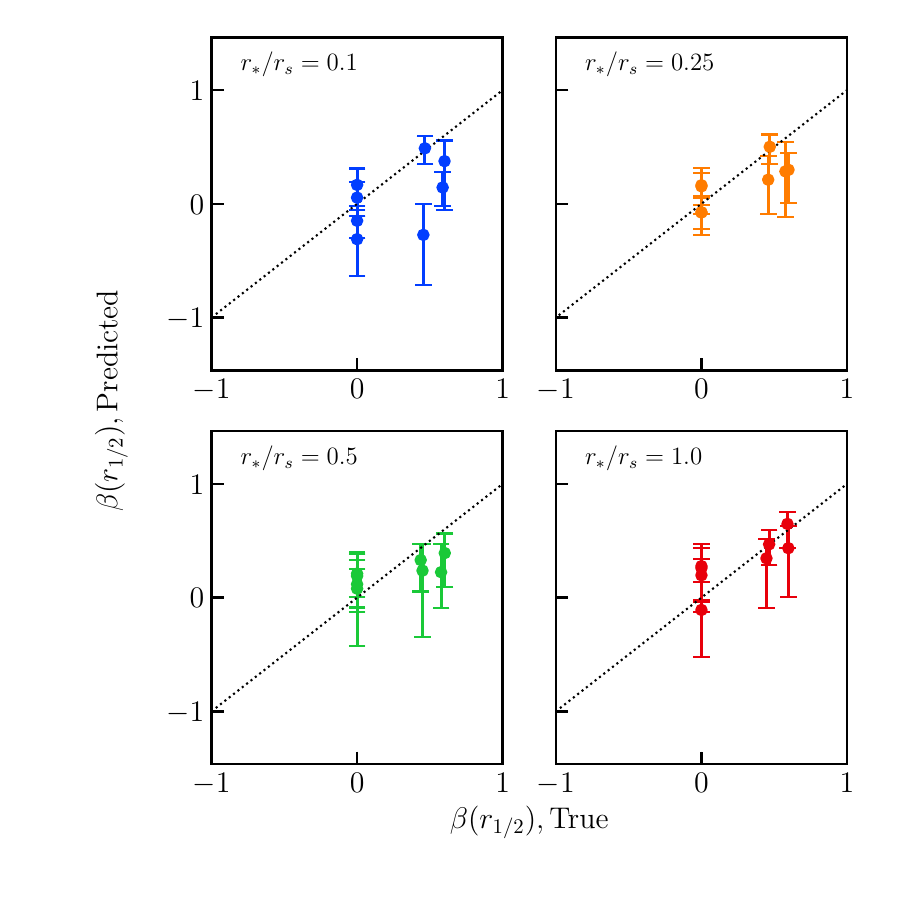}
\end{figure*}

Figure~\ref{fig:mock_beta_rhalf_diag} shows the true and predicted posteriors for the anisotropy parameter $\beta$ at the half-light radius for the mock data sets. Half of the mock data sets are anisotropic over their entire range, while the other half have rising $\beta$ profiles, with a true value between 0.4 and 0.6 at the half-light radius. For the isotropic data sets, the model predictions have median values centered near zero and with a range of -0.2 to 0.2. For the anisotropic data sets, the model tends to systematically underestimate $\beta$, except for least embedded data sets denoted in red.

\FloatBarrier
\section{Core Parameter Posteriors of the Observed Sample}
\label{app:core parameter}

\begin{figure}
    \centering
    \caption{Posteriors of the core parameter, $c = r_\mathrm{c}/r_\mathrm{s}$ for the bright MW dSphs. Only Fornax and Carina show a significant posterior above $c = 1$.}
    \label{fig:obs_core}
    \includegraphics[width=0.9\textwidth]{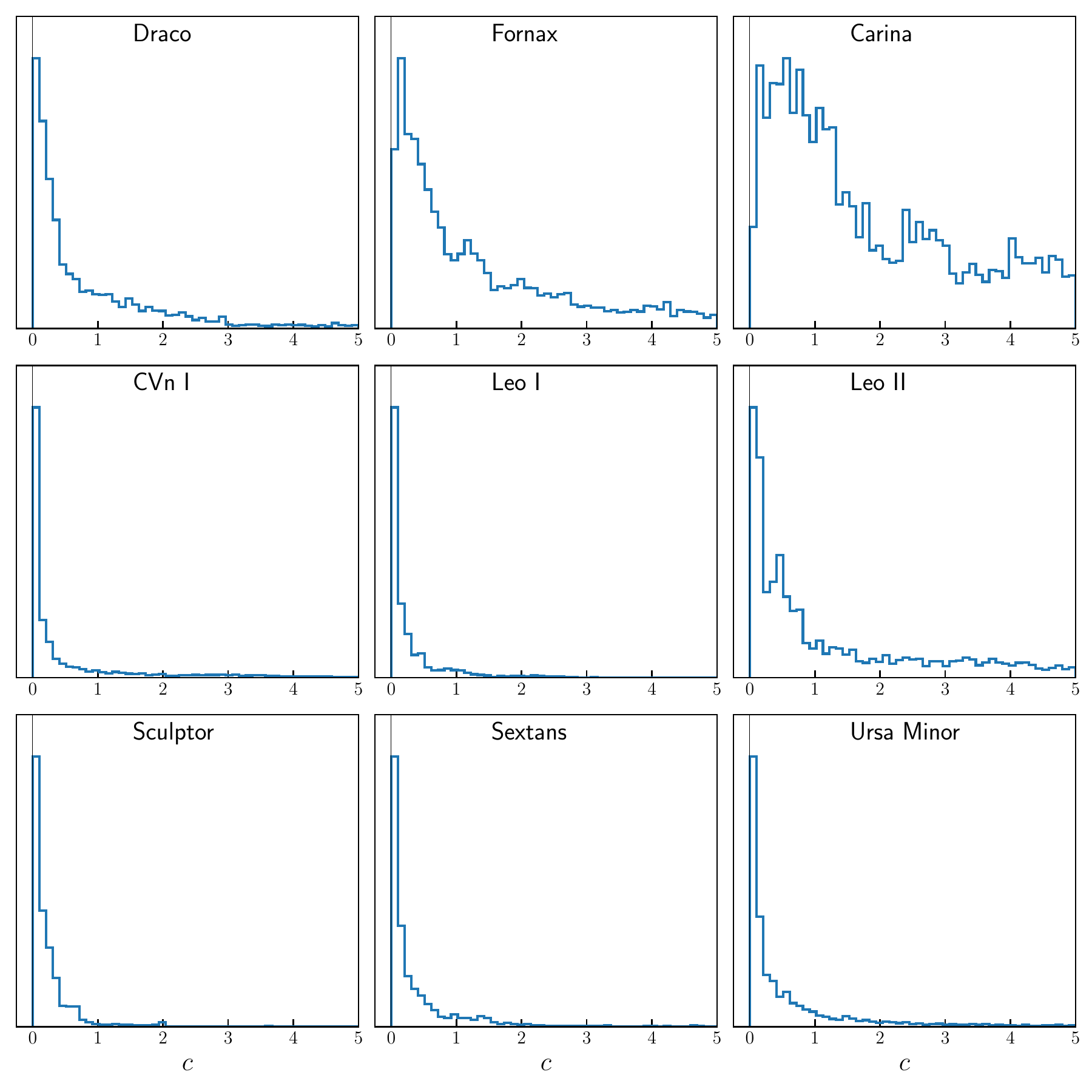}
\end{figure}

The posteriors for $c=r_\mathrm{c}/r_\mathrm{s}$ are shown in Figure~\ref{fig:obs_core}. Most of the sample shows a preference for nearly zero values of the core parameter, indicating a density profile that is close to the NFW profile. However, Fornax and Carina do have significant tails above $c=1$. In that area of parameter space, the scale radius $r_\mathrm{s}$ is smaller than the core radius $r_\mathrm{c}$, so that they can be said to switch roles in defining the shape of the profile (see Equation~\ref{eq:cNFW}).

\FloatBarrier
\section{Velocity Dispersion Anisotropy in the Observed Sample}
\label{app:anisotropy}

\begin{figure}
    \centering
    \caption{Posteriors of anisotropy parameter $\beta$ at the half-light radius for the bright MW dSphs.}
    \label{fig:obs_beta_rhalf}
    \includegraphics[width=0.9\textwidth]{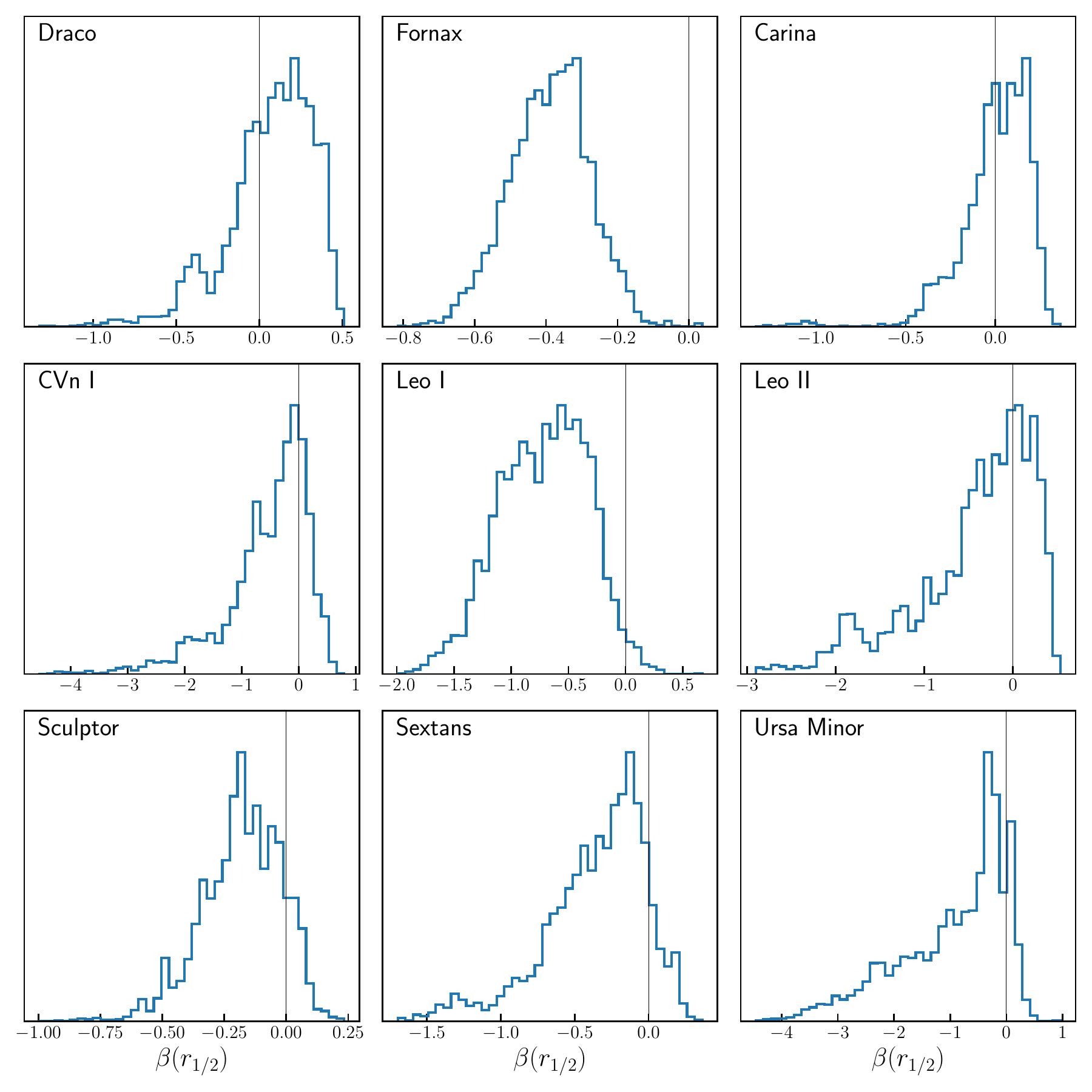}
\end{figure}

The posteriors for the anisotropy parameter $\beta(r_{1/2})$ for the observed sample are shown in Figure~\ref{fig:obs_beta_rhalf}. In the tests of the mock data in predicting this parameter, the inferences had limited accuracy and tended to understate the true value of $\beta$ where the value was positive, although the accuracy was better where the star populations were the least embedded. Here, the inferences for Draco and Carina are for positive anisotropy, and as neither of the two are particularly deeply embedded (see Figure~\ref{fig:obs embeddedness}), and given that a bias lower will cause their true anisotropy to be even higher, it seems likely that Draco and Carina are indeed likely to have positive anisotropy at their half-light radii.

\FloatBarrier
\section{Embeddedness of the Observed Sample}
\label{app:embeddedness}

\begin{figure}
    \centering
    \caption{The posterior inference for $r_{1/2}/r_\mathrm{s}$, which indicates the degree to which the stellar population is embedded in its DM halo, for the observed sample. The 68\% confidence intervals are shown.}
    \label{fig:obs embeddedness}
    \includegraphics[width=0.9\textwidth]{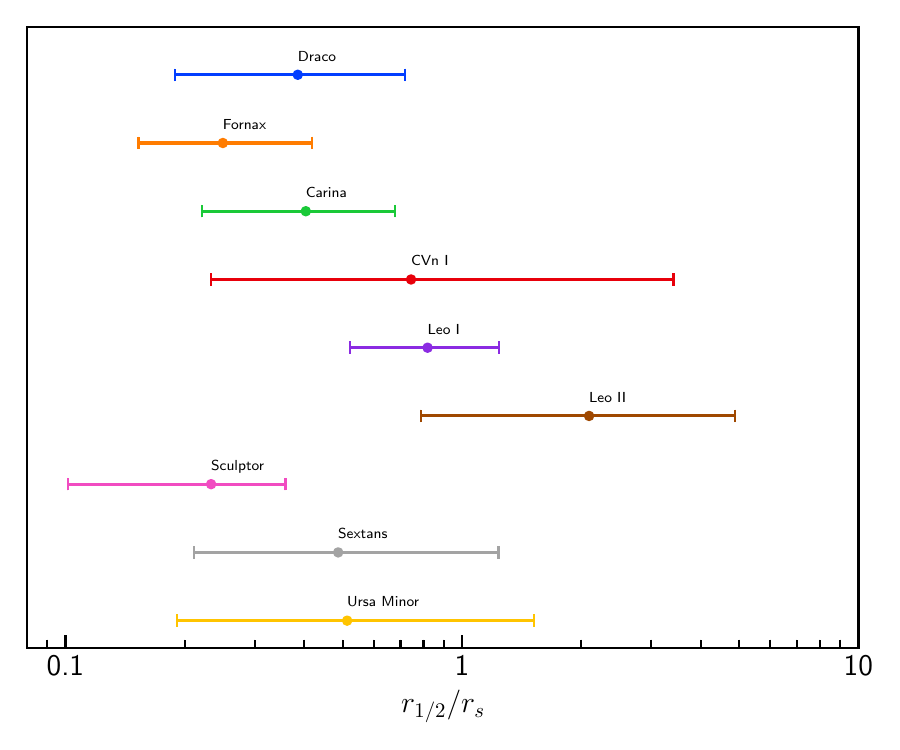}
\end{figure}

Because the degree to which the stellar population is embedded in the DM potential is an important factor for the accuracy of the model, we examine this in Figure~\ref{fig:obs embeddedness}, which shows $r_{1/2}/r_\mathrm{s}$ for the observed sample. Sculptor and Fornax have lowest inferences, with median values of 0.23 and 0.25, respectively. Though still less embedded than the most embedded mock data sets, the inferences for Sculptor and Fornax could be vulnerable to the types of biases seen in the most embedded mock data sets. 

\FloatBarrier
\section{Line Fit for Density vs. Pericenter}
\label{app:line fit}
\begin{figure}
    \centering
    \caption{Parameter posteriors and correlation for the best fit line of the form $\log_{10}[r_\mathrm{p} / \mathrm{kpc}]=f + g \log_{10}[\rho_{150} / (M_\odot \mathrm{kpc}^{-3})]$. The shaded regions indicate the 1, 2 and 3-$\sigma$ regions, respectively. The dotted lines indicate the the 15.9, 50, and 84.1-percentiles, respectively, from left to right.}
    \label{fig:f vs g triangle}
    \includegraphics[width=0.47\textwidth]{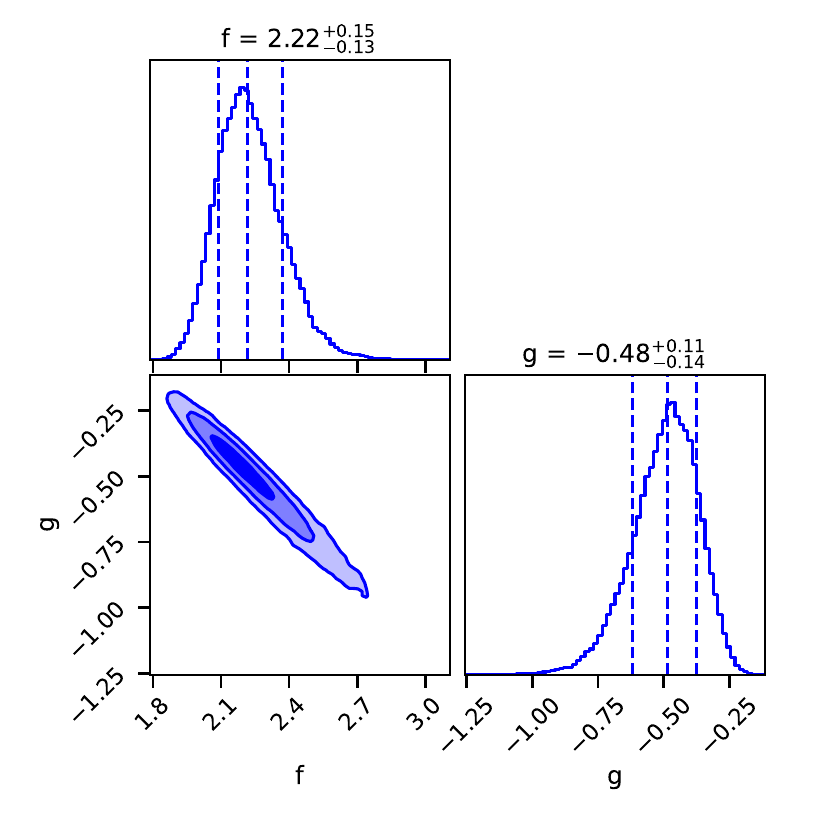}
\end{figure}

The power law fit for the density ($\rho_{150}$) vs. pericenter ($r_\mathrm{p}$) data was determined by fitting a line of the form $x=f + g y$, with f and g representing the intercept and slope, respectively. We have defined $x=\log_{10}[r_\mathrm{p} / \mathrm{kpc}]$, $y=\log_{10}[\rho_{150} / (10^7 M_\odot \mathrm{kpc}^{-3})]$, with $\delta \! x$ and $\delta \! y $ corresponding to the uncertainties on x and y, respectively.  The fit was determined according to the likelihood
\begin{eqnarray}
    \log(\mathcal{L}) =- \frac{1}{2} \sum_i \left( \frac{(x_\mathrm{i} -f - g y_\mathrm{i})^2}{\sigma_\mathrm{i}^2} +\log(2 \pi \sigma_\mathrm{i}^2) \right),
\end{eqnarray}
where $\sigma_\mathrm{i}^2 = g^2 \delta \! y_\mathrm{i}^2 + \delta \! x_\mathrm{i}^2$. The posteriors are shown in Figure~\ref{fig:f vs g triangle}.

\FloatBarrier
\section{Pericenter vs. Density Comparisons for all Phat Elvis Halos}
\label{app:Pericenter vs density for all PE halos}

\begin{figure*}
    \centering
    \caption{The DM density at 150 pc ($\rho_{150}$) inferred from the DF fits for the bright MW dSphs vs. orbital pericenter distance, in red, compared to the host halos from the Phat Elvis simulation. For the simulated halos, the black circles represent the 10 subhalos with the largest $V_\mathrm{peak}$ (i.e., the largest $\vmax$ since their infall) that are currently more than 50 kpc from galactic center. The gray circles denote the 10th through 20th $V_\mathrm{peak}$ subhalos.}
    \label{fig: Phat Elvis grid}
    \includegraphics[width=0.93\textwidth]{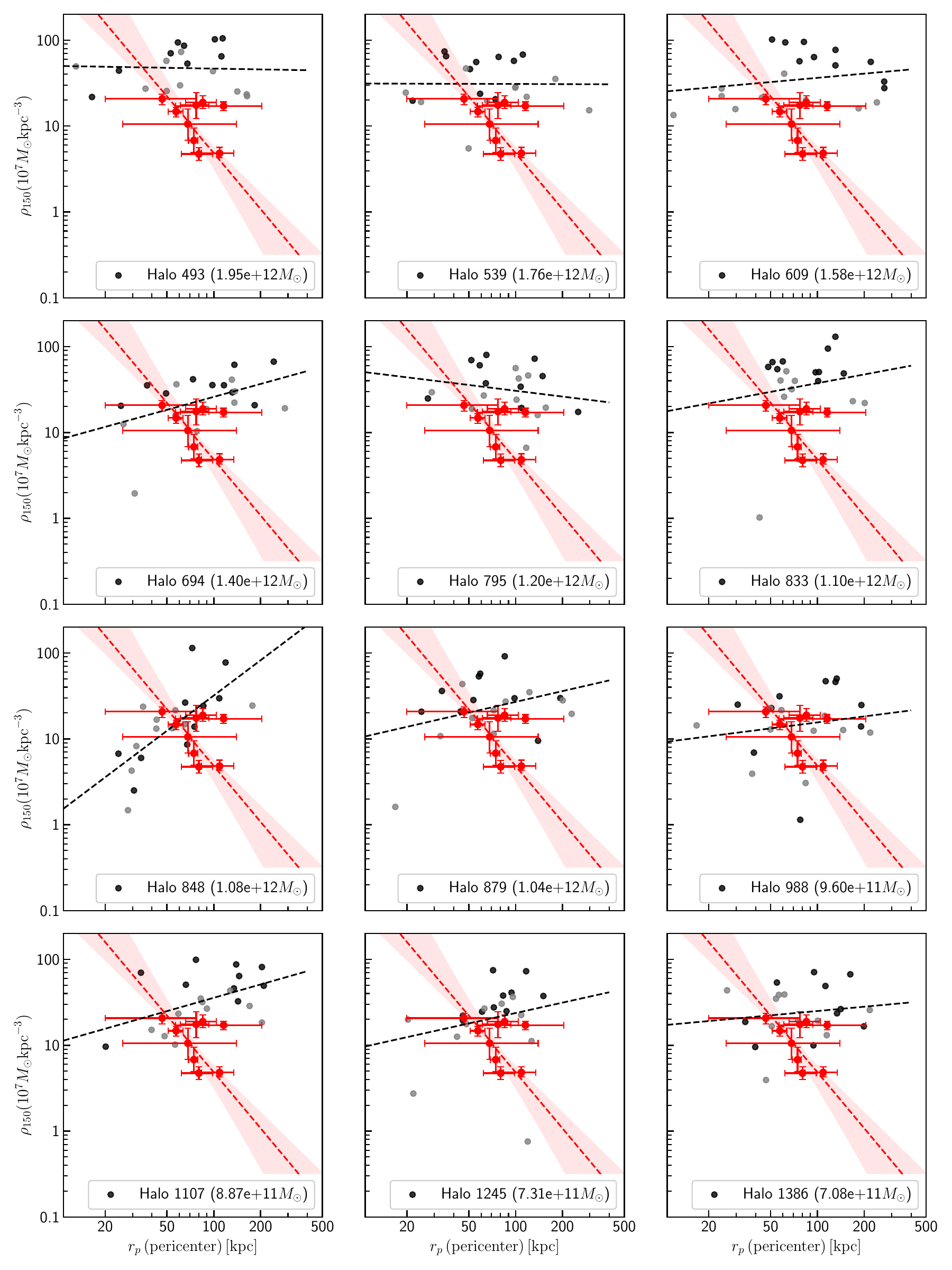}
\end{figure*}

Here we compare the DM density at 150 pc ($\rho_{150}$) versus orbital pericenter distance to each of the Phat Elvis halos \citep{Kelley2019}. We examine subhalos that are greater than 50 kpc from the galactic center, and show the 20 subhalos with the largest $V_{\textrm{peak}}$. Figure~\ref{fig: Phat Elvis grid} shows the regression for the bright MW dSphs in red, and the host halos from the Phat Elvis simulation in black and gray.

\FloatBarrier
\section{Comparison with Other Sources}
\label{app:comparisons}

\begin{figure}
    \centering
    \caption{DM density at 150 pc ($\rho_{150}$) inferred from the DF fits for the bright MW dSphs vs orbital pericenter distance ($r_\mathrm{p}$), in blue, similar to Figure~\ref{fig: PE pericenter comparison 1107 Battaglia} but now using the heavy MW model of \citet{Battaglia2022b} and comparing to host halo 609 of the Phat Elvis simulation \citep{Kelley2019}. The error bars indicate the 68\% confidence interval. The best fit line through the observations is shown in dashed blue, with the 68\% confidence interval in light blue. The black circles indicate the 10 subhalos with current radial positions greater than 50 kpc and with the largest $V_{\textrm{peak}}$ for host halo 609 of the Phat Elvis simulation. The gray circles denote the 10th through 20th largest $V_\mathrm{peak}$ subhalos. The best-fit regression line for the Phat Elvis points is shown as a black dashed line.}
    \includegraphics[width=0.55\textwidth]{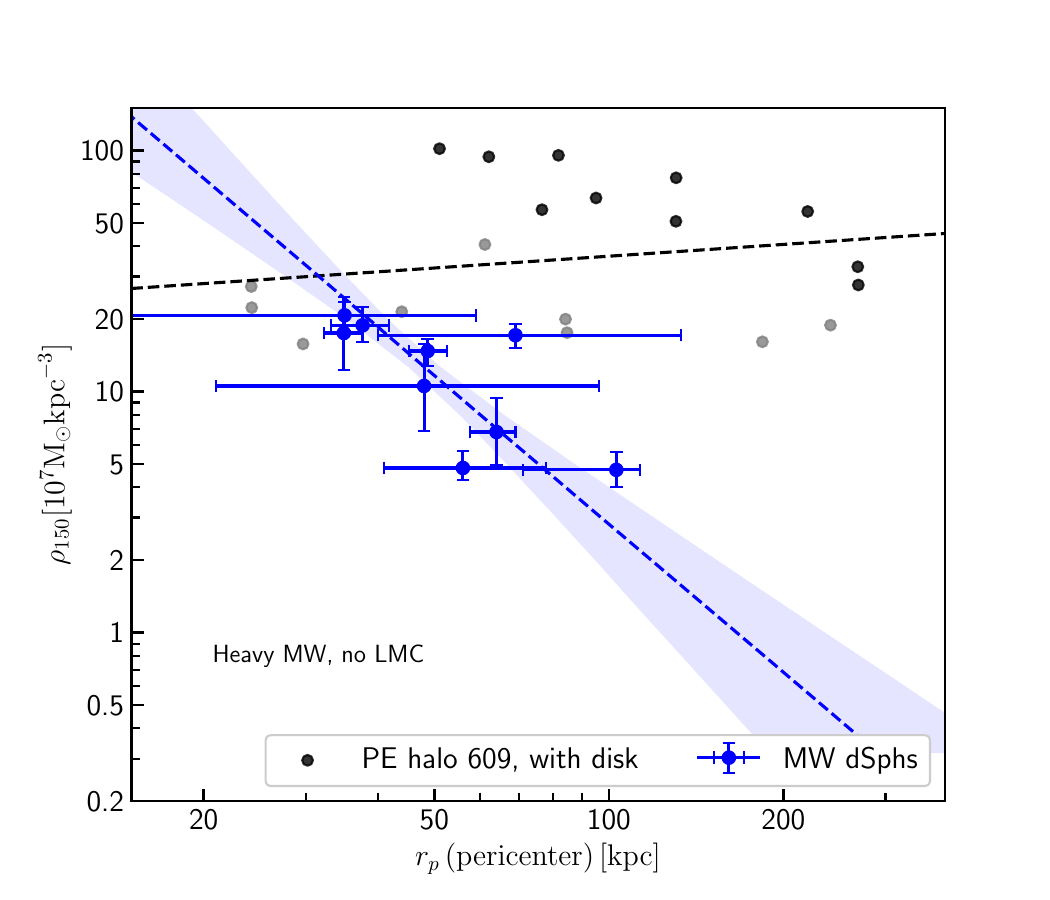}
    \label{fig: PE pericenter comparison 609 heavy}
\end{figure}

To address the concern that the heavier MW models of \citet{Battaglia2022b} might result in a different result as to the anti-correlation between density and pericenter, we examine the heavier case in Figure~\ref{fig: PE pericenter comparison 609 heavy}. In that figure we use the pericenters from the heavier MW model, with mass $\num{1.6e12} M_\odot$. The anti-correlation between $\rho_{150}$ and pericenter is evident. The black and grey dots in the figure are from Phat Elvis halo 609, which has a mass of $\num{1.58e12} M_\odot$, the most similar to that of the heavy MW model.

\begin{figure}
    \centering
    \caption{Comparison of recent orbital pericenter data sets. The values used in the main analysis of this work are those of \citet{Battaglia2022b}(light MW, no LMC) noted with **.}
    \includegraphics[width=0.6\textwidth]{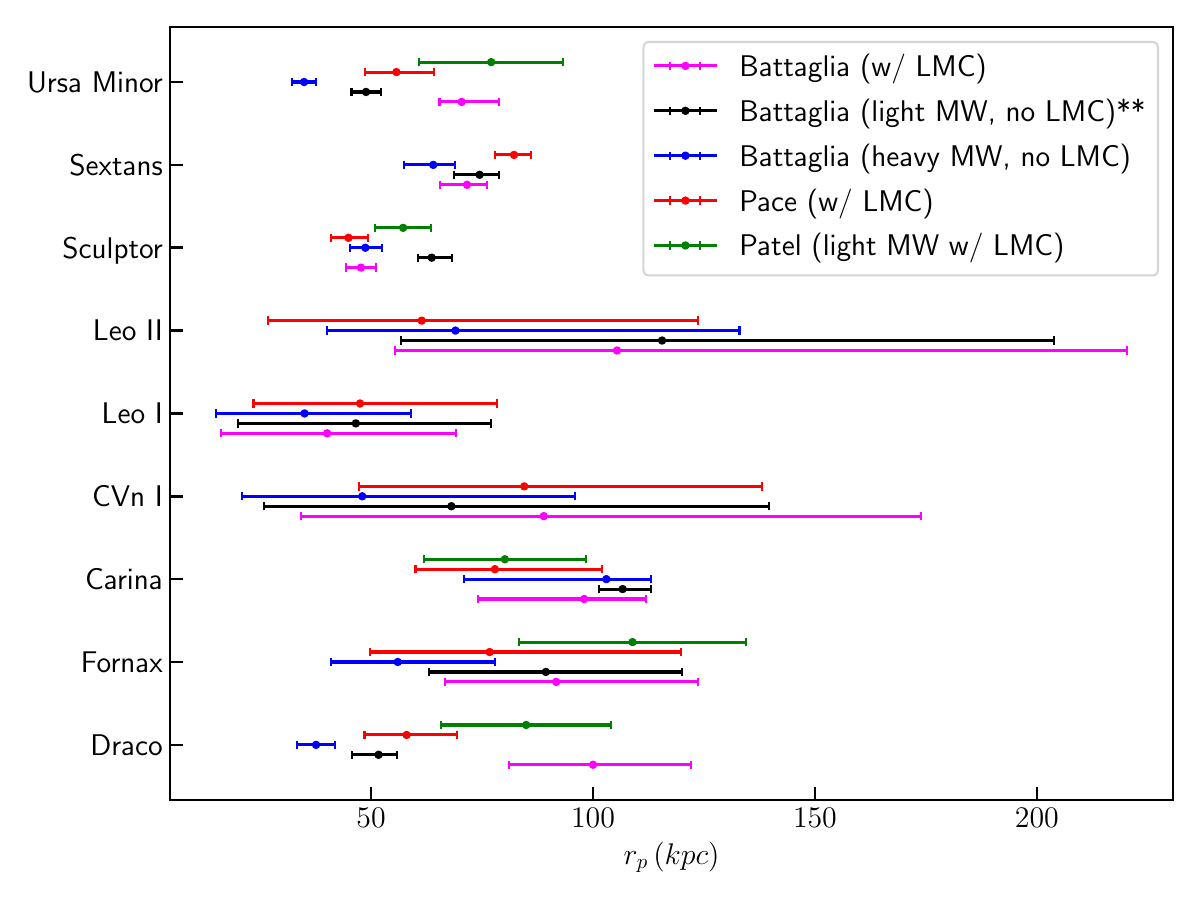}    
    \label{fig:peri data comparison}
\end{figure}

In Figure~\ref{fig:peri data comparison} we compare orbital pericenter data from \citet{Patel2020}, \citet{Battaglia2022b} and \citet{Pace2022}. We note the version we adopt in our main analysis in black. The data sets are fairly consistent given their stated uncertainties. The most tension appears in the projections for Sculptor, which has the smallest error bars of the nine dSphs.

In \citet{Cardona2023}, the authors closely examine the correlation of $\rho_{150}$ with pericenter between a variety of data sets. For central densities, they use the density results of \citet{Kaplinghat2019}, \citet{Read2019}, and \citet{Hayashi2020}. For pericenter distance they use the data of \citet{Fritz2018}, \citet{Battaglia2022b} and \citet{Pace2022}, some of which have different mass assumptions for the MW and may or may not attempt to account for the LMC. They conclude that the anti-correlation appears statistically significant in some combinations of data sets but not others. Specifically, they find that the \citet{Kaplinghat2019} density data yields a substantial correlation, the \citet{Hayashi2020} data lead to weak correlation, and that the uncertainties in the density are a key determinant of the strength of the correlation. This lends support to our results since the $\rho_{150}$ uncertainties are generally smaller than those of other results. Figure~\ref{fig:rho_150 comparison} compares those results, and we point out that we assess an uncertainty which is much smaller than that obtained by \citet{Hayashi2020}, which used a completely different method. The caveat here is that spherical symmetry needs to be a good working hypothesis for dSph DM halos, as our analysis assumes such symmetry but the analysis of \citet{Hayashi2020} does not. Furthermore, \citet{Cardona2023} find that the data is better described by models in which the central density $\rho_{150}$ decreases as function of $r_\mathrm{p}$, which contrasts with most of the Phat Elvis simulations (Figure~\ref{fig: Phat Elvis grid}).

\FloatBarrier
\section{Pericenter Reprojections for Phat Elvis Subhalos}
\label{app:Pericenter reprojection}

Here we examine the impact of using static, axisymmetric potentials and z=0 initial conditions to project subhalo pericenter distances, as is done in \citet{Patel2020}, \citet{Battaglia2022b} and \citet{Fritz2018}. Our subhalo sample is constructed by starting with the subhalos of all 12 host halos from the Phat Elvis simulation \citep{Kelley2019}, then selecting the 20 subhalos with the largest $V_\textrm{peak}$ from each host family that are within the host's virial radius. As a starting point for orbital integration, we use the positions and velocities of the subhalos at z=0, and project backwards in time to find pericenter. The orbital integrations are done in the \texttt{galpy} software package \citep{Bovy2015}. For each of the 12 host halos, the potential is based on the "MWPotential2014" potential of \texttt{galpy}, which was obtained by fitting to a wide variety of data on the MW. It is the sum of three components: (i) a "Power Spherical Potential with Cutoff" with mass $\num{4.5e9} M_\odot$, (ii) a Miyamoto Nagai Potential with mass $\num{6.81e10} M_\odot$, and (iii) an NFW potential. For our purposes, the NFW component's mass is adjusted so that the sum of the masses of the modeled potentials are the same as that of the corresponding Phat Elvis host halo. 

\begin{figure}
    \centering  
    \caption{Comparison of the true pericenter distances (y-axis) with projections of pericenter using a static, axisymmetric potential (x-axis) for the subhalos with the largest $V_\textrm{peak}$ in the 12 Phat Elvis hosts. The solid black line indicates equality, the dashed blue line indicates 25\% error, and the dotted red line indicates 50\% error.}
    \label{fig: pericenter reprojection comparison}
    \includegraphics[width=0.48\textwidth]{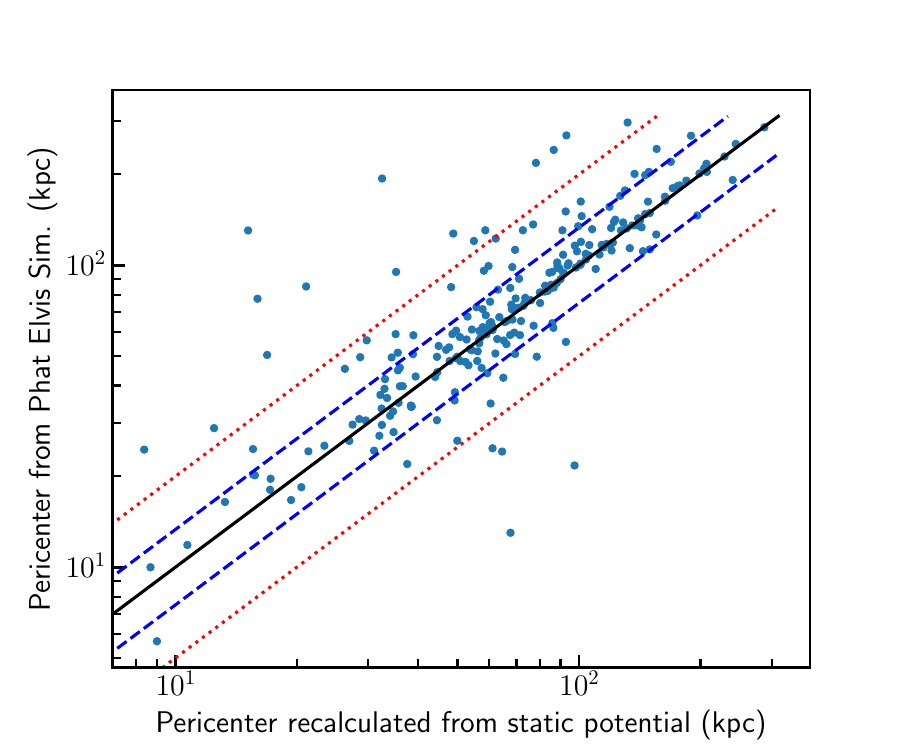}
\end{figure}

\begin{figure}
    \centering
    \caption{A histogram of the error between the reprojected pericenters and the true values from the Phat Elvis simulation, in dex (i.e., $\log_{10}[\textrm{peri}_\textrm{reproj} / \textrm{peri}_\textrm{true}]$). The 16th, 50th and 84th percentile values are indicated with vertical lines.}
    \label{fig: pericenter reprojection error}
    \includegraphics[width=0.48\textwidth]{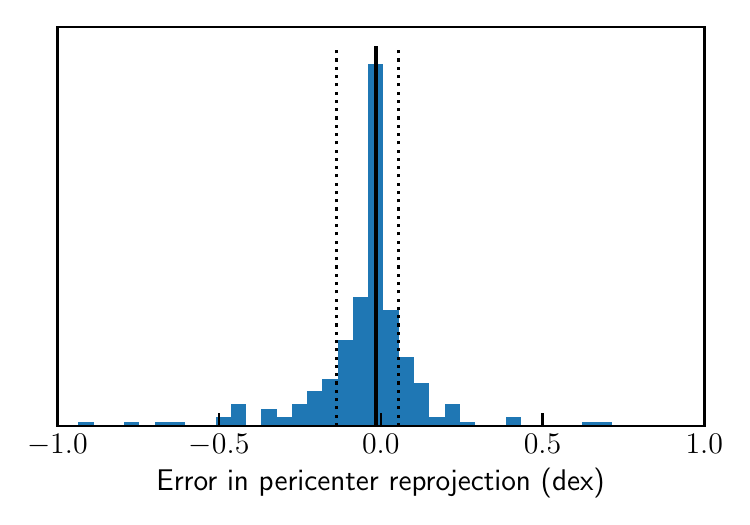}
\end{figure}

Figure~\ref{fig: pericenter reprojection comparison} compares the reprojections of pericenter using the method described above to the true values for the Phat Elvis subhalos. In general, there is good agreement with the true value, although the reprojections exhibit a mild tendency to be underestimated, and there are more outliers on the side of underprojection. The 16th, 50th, and 84th percentile values of the errors are -0.14 dex, -0.01 dex and 0.05 dex, respectively, as shown in the histogram of Figure~\ref{fig: pericenter reprojection error}.

\FloatBarrier
\section{Possible Effects from Tidal Truncation}
\label{app:Tidal truncation}

As a satellite galaxy falls into the potential well of our Galaxy, the outer part of the satellite's DM halo may become stripped by tidal forces, resulting in a DM profile that may not be well-modeled by the cNFW profile. To investigate the possible effects of this tidal truncation we calculate the radius of truncation based on the method in \citet{Jiang2021}, Equation S9, which is
\begin{eqnarray}
    r_\mathrm{t}^3 \approx r^3 \, \Big(m_{\mathrm{sat}}(r_\mathrm{t}) / M_{\mathrm{host}}(r)\Big)\Big/\left(2- d \ln{M_\mathrm{host}}/d \ln{r} + v^2_\mathrm{tan}/v^2_\mathrm{circ} \right) \ ,
\end{eqnarray}
where $r_\mathrm{t}$ is the tidal truncation radius of the satellite, $m_{\mathrm{sat}}(r_\mathrm{t})$ is the mass of the satellite within the tidal truncation radius, r is the distance of the satellite from the host center, $M_{\mathrm{host}}(r)$ is the mass of the host galaxy within radius r, $v_\mathrm{tan}$ is the tangential velocity of the satellite , and $v_\mathrm{circ}$ is the circular velocity of the host potential at the radius in question. For $m_{\mathrm{sat}}(r_\mathrm{t})$ we compute the mass using the cNFW profile. For $M_{\mathrm{host}}(r)$ we use the "MWPotential2014" model in the \texttt{galpy} software package \citep{Bovy2015}.

\begin{table}
\centering
	\caption{Calculated tidal radii for the bright MW dSphs,  as compared to $\rmax$ and in kpc. The median posterior value and the 68\% confidence intervals are indicated.}
	\renewcommand{\arraystretch}{1.3}
	\begin{tabular}{lcc} 
	\hline
	   MW dSph Name & $r_\mathrm{t}/\rmax$ & $r_\mathrm{t} (\mathrm{kpc})$  \\
	\hline
	Draco           & $1.40^{+0.89}_{-0.56}$ & $3.61^{+2.28}_{-1.36}$ \\
Fornax          & $1.48^{+0.66}_{-0.63}$ & $20.00^{+10.20}_{-8.59}$ \\
Carina          & $1.60^{+0.36}_{-0.47}$ & $10.69^{+3.60}_{-3.21}$ \\
CVn I           & $2.22^{+3.80}_{-1.65}$ & $5.04^{+6.63}_{-3.89}$ \\
Leo I           & $3.14^{+5.19}_{-2.20}$ & $4.44^{+7.54}_{-3.10}$ \\
Leo II          & $4.79^{+10.75}_{-3.47}$ & $3.62^{+7.81}_{-2.62}$ \\
Sculptor        & $1.66^{+0.53}_{-0.51}$ & $7.09^{+3.00}_{-1.76}$ \\
Sextans         & $1.29^{+0.77}_{-0.42}$ & $4.67^{+1.77}_{-1.20}$ \\
Ursa Minor      & $1.34^{+1.20}_{-0.59}$ & $3.73^{+2.66}_{-1.46}$ \\
	\end{tabular}
	\label{tab:tidal radius}
\end{table}

For our purposes, we wish to calculate the truncation radius at pericenter, as any satellite that has made at least one pericenter passage will have been truncated to the maximum extent. To obtain the tangential velocity at pericenter, we assume that the satellite's motion conserves angular momentum about the galactic center, so that $v_\mathrm{tan}(r_\mathrm{p}) = v_\mathrm{tan}(r_0) r_0 / r_\mathrm{p}$. For consistency, we use \citet{Fritz2018} Tables 2 and 3 for the values of $v_\mathrm{tan}(r_0)$, $r_0$ and $r_\mathrm{p}$, corresponding to their Milky Way mass model with mass $\num{0.8e12}  M_{\odot}$. The resulting truncation radii $r_\mathrm{t}$ and the ratios for $r_\mathrm{t} / \rmax$ for the observed sample are shown in Table~\ref{tab:tidal radius}. None of the satellites in the sample are severely truncated. Relative to their $\rmax$, the most truncation occurs in Sextans and Ursa Minor.

As a check on the possible impact of tidal truncation on our models, we modeled an abruptly truncated halo with truncation radius of 1.5 $\rmax$. That is, we define a new profile as follows:
\begin{equation}
\label{eq:tcNFW}
    \rho(r)=
    \begin{cases}
        \rho_\mathrm{NFW}(r), & r<=1.5\rmax, \\
		0, & r>1.5\rmax\\
	\end{cases}
\end{equation}
We chose $1.5\rmax$ as a test value because several of the observed dSphs in Table~\ref{tab:tidal radius} have truncation radii approximating that value. We reran the DF model using this density profile and its associated potential for three of the observed dSphs: Draco, Sextans and Ursa Minor. We found that the impact of the truncation is modest, with $\rmax$ increasing 0.1 to 0.2 dex, $\vmax$ increasing typically $\approx 0.1$ dex, and $\rho_{150}$ decreasing 0 to 0.2 dex. A more robust approach would be to allow the truncation radius to be a varying parameter in the model; we hope to do so in future work. At present, we are satisfied that abrupt tidal truncation at 1.5$\rmax$ does not seem to have a strong impact on our inferences.

\newpage 
\FloatBarrier
\section{Mock Data Characteristics}
In the Table~\ref{tab:mock data characteristics} we report the main ingredients characterizing the mock dataset analyzed in our study.
\label{app:mock data info}
\begin{table}
\caption{Mock data characteristics. The columns, from left, are (1) ID number, (2) ID string, (3) the number of stars in the data set, (4) DM profile type, (5) log slope of the inner stellar profile, (4) scale radius of the stellar profile, (6) anisotropy radius of the stellar profile, (7) 3D half-light radius, and (8) virial shape parameter of the data set.}
\label{tab:mock data characteristics}
\centering
\renewcommand{\arraystretch}{1.5}
{\footnotesize
\begin{tabular}{lcccccccccccccc} 
\hline
   Number & ID & w & DM Profile & $\gamma_*$ & $r_*$ (kpc) & $ r_a$ (kpc) & $r_{1/2}$ (kpc)  & VSP ($10^3 \; \mathrm{km}^4/\mathrm{sec}^4$) \\
\hline
    1 & aaaO-4639 & 4639 & cored & 0.1 & 0.1 &      1 & 0.122 & $   4.25_{-0.75}^{+1.59}$ \\
 2 & aabO-4941 & 4941 & cored & 0.1 & 0.1 &  10000 & 0.119 & $   3.08_{-0.43}^{+0.62}$ \\
 3 & abaO-1801 & 1801 & cored & 0.1 & 0.25 &      1 & 0.310 & $  25.87_{-3.95}^{+6.29}$ \\
 4 & abbO-5483 & 5483 & cored & 0.1 & 0.25 &  10000 & 0.300 & $  23.00_{-1.79}^{+2.20}$ \\
 5 & acaO-3904 & 3904 & cored & 0.1 & 0.5 &      1 & 0.596 & $  94.09_{-7.99}^{+10.67}$ \\
 6 & acbO-2607 & 2607 & cored & 0.1 & 0.5 &  10000 & 0.588 & $  84.75_{-9.30}^{+12.34}$ \\
 7 & adaO-1980 & 1980 & cored & 0.1 & 1.0 &      1 & 1.233 & $ 231.66_{-22.09}^{+25.55}$ \\
 8 & adbO-1441 & 1441 & cored & 0.1 & 1.0 &  10000 & 1.251 & $ 172.54_{-14.22}^{+15.29}$ \\
 9 & baaO-1826 & 1826 & cored & 1.0 & 0.1 &      1 & 0.093 & $   2.89_{-0.80}^{+2.26}$ \\
10 & babO-2156 & 2156 & cored & 1.0 & 0.1 &  10000 & 0.090 & $   1.80_{-0.40}^{+0.72}$ \\
11 & bbaO-1776 & 1776 & cored & 1.0 & 0.25 &      1 & 0.238 & $  17.31_{-3.04}^{+5.25}$ \\
12 & bbbO-3368 & 3368 & cored & 1.0 & 0.25 &  10000 & 0.227 & $  15.07_{-1.87}^{+2.67}$ \\
13 & bcaO-2107 & 2107 & cored & 1.0 & 0.5 &      1 & 0.463 & $  70.06_{-9.49}^{+12.83}$ \\
14 & bcbO-2349 & 2349 & cored & 1.0 & 0.5 &  10000 & 0.464 & $  50.70_{-5.63}^{+8.25}$ \\
15 & bdaO-2677 & 2677 & cored & 1.0 & 1.0 &      1 & 0.913 & $ 164.09_{-14.81}^{+17.59}$ \\
16 & bdbO-2456 & 2456 & cored & 1.0 & 1.0 &  10000 & 0.914 & $ 113.42_{-8.47}^{+8.75}$ \\
17 & aaaN-2358 & 2358 & NFW   & 0.1 & 0.1 &      1 & 0.121 & $   3.73_{-0.39}^{+0.45}$ \\
18 & aabN-3539 & 3539 & NFW   & 0.1 & 0.1 &  10000 & 0.122 & $   3.04_{-0.21}^{+0.23}$ \\
19 & abaN-2975 & 2975 & NFW   & 0.1 & 0.25 &      1 & 0.294 & $  12.40_{-1.04}^{+1.23}$ \\
20 & abbN-4239 & 4239 & NFW   & 0.1 & 0.25 &  10000 & 0.300 & $   9.06_{-0.47}^{+0.52}$ \\
21 & acaN-1088 & 1088 & NFW   & 0.1 & 0.5 &      1 & 0.600 & $  20.78_{-2.22}^{+2.64}$ \\
22 & acbN-550  &  550 & NFW   & 0.1 & 0.5 &  10000 & 0.603 & $  15.12_{-2.17}^{+2.46}$ \\
23 & adaN-1860 & 1860 & NFW   & 0.1 & 1.0 &      1 & 1.238 & $  30.86_{-2.96}^{+3.40}$ \\
24 & adbN-826  &  826 & NFW   & 0.1 & 1.0 &  10000 & 1.226 & $  23.92_{-2.76}^{+3.38}$ \\
25 & baaN-1533 & 1533 & NFW   & 1.0 & 0.1 &      1 & 0.096 & $   2.96_{-0.36}^{+0.49}$ \\
26 & babN-1491 & 1491 & NFW   & 1.0 & 0.1 &  10000 & 0.092 & $   2.43_{-0.28}^{+0.37}$ \\
27 & bbaN-1214 & 1214 & NFW   & 1.0 & 0.25 &      1 & 0.238 & $   6.68_{-0.65}^{+0.75}$ \\
28 & bbbN-1153 & 1153 & NFW   & 1.0 & 0.25 &  10000 & 0.224 & $   7.27_{-0.71}^{+0.76}$ \\
29 & bcaN-2054 & 2054 & NFW   & 1.0 & 0.5 &      1 & 0.453 & $  16.46_{-1.40}^{+1.61}$ \\
30 & bcbN-1222 & 1222 & NFW   & 1.0 & 0.5 &  10000 & 0.434 & $  14.89_{-1.38}^{+1.57}$ \\
31 & bdaN-2912 & 2912 & NFW   & 1.0 & 1.0 &      1 & 0.953 & $  25.33_{-1.80}^{+2.03}$ \\
32 & bdbN-1524 & 1524 & NFW   & 1.0 & 1.0 &  10000 & 0.925 & $  24.24_{-2.07}^{+2.18}$ \\
\hline
\end{tabular}
}
\end{table}


\label{lastpage}
\end{document}